\documentclass[10pt,journal,cspaper,compsoc]{IEEEtran}
\usepackage{graphicx}   % extended graphics package
\usepackage{amsmath}    % math package
\usepackage{cases}
\usepackage{hhline}     % extended styles for tables

\usepackage[font=footnotesize,labelfont=bf]{caption}

\makeatletter
\def\ps@headings{%
\def\@oddhead{\mbox{}\scriptsize\rightmark \hfil \thepage}%
\def\@evenhead{\scriptsize\thepage \hfil \leftmark\mbox{}}%
\def\@oddfoot{}%
\def\@evenfoot{}}

\newtheorem{theorem}{Theorem}%[section]

\begin{document}

\title{LIPS: A Light Intensity Based Positioning System For Indoor Environments}

\author{Bo Xie,
        Guang~Tan,~\IEEEmembership{Member,~IEEE,}
        Yunhuai~Liu,~\IEEEmembership{Member,~IEEE,}
        Mingming~Lu,~\IEEEmembership{Member,~IEEE,}
        Kongyang~Chen,
        and~Tian~He,~\IEEEmembership{Senior Member,~IEEE}% <-this % stops a space
\IEEEcompsocitemizethanks{\IEEEcompsocthanksitem Bo Xie, Guang Tan
and Kongyang Chen are with
SIAT, Chinese Academy of Sciences. E-mail: \{bo.xie, guang.tan, ky.chen\}@siat.ac.cn.\protect\\
\IEEEcompsocthanksitem Yunhuai Liu is with Third Research Institute of the Ministry of Public Security, China. E-mail: yunhuai.liu@gmail.com. \protect\\
\IEEEcompsocthanksitem Mingming Lu is with Central South University, China. E-mail: ming.lu@gmail.com. \protect\\
\IEEEcompsocthanksitem Tian He is with University of
Minnesota, U.S. E-mail: tianhe@cs.umn.edu.}% <-this % stops a space
\thanks{}}
%
%
%\numberofauthors{1}
%\author{
%\alignauthor Bo Xie$^1$, Guang Tan$^1$, Yunhuai Liu$^2$, Mingming Lu$^3$, Tian He$^4$\\
%\affaddr{$^1$SIAT, Chinese Academy of Sciences, China.}\\
%\affaddr{$^2$Third Research Institute of the Ministry of Public Security, China} \\
%\affaddr{$^3$Central South University, China, $^4$University of
%Minnesota, U.S.}\\\affaddr{ \{bo.xie,guang.tan\}@siat.ac.cn,
%\{yunhuai.liu,ming.lu\}@gmail.com, tianhe@cs.umn.edu}}

\maketitle

\begin{abstract}

This paper presents {\em LIPS}, a Light Intensity based Positioning
System for indoor environments. The system uses off-the-shelf LED
lamps as signal sources, and uses light sensors as signal receivers.
The design is inspired by the observation that a light sensor has
{\em deterministic sensitivity} to both distance and incident angle
of light signal, an under-utilized feature of photodiodes now widely
found on mobile devices. We develop a stable and accurate light
intensity model to capture the phenomenon, based on which a new
positioning principle, {\em Multi-Face Light Positioning} (MFLP), is
established that uses three collocated sensors to uniquely determine
the receiver's position, assuming merely a {\em single} source of
light. We have implemented a prototype on both dedicated embedded
systems and smartphones. Experimental results show average
positioning accuracy within 0.4 meters across different
environments, with high stability against interferences from
obstacles, ambient lights, temperature variation, etc.
\end{abstract}

\section{Introduction}\label{sec:introduction}

%
%The radio signal propagation is dependent on the physical
%environment and varies dramatically from location to location.
%Constructive and destructive interference adds further errors to any
%static RSS measurements.

%Being mainly line of sight, the RSS quite predictably attenuates
%according to a square law in free space.  The received signal also
%does not suffer from constructive and destructive interference.

The prospect of indoor location based services (LBS) and the
unfortunate unavailability of GPS signals indoors have fueled much
interest in indoor positioning techniques recently. Among the
numerous approaches to achieving such a service, the radio-frequency
(RF) based positioning technique has attracted perhaps the most
attention, due to the wide deployment of WiFi access points. A major
challenge faced by this approach is that RF is subject to serious
multipath effect and is vulnerable to environmental interferences.
This makes it difficult to establish an accurate propagation model
that allows accurate distance estimation. Recent work has focused on
the RF fingerprint approach, which obviates the need of a
propagation model, but then requires manual efforts to establish a
fingerprint database in support of mapping from signal strengths to
positions. The fingerprint collection process is often laborious,
leading to various research efforts to reduce the cost
(e.g.,~\cite{no-pain,lifs}). Despite significant advances made in
this direction, a fully automatic solution for general indoor
environments has remained open.

\begin{figure}[tb]
    \centering
    {\footnotesize
    %\quad
    \shortstack{
            \includegraphics[width=0.21\textwidth]{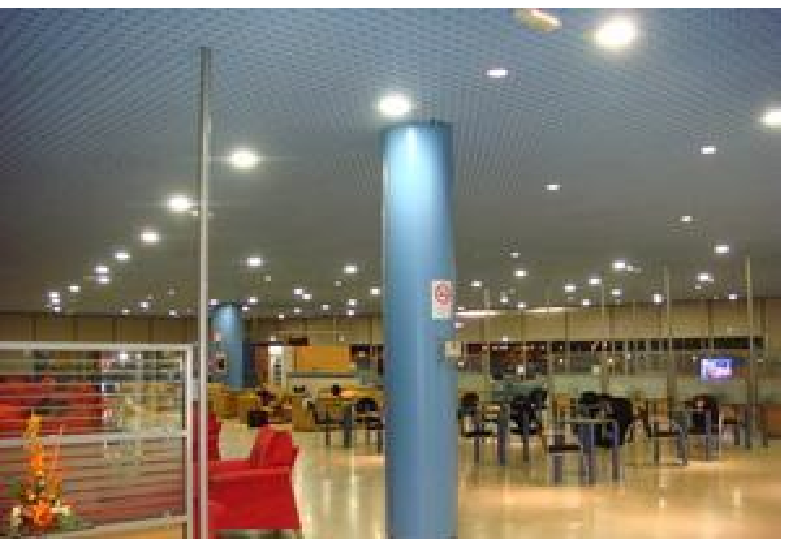}\\
            {(a)  }
    }\hspace{5pt}
    \shortstack{
            \includegraphics[width=0.2\textwidth]{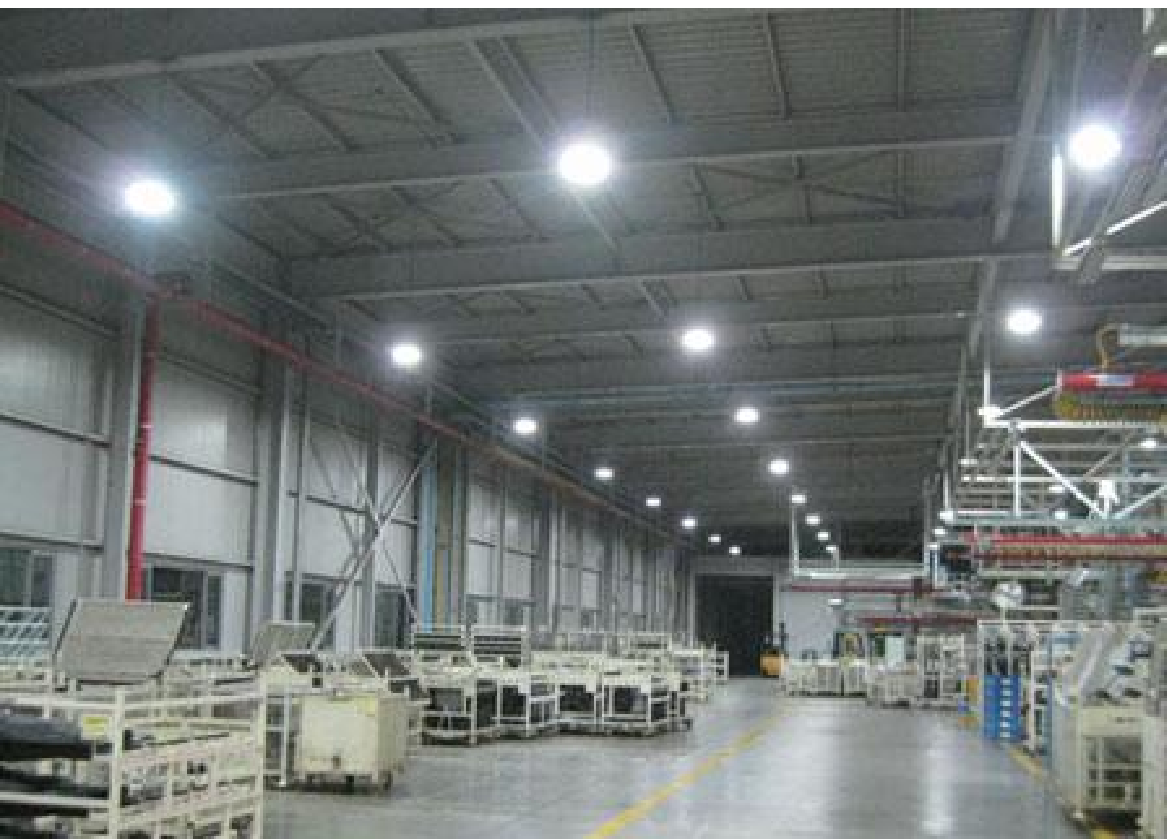}\\
            {(b) }
    }
    \shortstack{
            \includegraphics[width=0.15\textwidth]{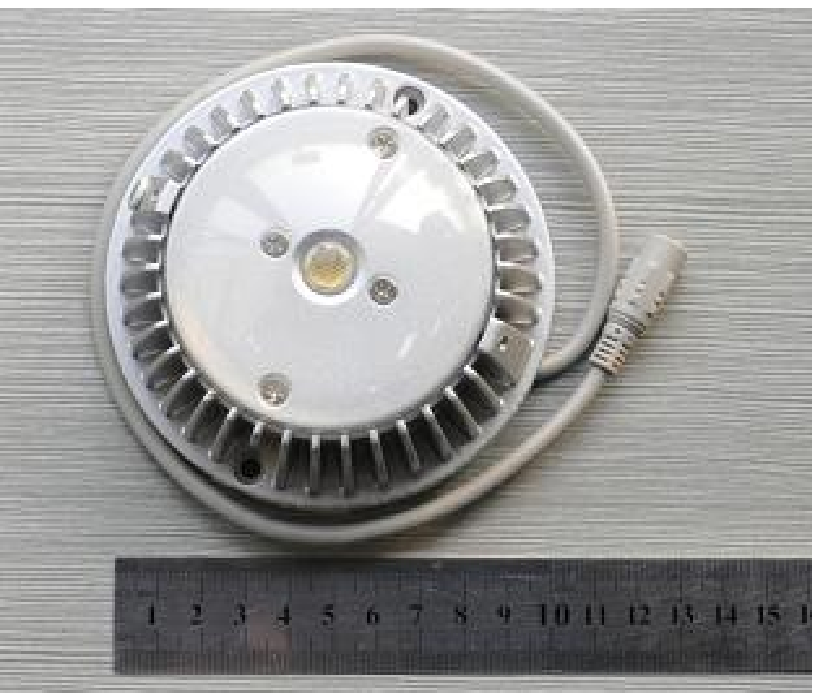}\\
            {(c) }
    }
    %\quad
    \shortstack{
            \includegraphics[width=0.15\textwidth]{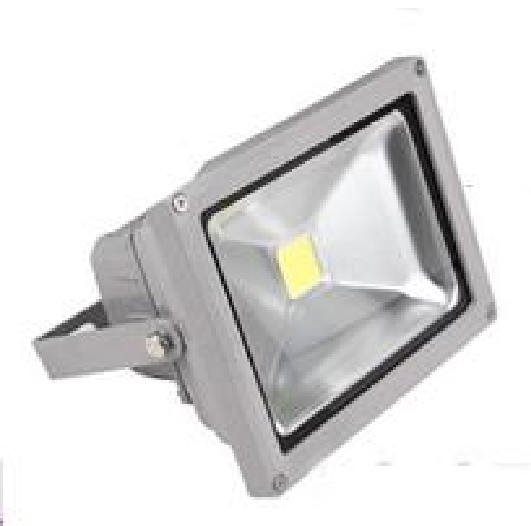}\\
            {(d) }
    }\hspace{-5pt}
    \shortstack{
            \includegraphics[width=0.13\textwidth]{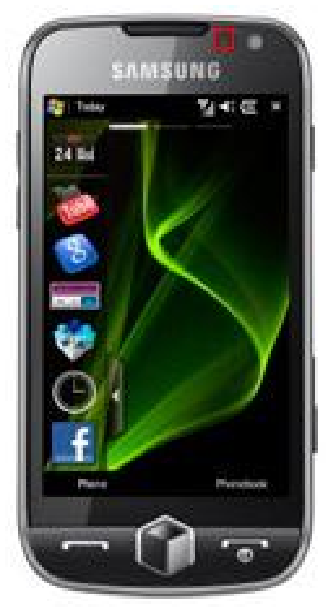}\\
            {(e) }
    }
    \caption{LED lamps and light sensors. (a) LED lighting in an airport
    terminal. (b) LED lighting in a large warehouse. (c) An infrared
    LED lamp, 8 watts. (d) A high-power visible light LED lamp, 100 watts. (e) A smartphone with a
    light sensor (enclosed by red square).}
    \label{fig:lights-sensors}
    }
\end{figure}

In this paper we explore an alternative approach, using visible or
infrared (IR) light signal rather than RF, to achieving indoor
positioning. The main advantage of light signal over RF is that the
propagation of light is more predictable than that of RF. This is
because on ordinary objects such as walls and furniture, the light
signal experiences only insignificant reflection, thus the signal at
a receiver is subject to negligible multipath effect. This largely
eliminates the uncertainty in characterization of received signal
strength (RSS), laying a sound basis for further modeling and
derivation of position.

Our system, called {\em LIPS}, uses commodity LED lamps (see
Figure~\ref{fig:lights-sensors}) as signal sources, and uses light
sensors available on mainstream mobile devices as signal receivers.
A low-end microcontroller is used to make the LED lamp switch on and
off at specified frequencies, so that the light signals from sources
of interest can be separated from ambient ones in the frequency
domain. The recovered light signal strength on the light sensor
reliably reflects the sensor's distance and orientation with respect
to the lamp. This allows one to establish an accurate RSS model
which paves the way to {\em fingerprint-free} positioning. From a
practical point of view, the LIPS design could re-use the existing
lighting infrastructure for indoor positioning in many public
environments, such as airport terminals
(Figure~\ref{fig:lights-sensors}(a)), warehouses
(Figure~\ref{fig:lights-sensors}(b)), shopping malls, and hospitals,
where lamps are extensively deployed. These lamps can conveniently
serve as positioning references, provided they are distinguishable
in the frequency domain with different flashing rates and have known
positions.

Compared to the RF-based approach, LIPS essentially trades off
obstacle penetration ability for improved predictability of signal
propagation. The design is centered around two questions: (1)
\textit{how accurate and stable is a light sensor in producing
position-related information}, and (2) \textit{how to exploit that
information for positioning while minimizing the line-of-sight
limitation of light signal?} To explore these issues we make three
contributions.

% \footnote{It is easier to control the lighting
%frequency of LED lamps (operating on DC) than to control fluorescent
%lamps (operating on AC), though the latter is technically possible.}
First, we conduct comprehensive experiments showing that a light
sensor can be used to infer not only distance, but also {\em
angular} information from light signal, with a highlight on its {\em
sensitivity} and {\em stability}. The angular information turns out
to be very useful for obtaining the position of a light sensor.
%
%The light sensor had been used for quite restricted purposes, such
%as user context classification and light frequency
%detection~\cite{flight}. Here we show that the sensor can provide
%richer information than shown in prior work, opening opportunities
%for new applications.

Second, we develop novel methods for positioning. In particular, we
propose a {\em Multi-Face Light Positioning} (MFLP) method, which
uses three collocated sensors to uniquely determine the receiver's
position, assuming merely a {\em single} source of light. This
single-source positioning method alleviates the concern of possible
high deployment density of light sources, especially in a complex
environment, where the light signal's line-of-sight restriction
makes it expensive to circumvent obstacles. We show that in such an
environment, MFLP requires far less than 1/3 (the theoretical ratio)
of the lights as required by trilateration for the same degree of
coverage, thus slashes the deployment cost to more accessible
levels. In a real environment, different lamps flash at different
rates, so they can be distinguished in the frequency domain.

Last, we present two designs for indoor positioning, one for a
dedicated receiver and the other for smartphones. We evaluate the
systems in various realistic environments and show that LIPS can
produce positioning accuracy below 0.4 meters on the average, and
that is stable across significant environmental variations.
%
%The next section establishes the light intensity model;
%Section~\ref{sec:mflp} describes the positioning principle;
%Section~\ref{sec:lips-embedded} and Section~\ref{sec:prototype}
%describe the implementation of two prototype systems;
%Section~\ref{sec:experiments} presents the experimental results;
%Section~\ref{sec:discussion} discusses possible extensions of the
%design, Section~\ref{sec:related} documents related work, and
%Section~\ref{sec:conclusion} concludes the paper.

\section{Light sensor characteristics}\label{sec:model}

\begin{figure}[tb]
    \centering
    {\footnotesize
    %\quad
    \shortstack{
            \includegraphics[width=0.21\textwidth]{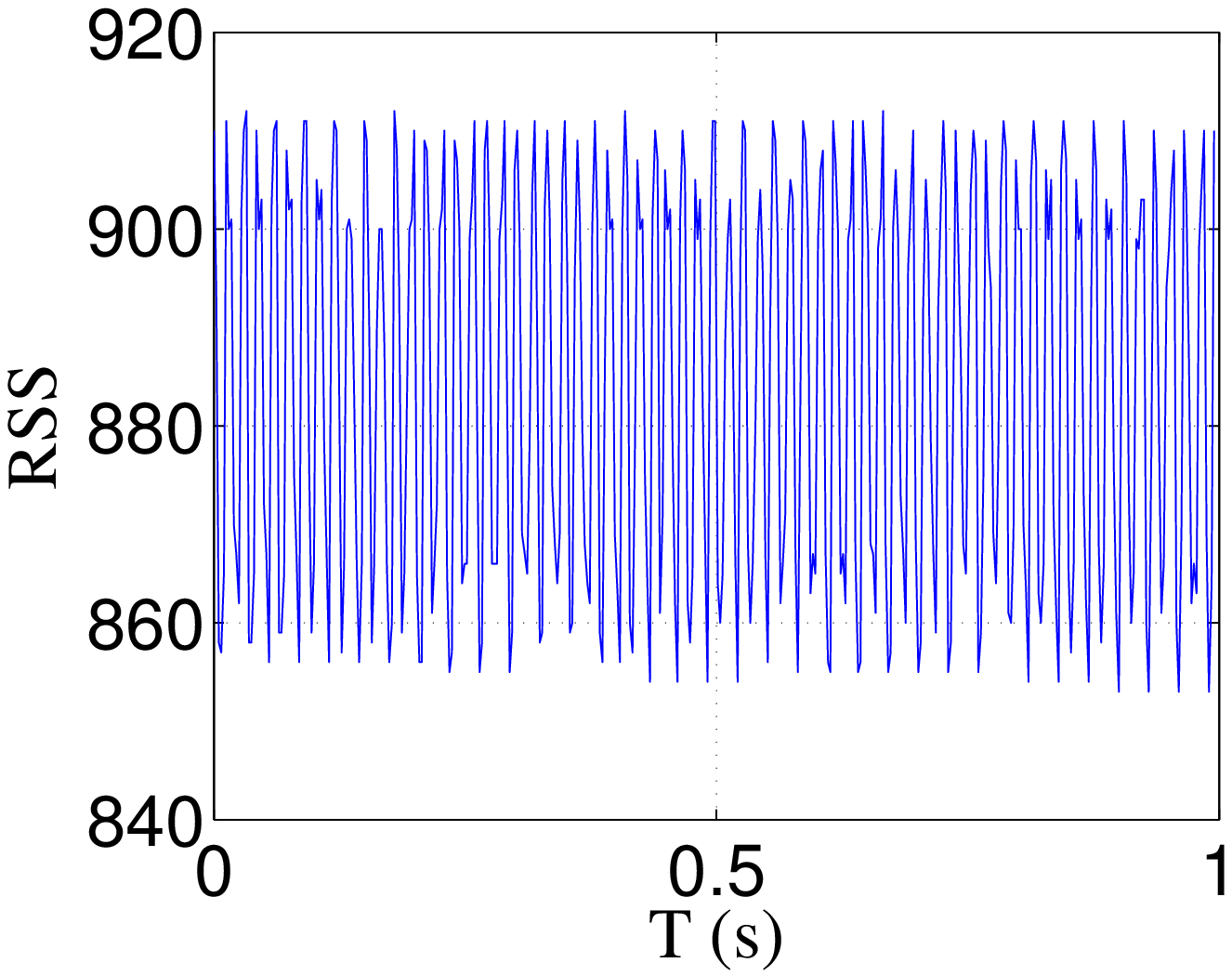}\\
            {(a) Time domain}
    }
    %\quad
    \shortstack{
            \includegraphics[width=0.23\textwidth]{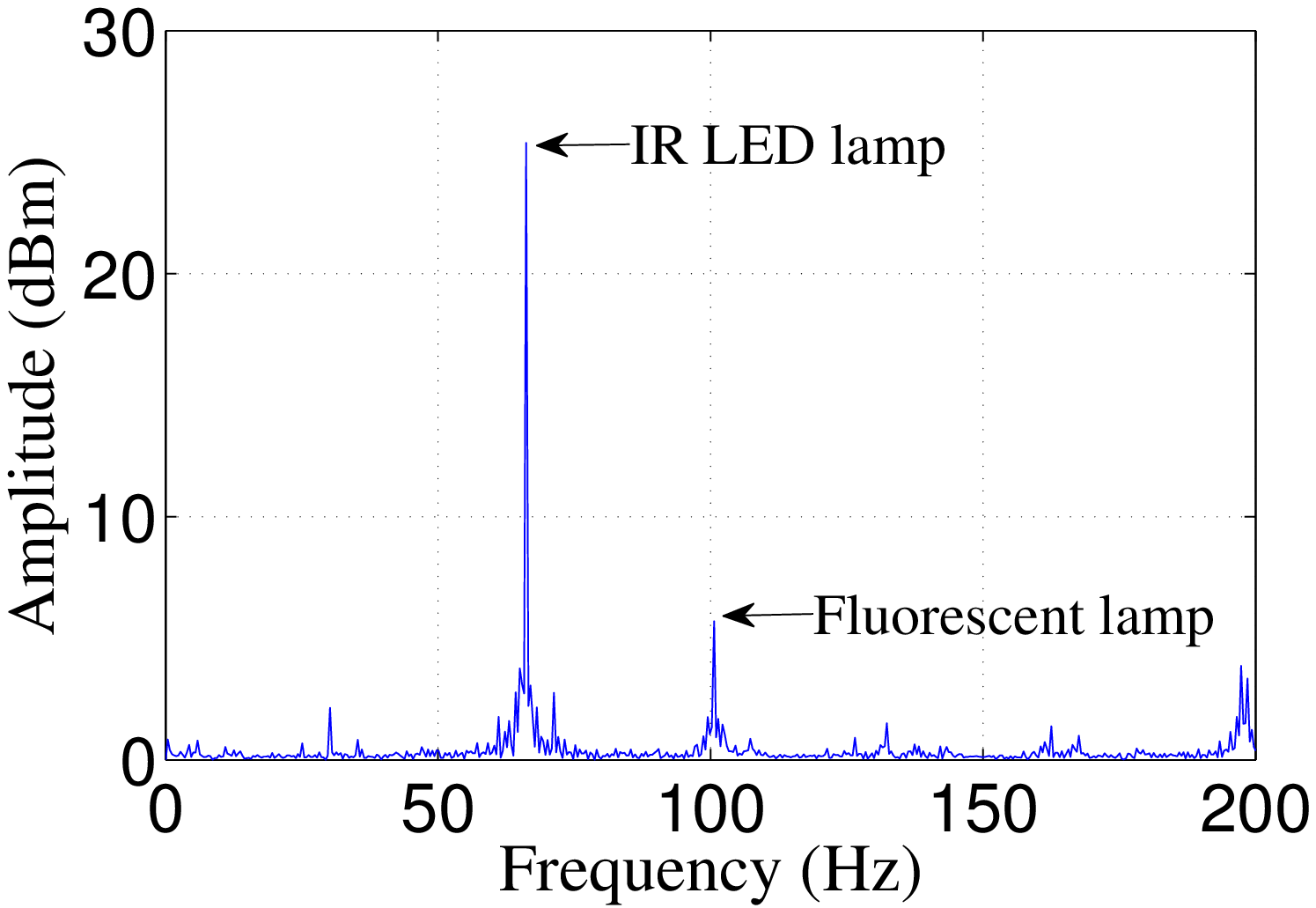}\\
            {(b) Frequency domain}
    }
    }
    \caption{RSS in the time domain and frequency domain (excluding the DC component). The LED lamp flashes at
    a frequency of 65 Hz, and the nearby fluorescent lamps flash at a
    frequency of 100 Hz.}
    \label{fig:time-freq}
\end{figure}

\begin{figure}[tb]
    \centering
    \includegraphics[width=0.34\textwidth]{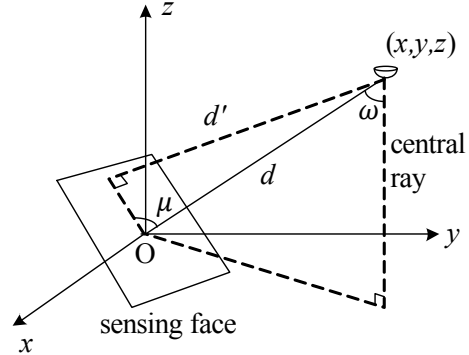}
    \caption{RSS model of the light sensor. The light sensor's surface (sensing face)
    is centered at $(0,0,0)$, while the lamp is located at $(x,y,z)$.
    The RSS of the sensor, $s$, is a function of distance $d$, incident
    angle $\mu$, and emitting angle $\omega$.}
    \label{fig:model}
\end{figure}

In order to examine the characteristics of a light sensor under
different light frequencies and power supplies, we considered two
typical types of LED lamps,

\begin{itemize}
  \item An IR LED lamp, with a 8 watts power rating and a 12 volts DC
  power supply (see Figure~\ref{fig:lights-sensors}(c)), as well as an
illuminating angle of 120 degrees. It has a sensing range of 7.5
meters for the sensor, covering an open area of 130 m$^2$ on the
ground when hung on a 3-meter high ceiling.
  \item A visible light LED lamp, with a 100 watts power rating
(see Figure~\ref{fig:lights-sensors}(d)), and a sensing range of 30
meters, covering 2000+ m$^2$ in open space when hung at a proper
height.
\end{itemize}

%Presently the two lamps cost about US\$30 and US\$80, respectively.
Each lamp has a small illuminating chip of size around $1.5$cm
$\times$ $1.5$cm, thus can be viewed as a point source of light from
a distance of a few meters. For space reasons we focus on the IR
lamp and only briefly report on the visible light one. In practice,
IR lamps can be deployed in environments where there already exist
lighting devices and extra visible lights are undesirable.

The light sensor used is an Intersil ISL29023\footnote{The ISL29023
is an integrated ambient and infrared light to digital
converter~\cite{isl29023}. The same family of sensors are also found
on other phones such as Motorola XT882.}, which is used by Samsung
Omnia II GT-I8000 smartphone. We used a stand-alone sensor connected
to a microcontroller for the experiment.
%
%have considered both stand-alone sensors (connected to a
%microcontroller) and smartphones. The results are similar, except
%that the smartphone permits a smaller effective incident angle, due
%to the sensor's being embedded into the phone body which blocks some
%light. In this section, we report only the results of a stand-alone
%sensor.

In reality, what the sensor receives is a mixture of the light
signals from the sources of interest and the background, including
daylight and artificial lights. (Note that visible lights also
contain IR signals.) The ambient light could be so strong that the
useful signals are completely overwhelmed. In order to isolate the
useful light signals, we make the LED lamp switch on and off with a
specified frequency using a low-end microcontroller. We then use FFT
and inverse FFT to extract the signal strength at a particular
frequency. Figure~\ref{fig:time-freq} shows a sequence of raw
measurements of IR intensity and the FFT result. In this test, the
IR LED lamp flashed at a frequency of 65 Hz, and the nearby
fluorescent lamps at the standard 100 Hz, with daylight imposing a
strong intensity on the sensor with reading about 850, an order of
magnitude higher than the lamp's effect on the sensor. Two spikes
corresponding to the IR lamp and the fluorescent lamps can be
clearly identified from the figure, which allows us to recover the
intensity of the IR signal faithfully. In the following, when we say
a light signal sensed by a light sensor, we mean the signal after
the FFT processing.

\begin{figure}[tb]
    \centering
    \includegraphics[width=0.23\textwidth]{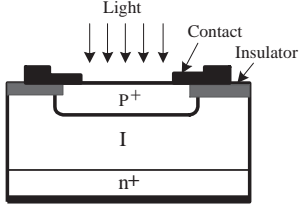}
    \caption{The PIN structure of a photodiode. }
    \label{fig:pin-struct}
\end{figure}

The model for light intensity, or receive signal strength (RSS), is
similar to a conventional light propagation model (e.g., the model
in~\cite{vc-multi-optical} or the Lambertian model~\cite{pharos}),
with a focus on its stability in a realistic environment. The RSS,
denoted by $s$, on a light sensor is mainly determined by three
factors: the distance of light sensor to the light bulb $d$, the
incident angle $\mu$ of light, and the emitting angle $\omega$. It
is well known that light intensity attenuates with increasing
distance $d$ according to an inverse square law. The incident angle
$\mu$ plays an important role here, due to the working principle of
the photodiode, which generates current under the striking of
photons. When the flat contact layer is not perpendicular to the
light, the energy with which the photons strike the contact layer
decreases, and thus the received light energy drops; see
Figure~\ref{fig:pin-struct} for an illustration. Normally, the
larger the deviation to the perpendicular orientation, the more loss
to the light intensity~\cite{isl29023}. Finally, $s$ decreases with
increasing $\omega$, following the characteristics of light emitting
diodes that behave in the same way. We call the direction with
$\omega=0$ the {\em central ray}. Furthermore, we say that the lamp
is {\em vertically oriented} when the central ray is vertical, and
{\em horizontally oriented} when the central ray is horizontal.
Figure~\ref{fig:model} shows the RSS model of a vertically oriented
lamp, in which the tilted rectangle represents the sensor's surface,
called its {\em sensing face}.

We need to determine three functions that respectively represent the
influences of the three factors on RSS.
\begin{itemize}
  \item $f_d(d)$, representing the impact of $d$ on $s$, is obtained by
varying $d$ while fixing $\mu=\pi/2$ and $\omega=0$, that is, making
the sensing face perpendicular with the central ray.
  \item $f_{\mu}(\mu)$, representing the impact of $\mu$ on $s$, is obtained
by varying the angle of the sensing face at a fixed $d$, with
$\omega=0$.
  \item $f_{\omega}(\omega)$, representing the impact of $\omega$ on $s$, is obtained
by moving the sensor along a circle while keeping $\mu=0$, that is,
making the sensing face perpendicular with the emitting ray.
\end{itemize}

In fact, $f_d(d)$ captures the light signal's propagation law,
$f_{\mu}(\mu)$ depends on the physical nature of a photodiode (at
the receiver side), and $f_{\omega}(\omega)$ reflects the optical
properties of the lamp's cover (at the transmitter side). The light
intensity function is then modeled as:
\begin{equation}\label{eq:fs}
s = f_d(d) \cdot f_{\mu}(\mu) \cdot f_{\omega}(\omega).
\end{equation}

\begin{figure}[tb]
    \centering
    {\footnotesize
    %\quad
    \shortstack{
            \includegraphics[width=0.23\textwidth]{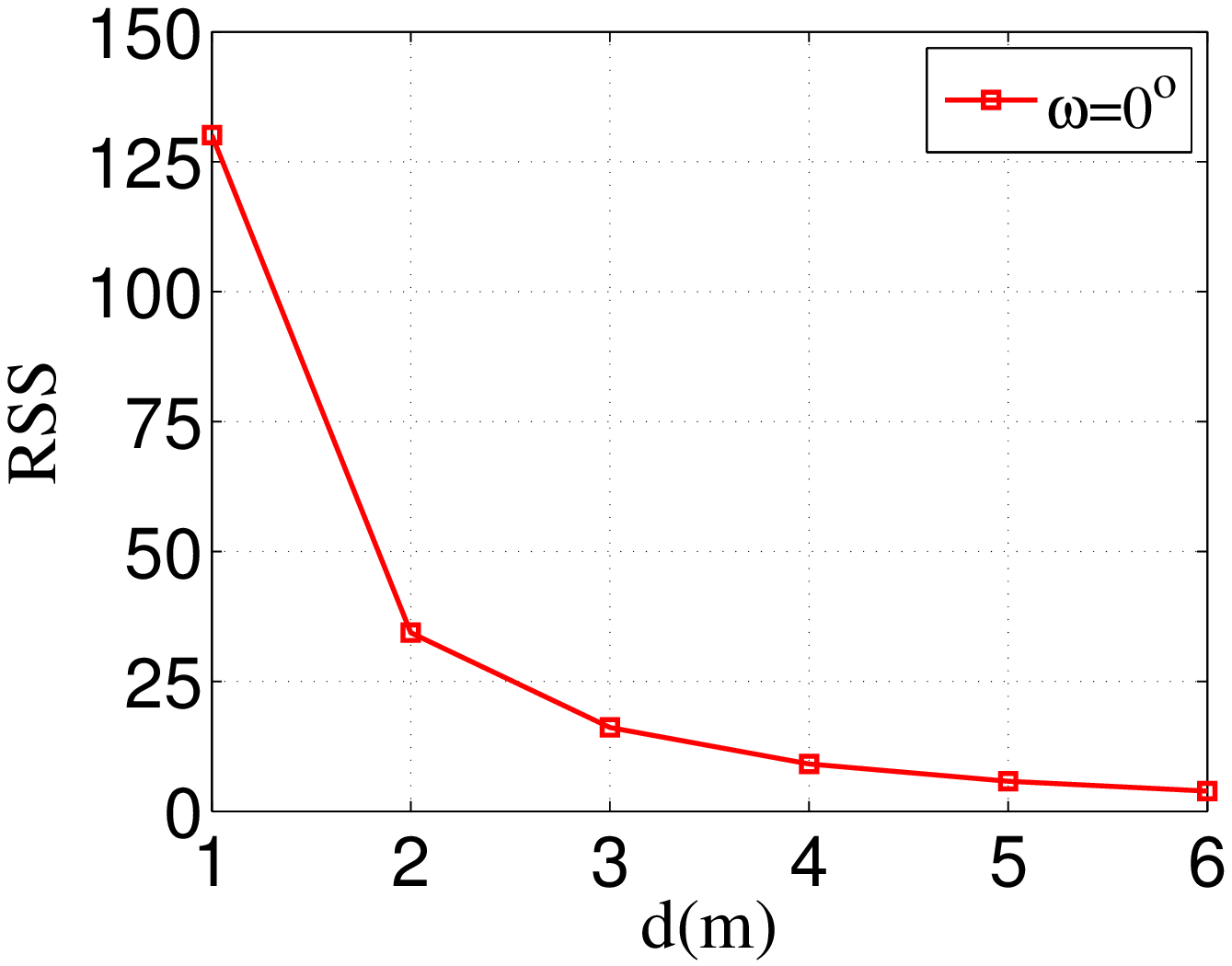}\\
            {(a) $f_d(d)$, no interference}
    }
    %\quad
    \shortstack{
            \includegraphics[width=0.23\textwidth]{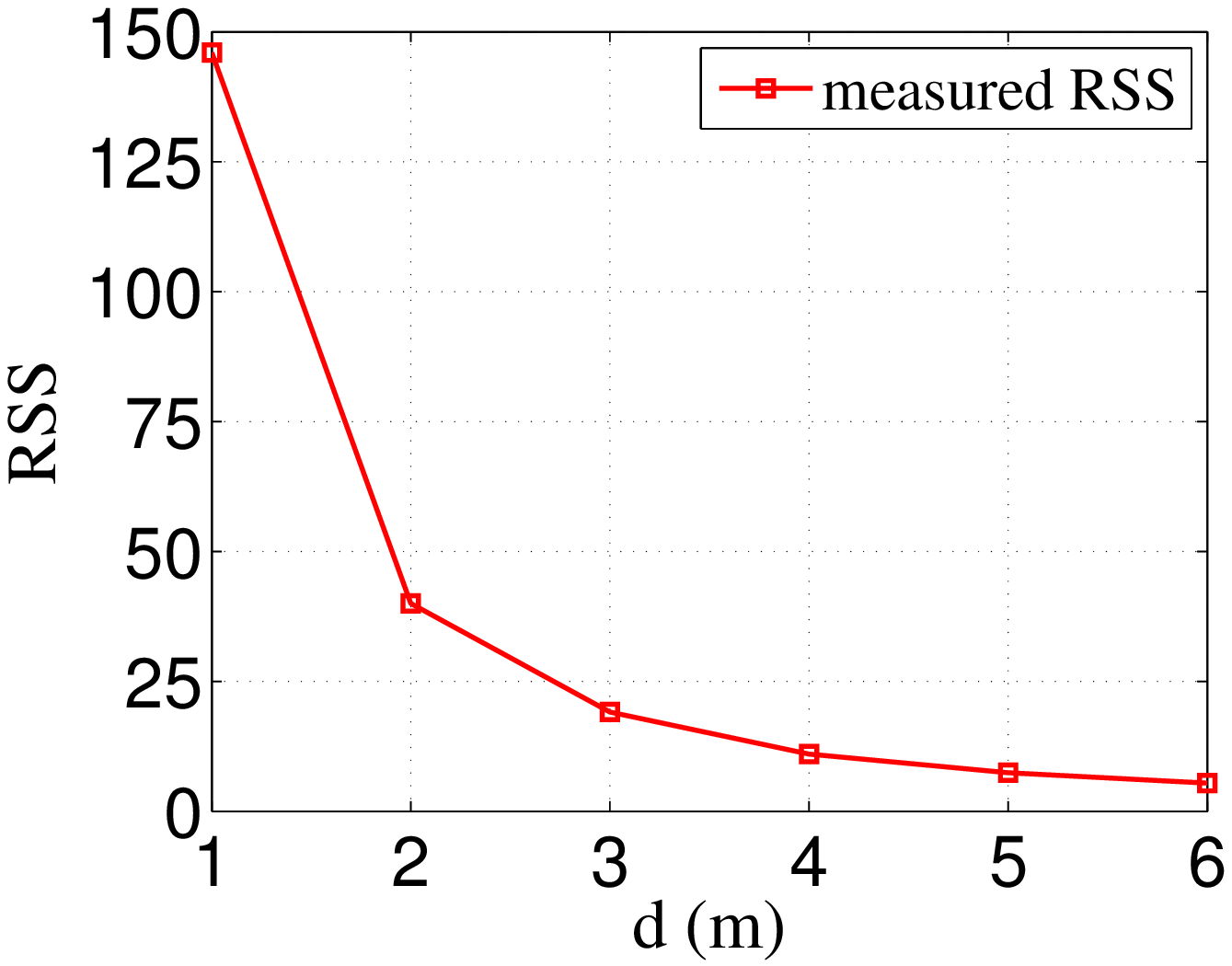}\\
            {(b) $f_d(d)$, with interference}
    }
    \shortstack{
            \includegraphics[width=0.23\textwidth]{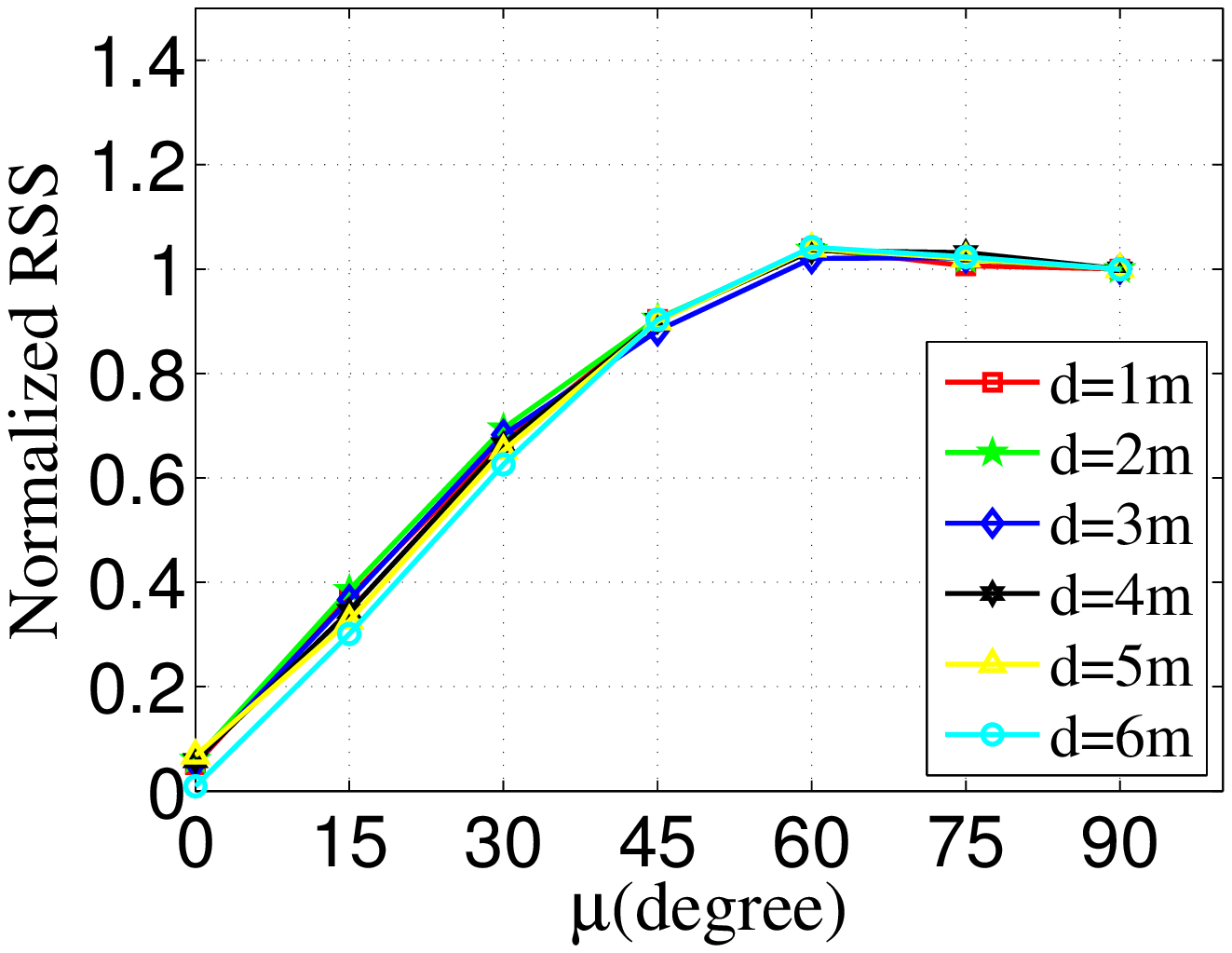}\\
            {(c) $f_{\mu}(\mu)$, no interference}
    }
    \shortstack{
            \includegraphics[width=0.23\textwidth]{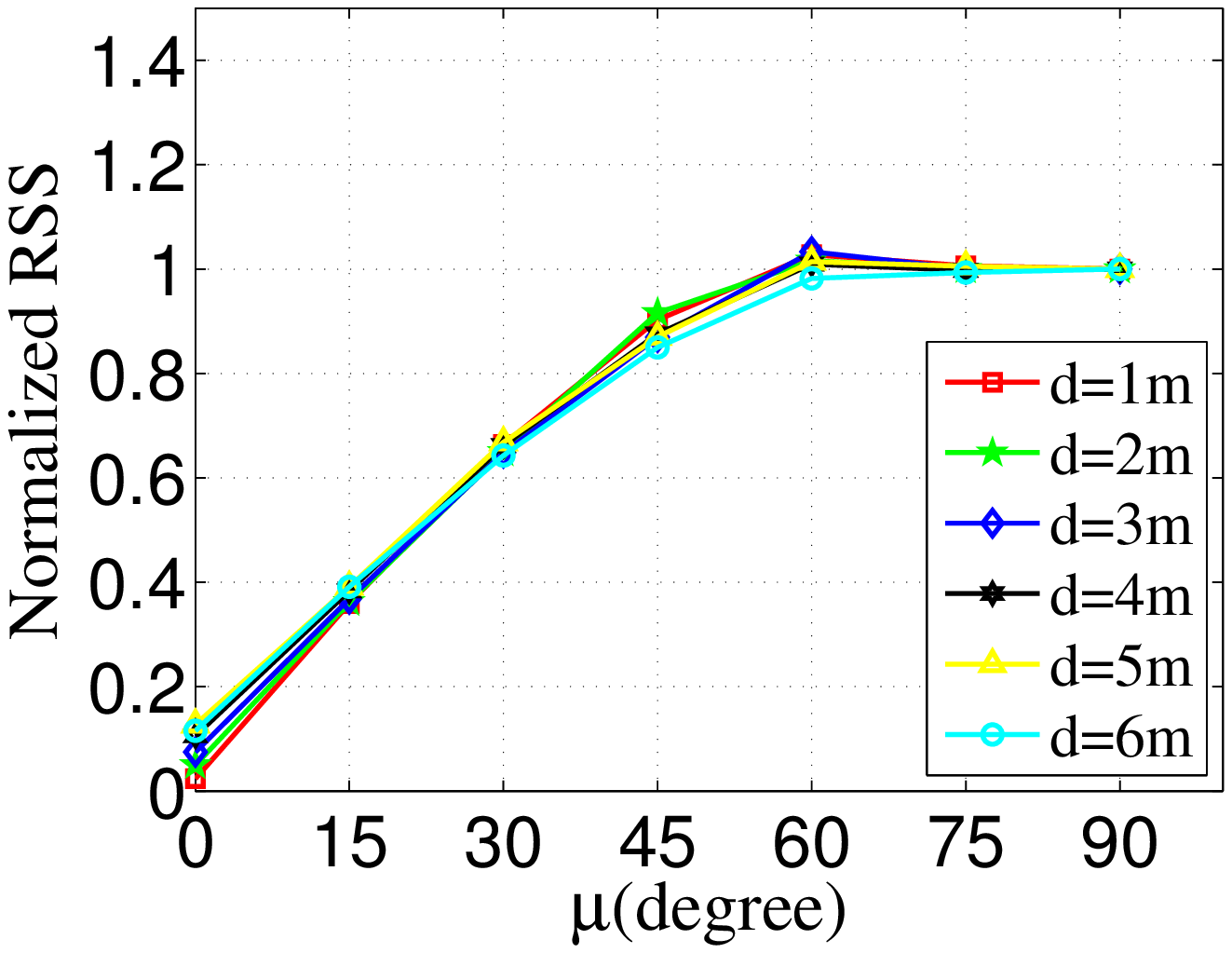}\\
            {(d) $f_{\mu}(\mu)$, with interference}
    }
    \shortstack{
            \includegraphics[width=0.23\textwidth]{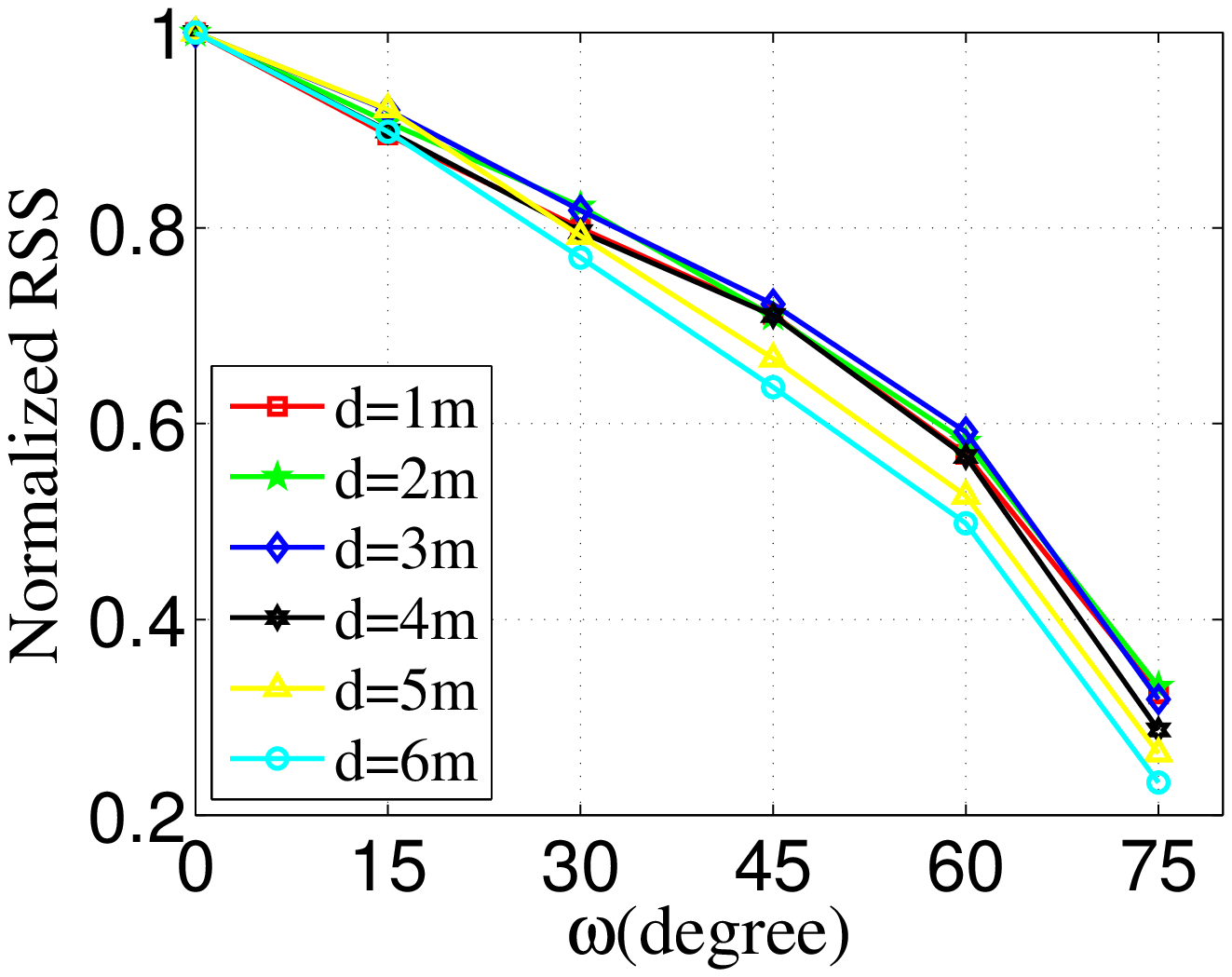}\\
            {(e) $f_{\omega}(\omega)$, no interference}
    }
    \shortstack{
            \includegraphics[width=0.23\textwidth]{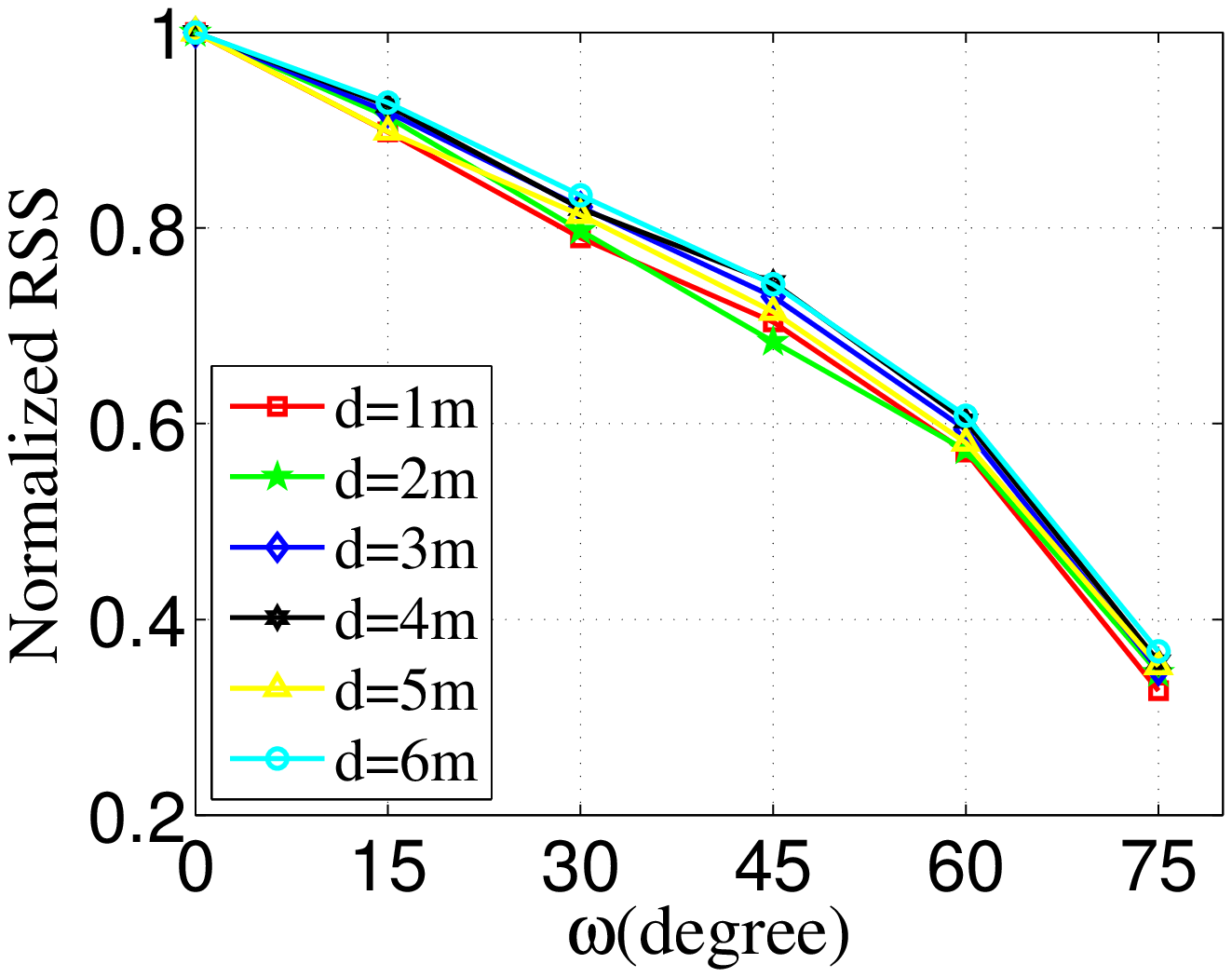}\\
            {(f) $f_{\omega}(\omega)$, with interference}
    }
    \caption{\label{fig:RSS} The light sensor's reading as a function of $d$, $\mu$, and $\omega$ in
    two scenarios, for the IR lamp. }
    }

\end{figure}

\begin{figure}[tb]
    \centering
    {\footnotesize
    \shortstack{
            \includegraphics[width=0.23\textwidth]{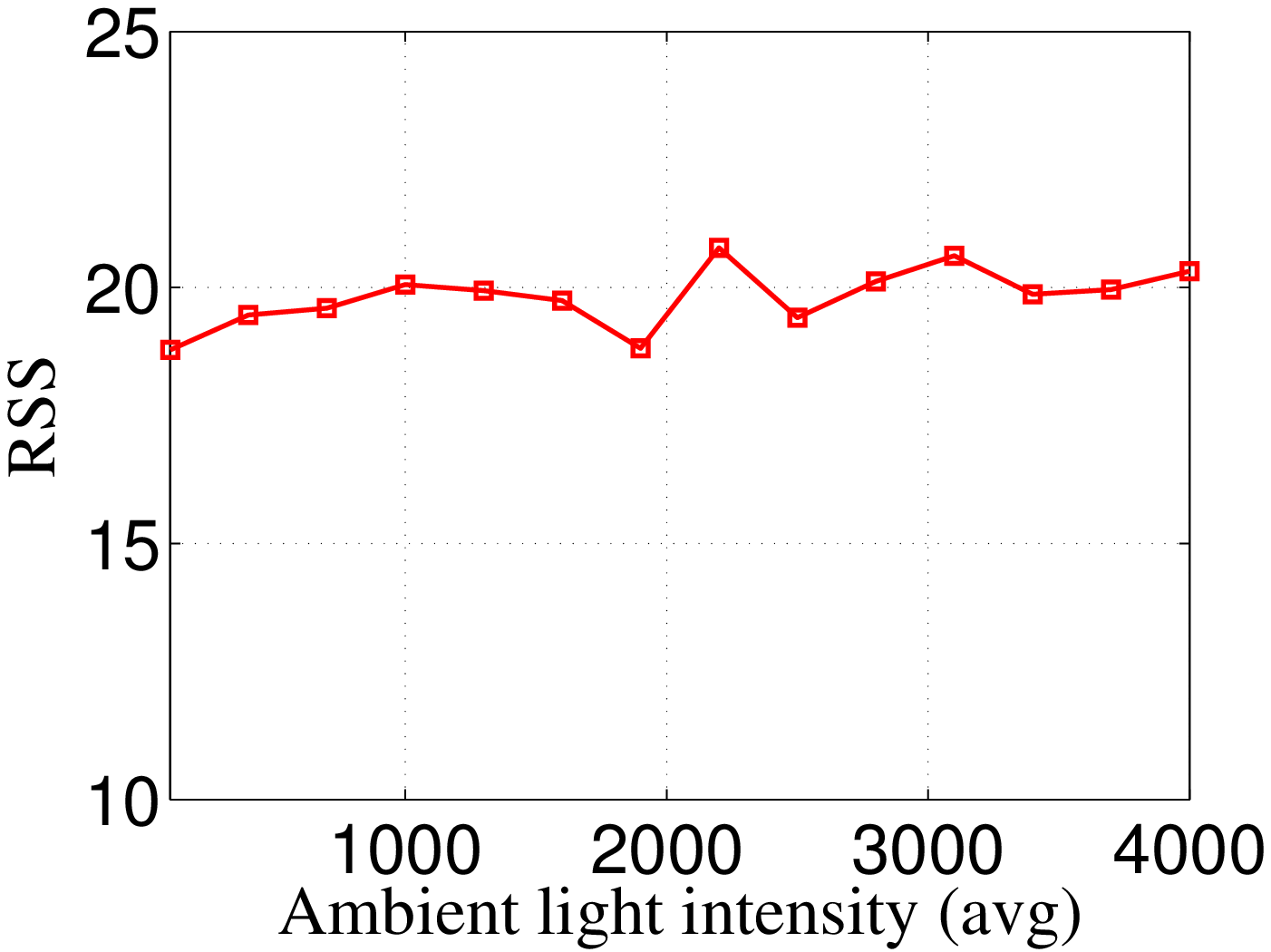}\\
            {(a)}
    }
    \shortstack{
            \includegraphics[width=0.23\textwidth]{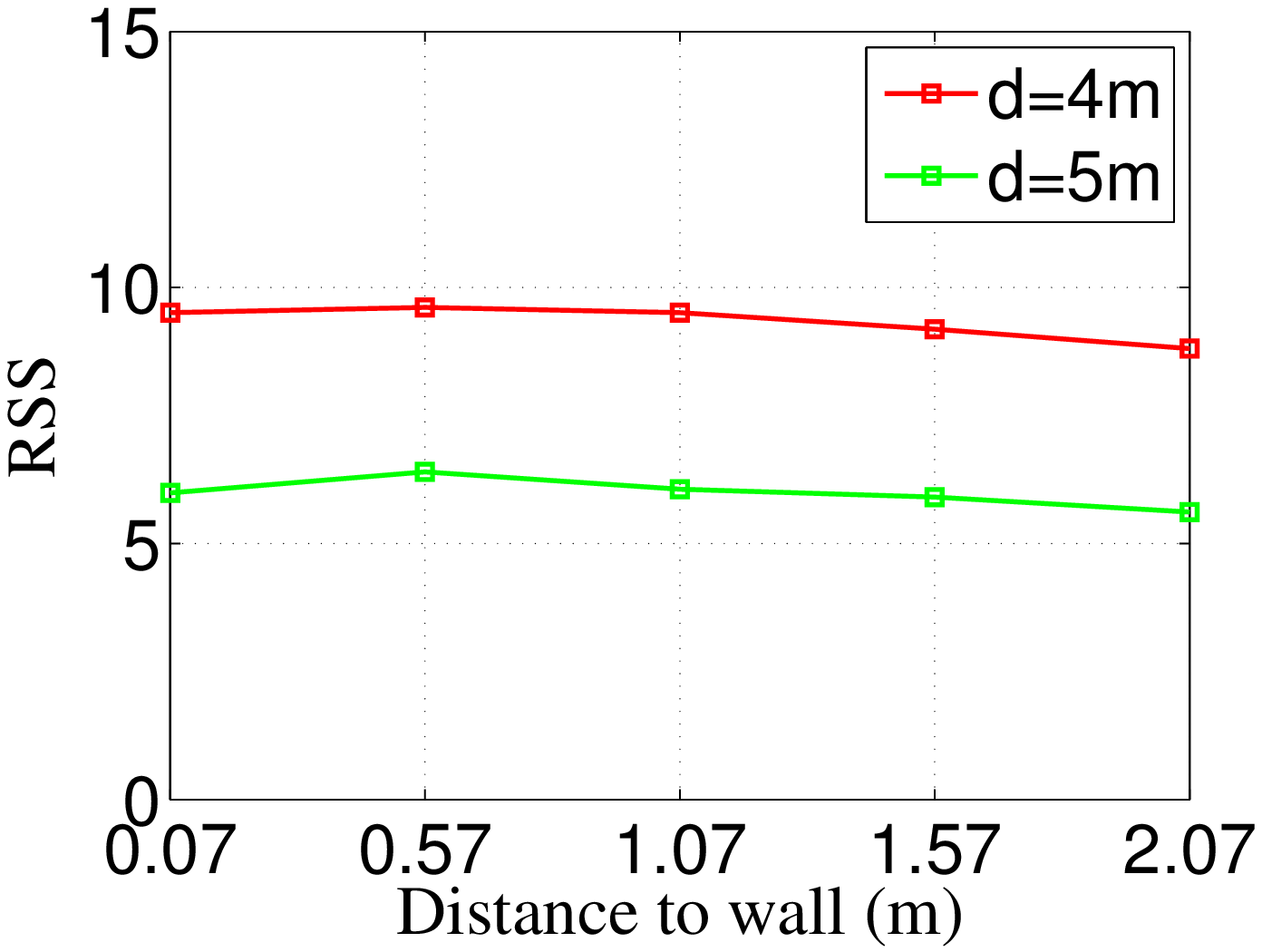}\\
            {(b)}
    }
    \shortstack{
            \includegraphics[width=0.23\textwidth]{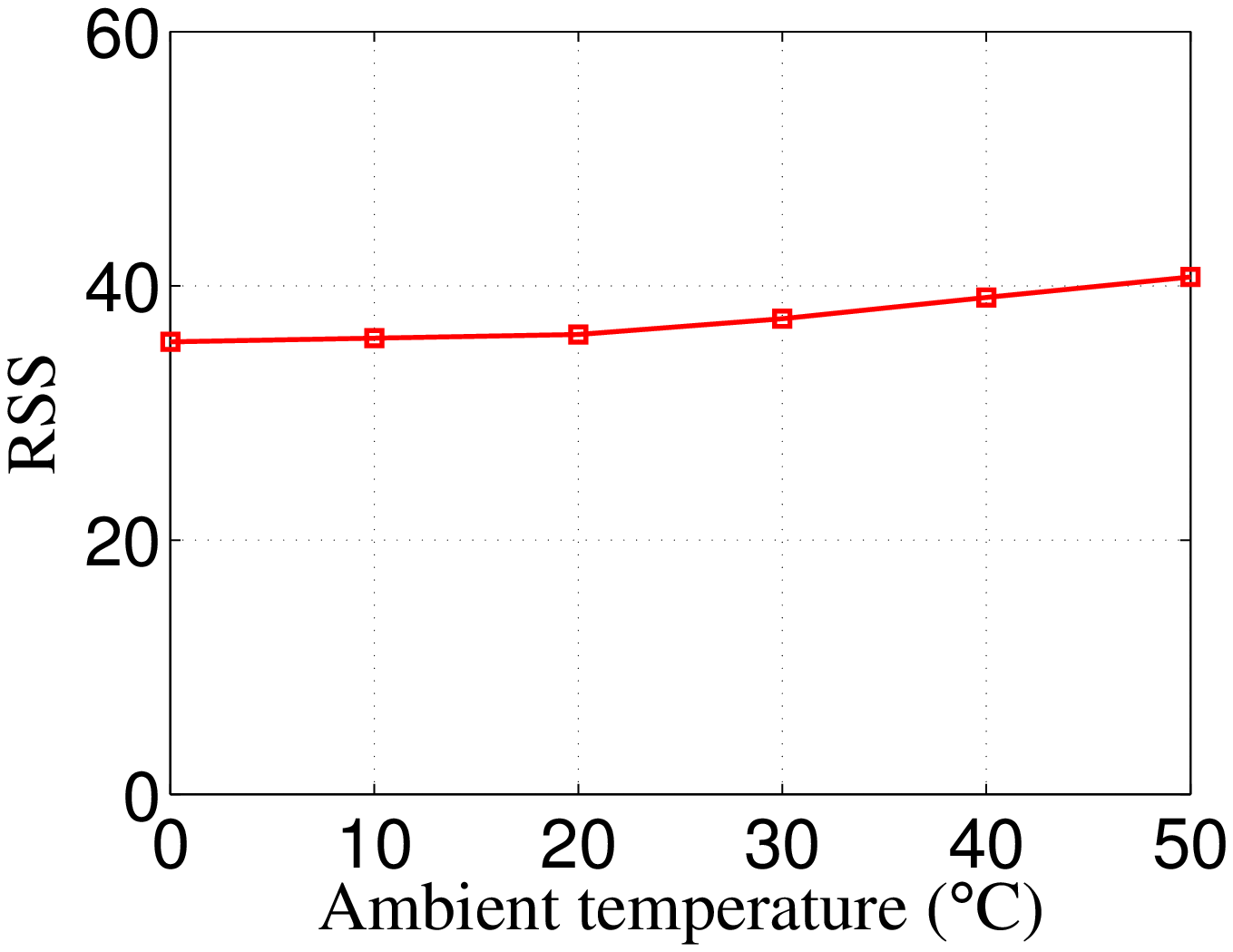}\\
            {(c)}
    }
    \shortstack{
            \includegraphics[width=0.23\textwidth]{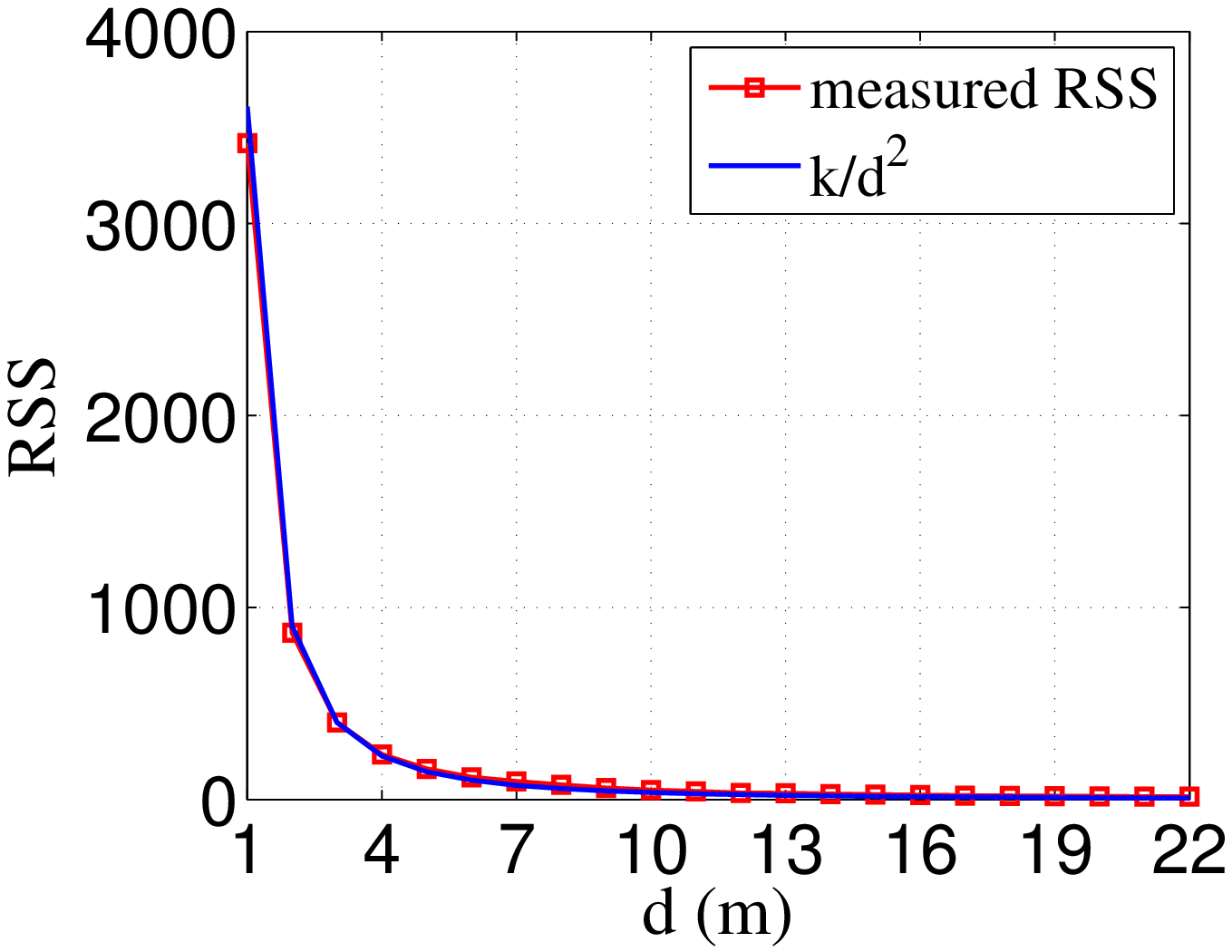}\\
            {(d)}
    }
    \shortstack{
            \includegraphics[width=0.23\textwidth]{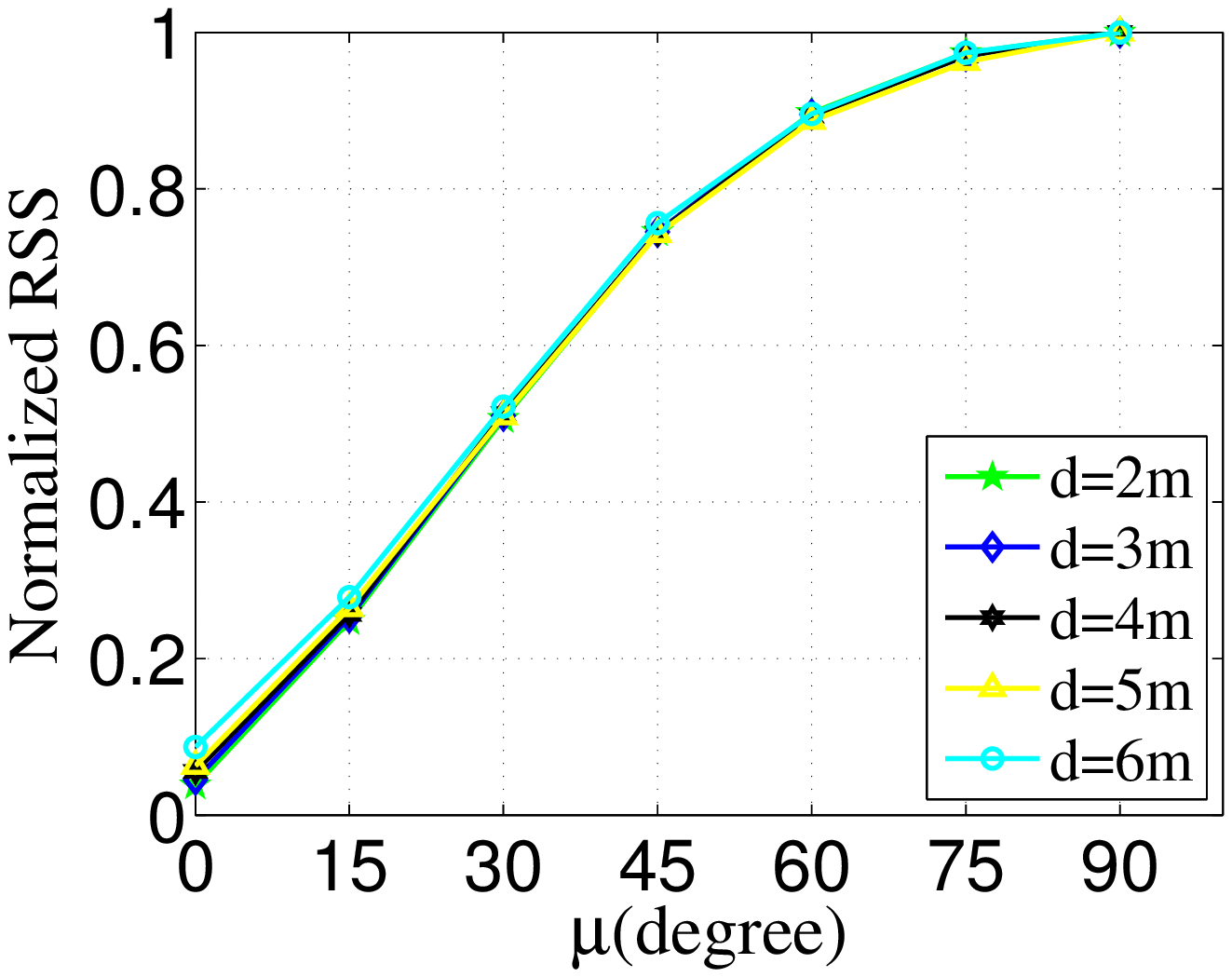}\\
            {(e)}
    }
    \shortstack{
            \includegraphics[width=0.23\textwidth]{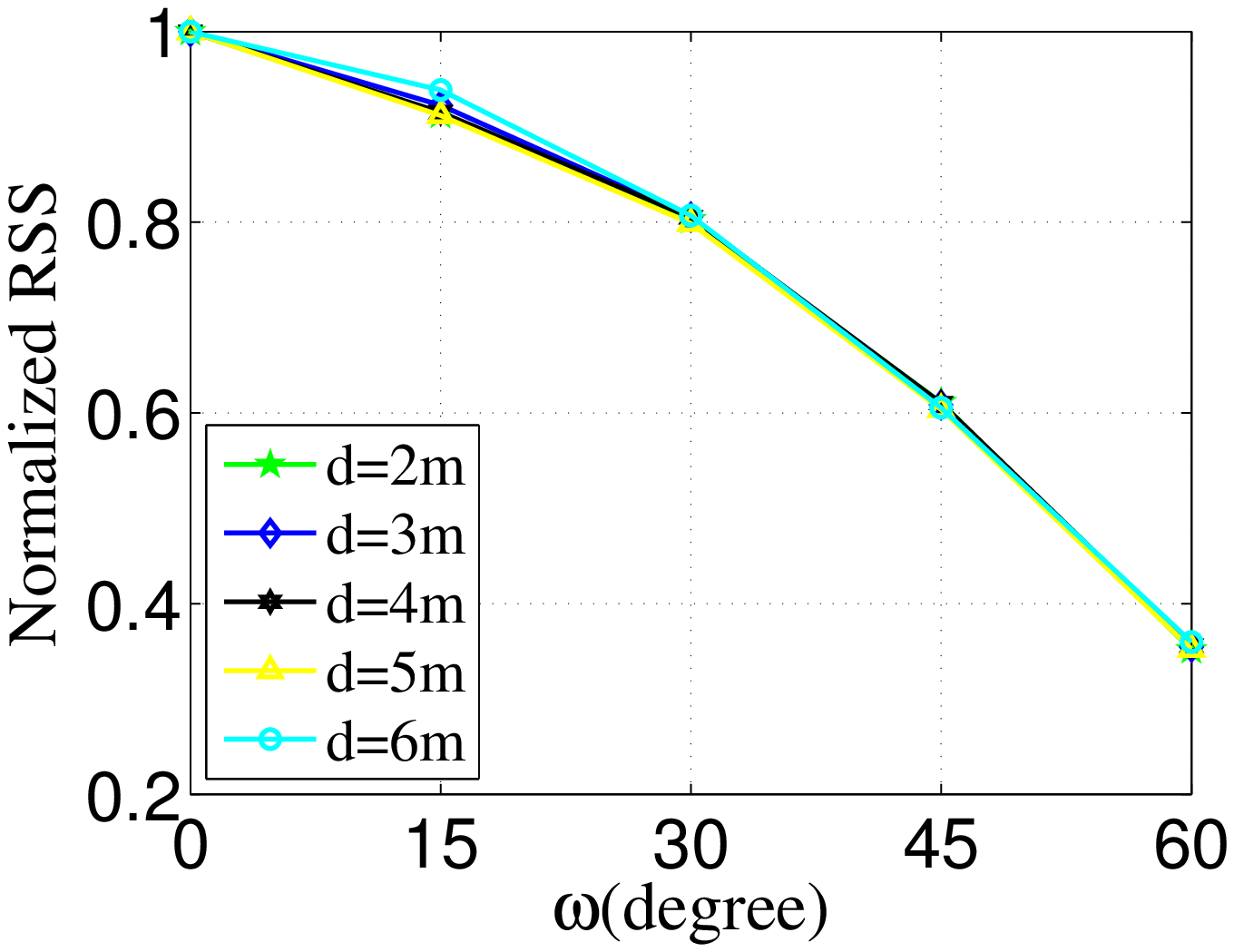}\\
            {(f)}
    }
    \caption{Properties of a light sensor under various
    conditions. (a) Sensitivity to ambient light intensity. (b)
    Sensitivity to light reflection. (c) Sensitivity to ambient
    temperature. (d) Sensitivity to $d$ under the visible light
    lamp. (e) Sensitivity to $\mu$ under the visible light
    lamp. (f) Sensitivity to $\omega$ under the visible light
    lamp.}\label{fig:interferences}
    }
\end{figure}

\vspace{5pt}\textbf{Interference-free scenario.} First, we examined
the property of the light sensor in a dark room at nighttime, where
no ambient light is present. The first column of
Figure~\ref{fig:RSS} shows the impacts of the three factors, $d$,
$\mu$, and $\omega$, on the RSS. It can be seen that $f_d(d)$ quite
closely follows the well-known inverse square law for light
intensity. (Small errors exist between the practical measurement and
theory due to the physical properties of a photodiode, such as
responsivity, dark current, etc~\cite{isl29023}. These could be
accounted for by calibration, but are ignored in our current
design.) We can also see that $f_{\mu}(\mu)$ decreases with $\mu$
with fairly high predictability. The trends can be roughly captured
by a $\sin$ function. For $f_{\omega}(\omega)$, the trend is also
very deterministic,
which can be modeled with a polynomial function.%\footnote{There
%exists a slight correlation between the RSS and distance that could
%be captured, but for simplicity we ignore this factor, which only
%adds to the error of positioning.}.

\vspace{5pt}\textbf{Impact of ambient lights.} This set of
experiment was conducted in the morning, when the daylight imposed
an IR intensity reading of 800 on the sensor. In addition, three
fluorescent lamps were turned on, emitting periodic IR signals with
frequency 100 Hz. The intensity of ambient IR light was strong
enough to overwhelm the RSS from the lamp at a distance of a few
meters, however the frequency domain treatment can successfully
extract the component of light intensity that we are interested in.
The second column of Figure~\ref{fig:RSS} shows the impacts of the
three factors, $d$, $\mu$, and $\omega$, on the RSS. It can be seen
that the results in both interference and interference-free
environments are quite consistent, with differences normally within
10\% of each other.

Figure~\ref{fig:interferences}(a) shows the impact of ambient light
intensity as experienced by the sensor at different times (e.g.,
nighttime and daytime) and at various places in a room. The maximum
value 4000 corresponds to the RSS near an open window at noontime of
a sunny day. It can be seen that the extracted IR intensity of the
LED lamp remains relatively stable, with a standard deviation as
small as 0.58.

\vspace{5pt}\textbf{Impact of light reflection.} This experiment
examines how light reflection from surrounding objects affects the
RSS. In a dark room, we kept the lamp horizontally oriented, with
its central ray parallel to a wall and the floor, and vary the ray's
distance to wall. The light sensor was placed at a certain distance
from the lamp with sensing surface perpendicular to the central ray.
Figure~\ref{fig:interferences}(b) shows how the RSS changes with the
distance to wall. It can be seen that the RSS experiences only
insignificant changes. In our daily life, most materials (except
glasses, polished metals, etc) give no more than a few percent
specular (i.e., mirror-like) reflection~\cite{diffuse}; that is,
most of the light, upon hitting the surface of an object, is
scattered in all directions, leaving only a small portion of
reflected energy on the sensor. This explains the robustness of the
RSS against wall reflection.

\vspace{5pt}\textbf{Impact of ambient temperature.}
Figure~\ref{fig:interferences}(c) shows how the RSS changes with
ambient temperature. The low temperatures were produced by placing
ice cubes around the sensor, and the high temperatures were
generated by blowing at the sensor using a hair drier. We can see
that the RSS increases with temperature, which agrees with the
property of the photodiode reported in~\cite{isl29023}. The trend is
very mild, suggesting that only small errors are introduced by the
temperature factor. In addition, the trend is monotonic, so the
error could be compensated for with simple calibration. %We leave
%this to future refinement of the design.

\vspace{5pt}\textbf{High power LED lamp.} We repeated the above
experiments with the visible light lamp.
Figure~\ref{fig:interferences}(d) shows the relationship between $d$
and RSS, which closely matches the baseline curve of function
$k/d^2$. Due to the much increased power, the lighting range extends
to nearly 30 meters, and for the same $d$, the RSS is much higher
than with the IR lamp. In this test, the lamp was horizontally
oriented and placed within a narrow corridor (about 2m wide)
surrounded by wall, floor, and wooden boards, which presented
complex conditions for light reflection. Compared to the low power
IR lamp, the increased power causes more noticeable reflection
effect. However, the variability is still below around 10\% of the
baseline. In a practical system, the lamps will be hung on ceilings
and sensor mostly oriented upward, so the sensor is unlikely to be
exposed to as strong reflection. It is thus reasonable to assume
that the reflection effect does not fundamentally invalidate our RSS
model in typical environments.

Figures~\ref{fig:interferences}(e)(f) show that the two functions
$f_{\mu}(\mu)$ and $f_{\omega}(\omega)$ remain highly deterministic
and consistent at different positions, though their particular forms
differ from those of the IR lamp, due to the different photoelectric
effects of the light sensor under IR and visible lights, and also
because of the different scattering effects of the lamp covers.

\begin{figure*}[tb]
    \centering{\small
    \shortstack{
            \includegraphics[width=0.18\textwidth]{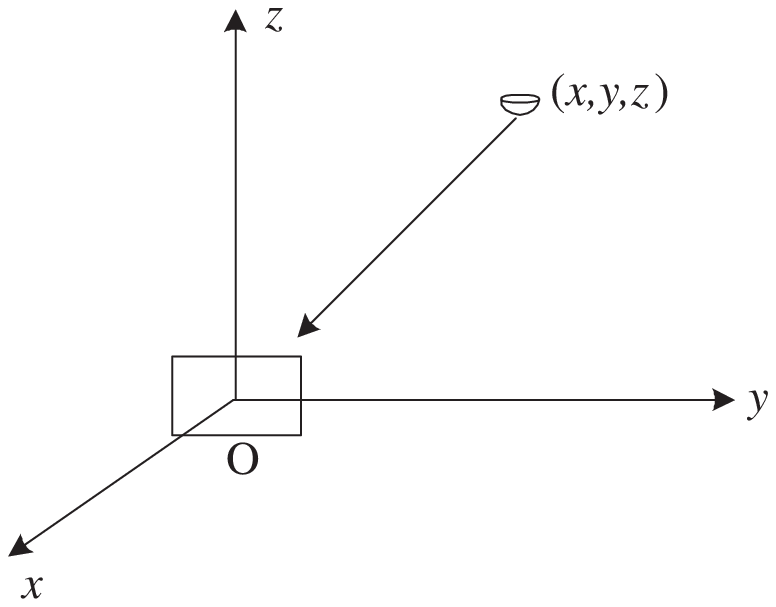}\\
            {(a) Sensing plane $x=0$. }
    }
    \shortstack{
            \includegraphics[width=0.18\textwidth]{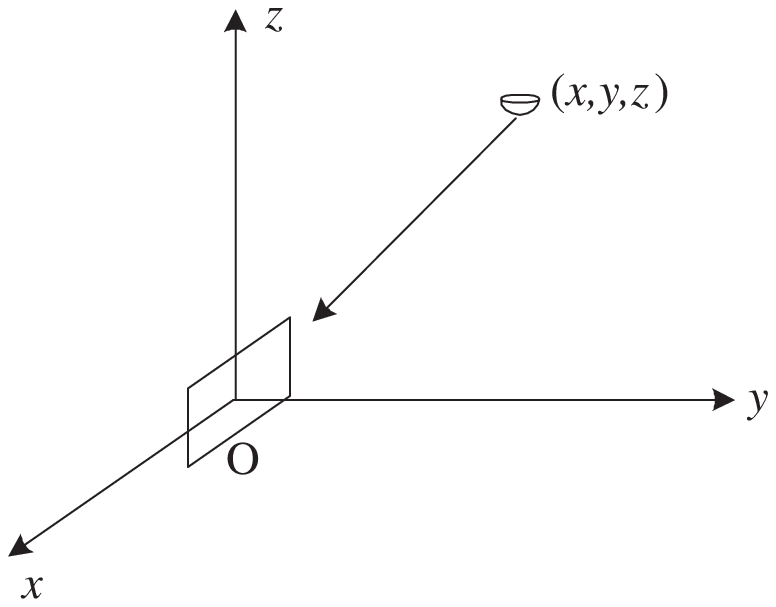}\\
            {(b) Sensing plane $y=0$.}
    }
    \shortstack{
            \includegraphics[width=0.18\textwidth]{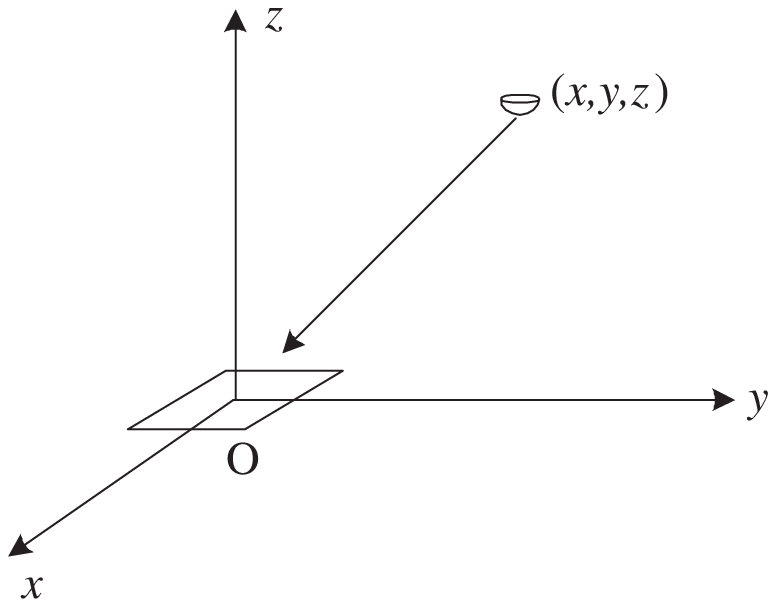}\\
            {(c) Sensing plane $z=0$.}
    }
    \shortstack{
            \includegraphics[width=0.18\textwidth]{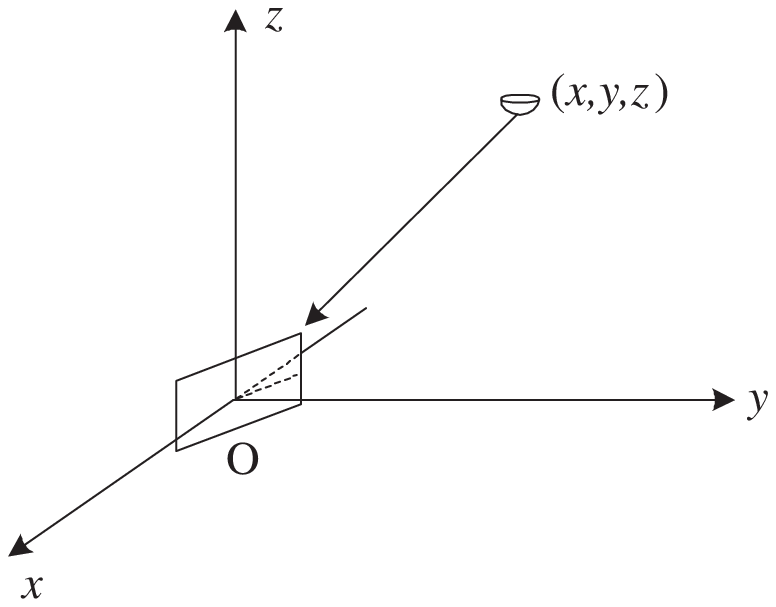}\\
            {(d) Sensing plane $x+2y=0$.}
    }\\
    \shortstack{
            \includegraphics[width=0.22\textwidth]{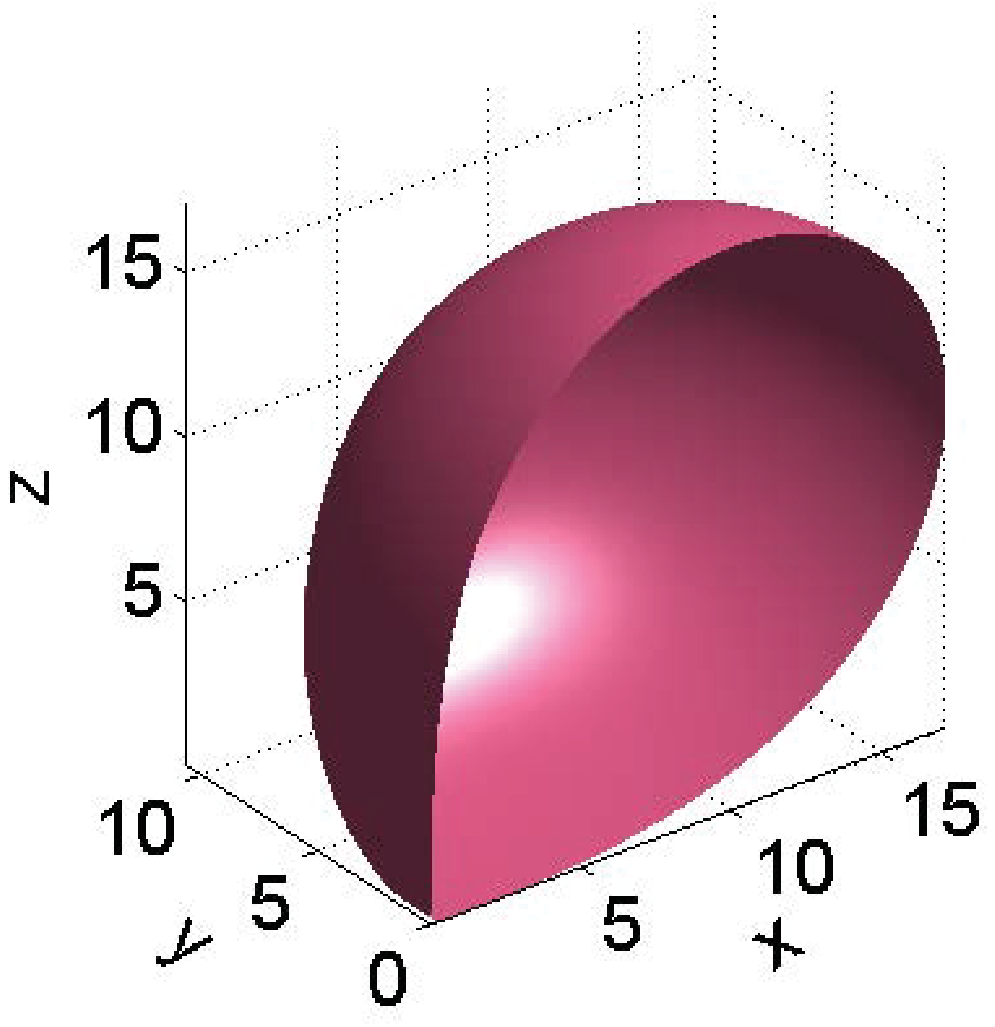}\\
            {(a') Solution set for (a).}
    }
    \shortstack{
            \includegraphics[width=0.22\textwidth]{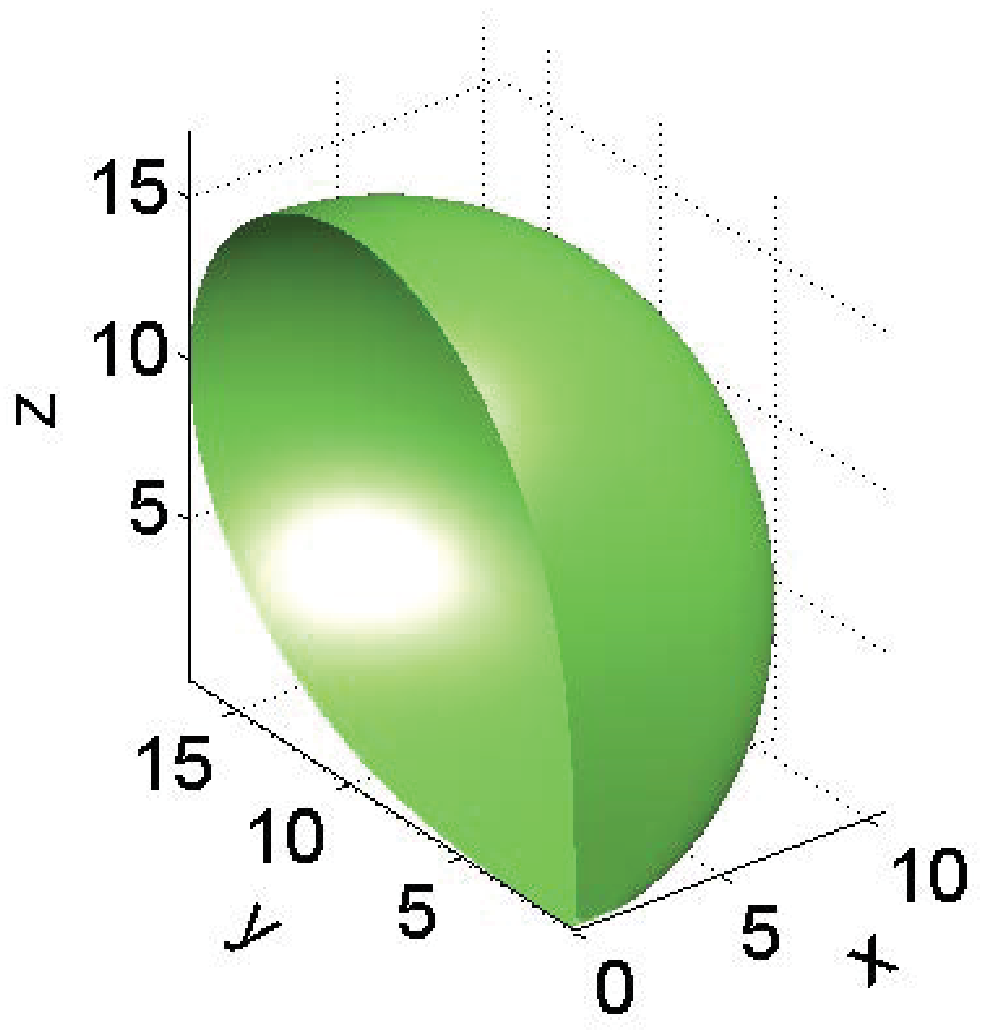}\\
            {(b') Solution set for (b).}
    }
    \shortstack{
            \includegraphics[width=0.21\textwidth]{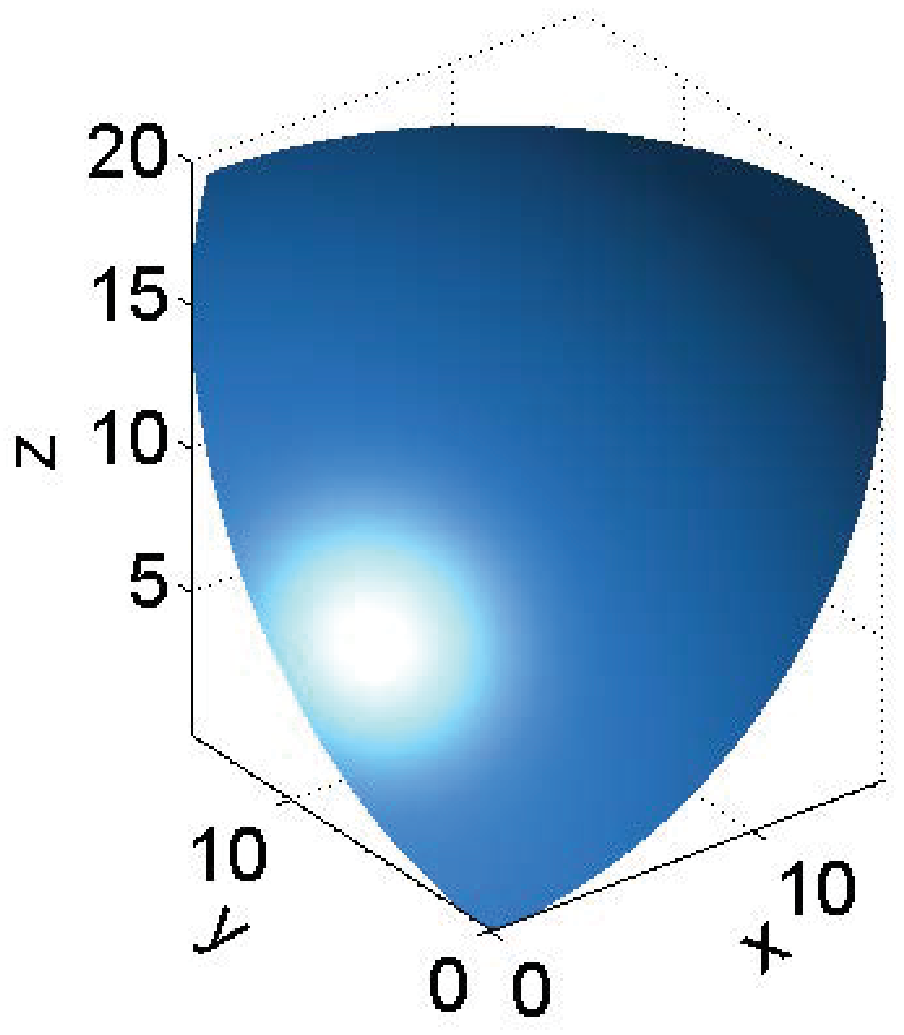}\\
            {(c') Solution set for (c).}
    }
    \shortstack{
            \includegraphics[width=0.26\textwidth]{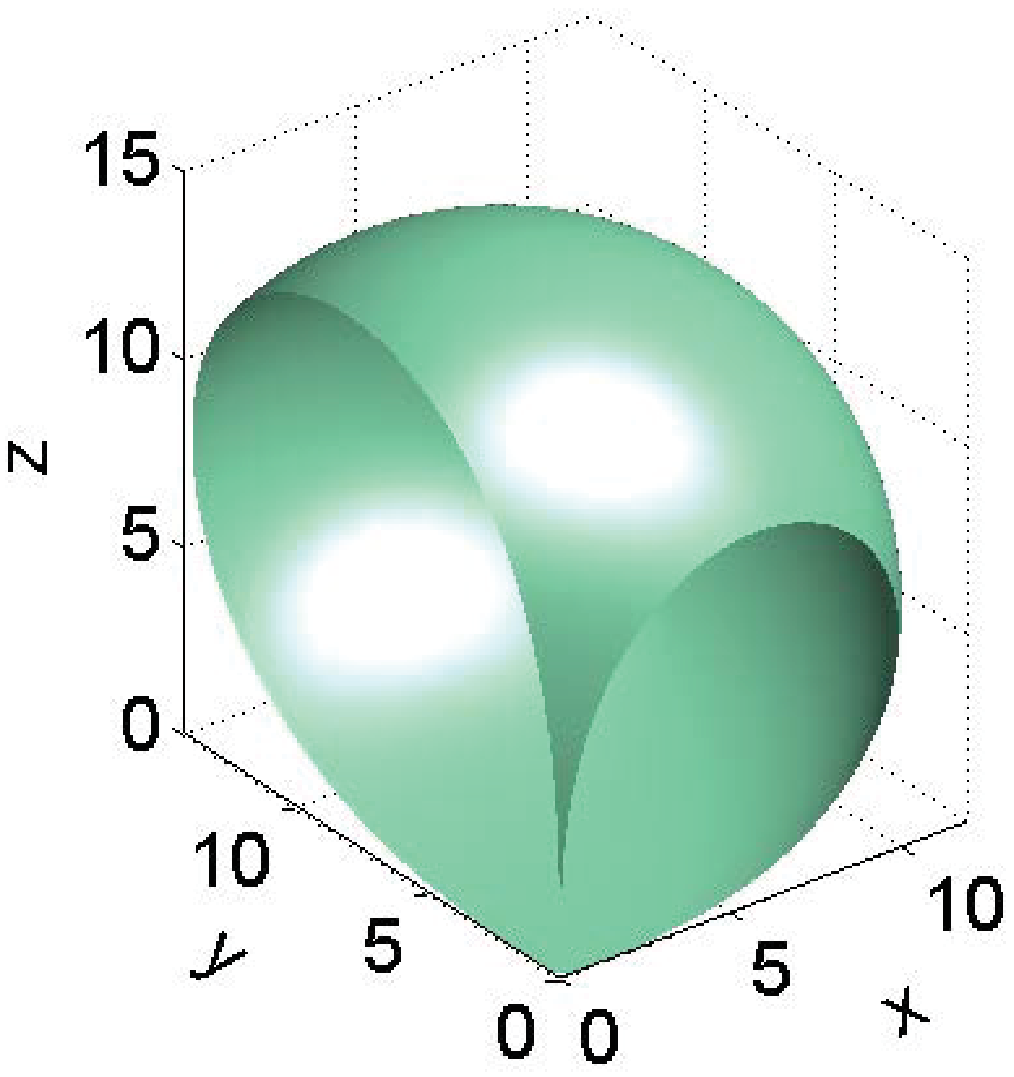}\\
            {(d') Solution set for (d).}
    }}
    \caption{The solution sets corresponding to different sensing faces.}
    \label{fig:mflp}
\end{figure*}

\section{Light positioning principle}\label{sec:mflp}

In this section we describe two principles of light sensor
positioning, one called {\em Multi-Face Light Positioning} (MFLP),
which is our emphasis, and the other following the classic
trilateration method.

The basic idea of MFLP is to have three (or more) properly oriented
sensors to collect signal strengths as well as the sensors'
orientation measures. Along with the pre-defined RSS model, the
measured data provides sufficient spatial constraints to locate the
receiver. We call the top contact layer of the photodiode the {\em
sensing face} of a light sensor, and the plane containing it the
sensor's {\em sensing plane}.

\subsection{Multi-Face Light Positioning}
Assume the considered sensing face is centered at the origin
$O=(0,0,0)$, and the point source of light is located at
$X=(x,y,z)$, where $x>0,y>0,z>0, \mu \in (0,\pi/2), \omega \in
(0,\pi/2)$. The corresponding sensing plane has the form
$Ax+By+Cz=0$, where $A,B,C$ are determined by the sensor's tilt and
heading. Then, the distance between $X$ and $O$ is $$d =
\sqrt{x^2+y^2+z^2},$$ and the distance between $X$ and the sensing
plane is $$d' = \frac{|Ax+By+Cz|}{\sqrt{A^2+B^2+C^2}}.$$ Following
Eq.~\ref{eq:fs}, let $f_d(d)=k/d^2$, $f_{\mu}(\mu)=\sin(\mu) =
d'/d$, and $f_{\omega}(\omega) = f_{\omega}(\arccos (z/d))$, where
$f_{\omega}(\cdot)$ is a monotonically decreasing function of
$\omega$ (hence $z$).%\footnote{Although the function $f_{\mu}$ can
%be well fitted with the $\sin$ function, experiments show that other
%fitting functions work as well, without affecting the uniqueness of
%positioning solution.}
Therefore,
 \begin{equation}\label{eq:rss-model}
    s = \frac{k}{d^3} \frac{|Ax+By+Cz|}{\sqrt{A^2+B^2+C^2}} f_{\omega}\Big(\arccos \frac{z}{d}\Big).
    \end{equation}
where $d = \sqrt{x^2+y^2+z^2}, x>0, y>0, z>0, k>0$, and $s$ is a
real number between 0 and some maximum reading value $s_m>0$.

\begin{theorem}\label{theo:unique} When no measurement errors occur,
three linearly independent sensing planes that pass through the
origin and that satisfy the RSS model as specified by
Eq.~\ref{eq:rss-model} determine a unique solution of $X=(x,y,z)$.
\end{theorem}

The system of equations generated by the mentioned three sensing
plane is a high-order and nonlinear one, whose properties are in
general not easy to obtain. Fortunately, the structure of $A, B, C$
is simple enough to enable reduction among the equations, which
makes it possible to establish an exact relationship between the
solvability and the linear independence property of $(A_i, B_i,
C_i)$. The proof is provided in Appendix A.

Theorem~\ref{theo:unique} suggests that if we can create three
linearly independent sensing planes on a receiver, and that these
sensors can simultaneously `see' the light source (i.e., in line of
sight), then one can determine a position of the light source. With
a bit of coordinate transformation, we can determine the position of
the receiver provided the position of the light source.

When measurement errors exist, three linearly independent sensing
faces may not lead to a solution. In this case, what we look for is
a least square solution that minimizes the sum of the squares of the
errors between each measured $s$ and the
calculated $s$ from the corresponding equation. %The three sensing
%planes can be realized in two ways: using three properly placed real
%sensors, and using a single sensor with manual rotation. We describe
%the two ways of implementation in Section~\ref{sec:lips-embedded}
%and Section~\ref{sec:lips-phone}.

\subsection{Why Linear Independence of
Faces}\label{subsec:linear-indep}

Since the geometric structure of the problem is not immediately
intuitive, we now give a hypothetical example of MFLP to illustrate
the necessity of the faces being linearly independent, which
provides a key guideline for our system design. For the purposes of
demonstration, we assume an simplified RSS model, in
which $f_d(d)= 1/{d^2},$ $f_{\mu}(\mu) = \sin(\mu),$ and
$f_{\omega}(\omega) = \cos(\omega).$ Then Eq.~\ref{eq:rss-model} can
be rewritten as
$$s = \frac {z|Ax + By + Cz|}{\sqrt{A^2+B^2+C^2}(x^2+y^2+z^2)^2},$$ where $x>0,
y>0,z>0$.

Consider four sensing faces centered at the origin, whose sensing
planes are $x=0$, $y=0$, $z=0$, and $x+2y=0$, as shown in
Figures~\ref{fig:mflp}(a)(b)(c)(d), respectively. The lamp is
located at $X=(10,10,10)$ and imposes light intensity $s_1$, $s_2$,
$s_3$, and $s_4$ on the sensors, respectively. Given the known
position of the lamp, the theoretical light intensities should be
$s_1' = s_2' = s_3' = 1/900$, and $s_4'=1/300\sqrt{5}$. Assume no
errors occurring from the light propagation and measurement
processes, the RSS seen on the sensors should be $s_i = s_i'$.

\begin{figure}[tb]
    \centering{\small
    \shortstack{
            \includegraphics[width=0.24\textwidth]{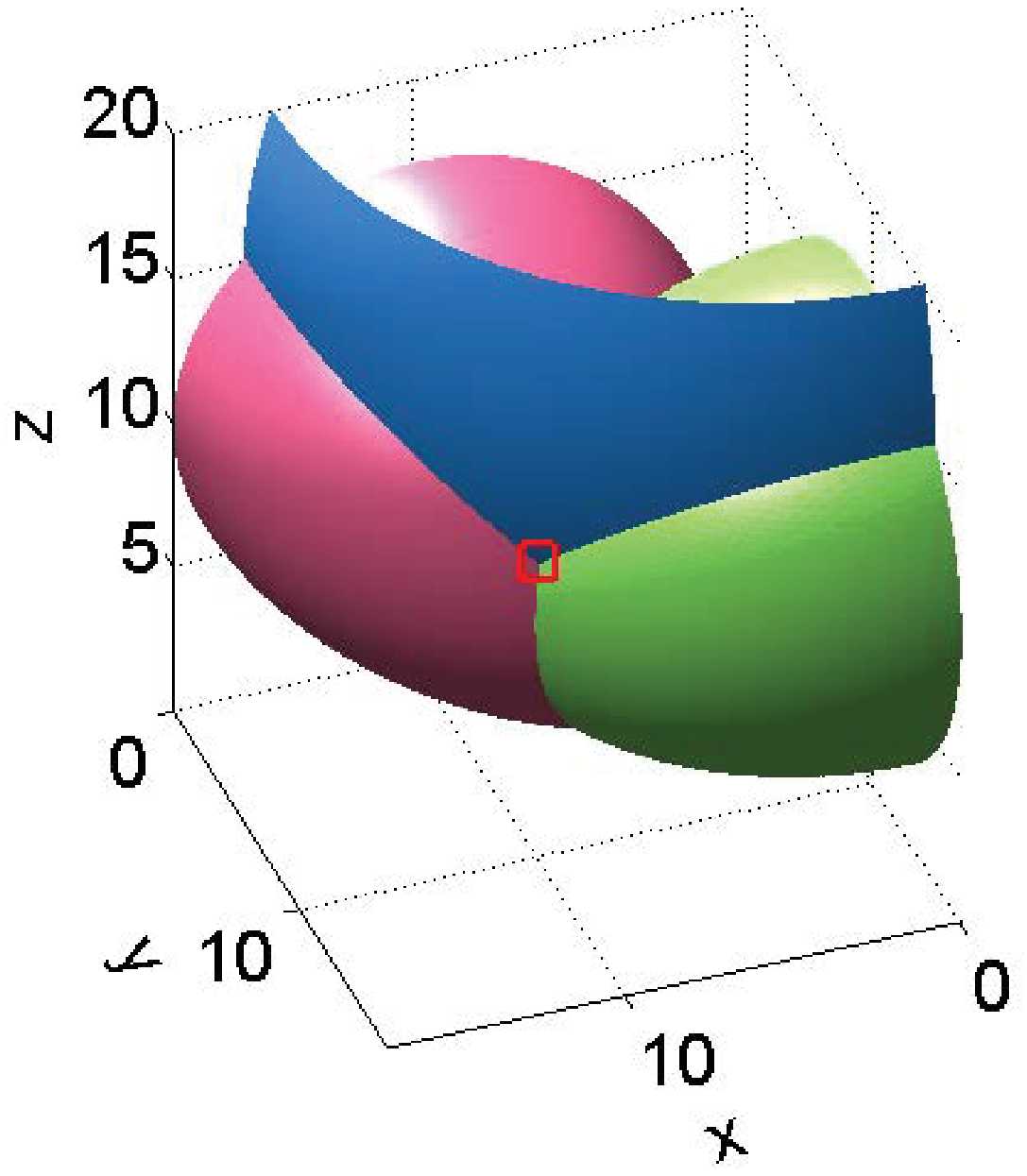}\\
            {(a)}
    }\hspace{-20pt}
    \shortstack{
            \includegraphics[width=0.26\textwidth]{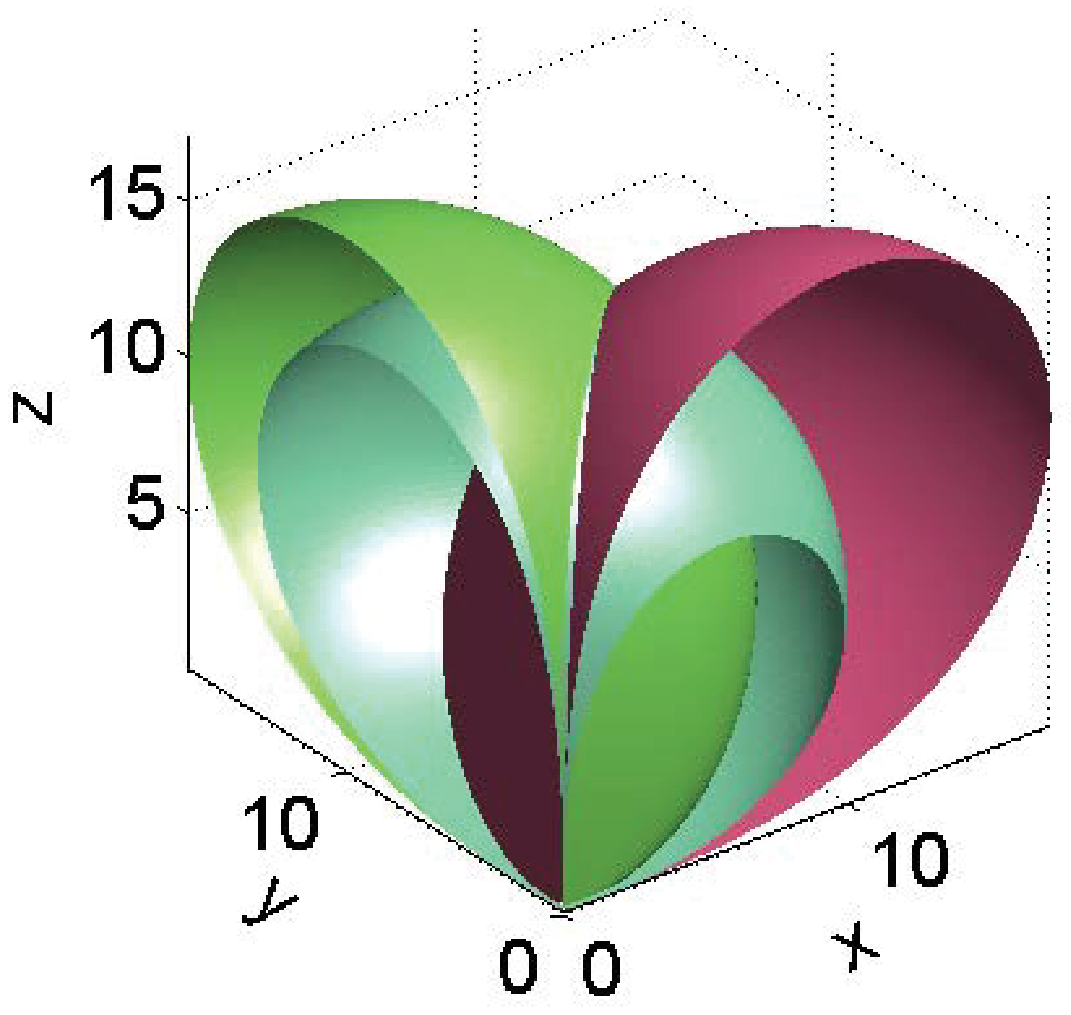}\\
            {(b)}
    }}
    \caption{The uniqueness of position solution. (a) The solution sets
    for the three sensing faces $x=0,y=0,z=0$ intersect at a single point (red box).
    (b) The solution sets
    for the three sensing faces $x=0,y=0,x+2y=0$ intersect at a curve (partly shown).}
    \label{fig:mflp-uniquess}
\end{figure}

Now we choose the first three sensing faces, $x=0,y=0,z=0$, which
are \textit{linearly independent}. These faces lead to a system of
equations $h \cdot z/(x^2+y^2+z^2)^2 = 1/900$, where $h = x, y$ or
$z$. Each of these equations will generate a solution set, as
depicted by the curved surfaces in
Figures~\ref{fig:mflp}(a')(b')(c'), respectively.
Figure~\ref{fig:mflp-uniquess}(a) shows the intersection of these
three solution set. It turns out that these surfaces intersect at a
single point $(10,10,10)$, which matches the true location.

%
%the following system of equations:
%\begin{numcases}{}
%   \frac {zx}{(x^2+y^2+z^2)^2} = \frac{1}{900}\label{example_eq_1}\\
%   \frac {zy}{(x^2+y^2+z^2)^2} = \frac{1}{900} \label{example_eq_2}\\
%   \frac {z^2}{(x^2+y^2+z^2)^2} = \frac{1}{900} \label{example_eq_3}
%\end{numcases}
%
%Each of Eq.~\ref{example_eq_1}, Eq.~\ref{example_eq_2}, and
%Eq.~\ref{example_eq_3}

Next, consider an alternative set of three sensing faces,
$x=0,y=0,x+2y=0$, which are \textit{linearly dependent}.
Figure~\ref{fig:mflp-uniquess}(b) shows the intersection of their
corresponding solution sets. Different from the first case, these
sets do not intersect at a single point, but instead produce a curve
segment (only partly shown due to blocking of surfaces), meaning
infinitely many valid solutions for the three sensing faces. This
comparison shows why linear independence of the sensing faces is
necessary for unique solution of positioning.

\subsection{MFLP vs. Trilateration}

\begin{figure}[htb]
    \centering
    \includegraphics[width=0.35\textwidth]{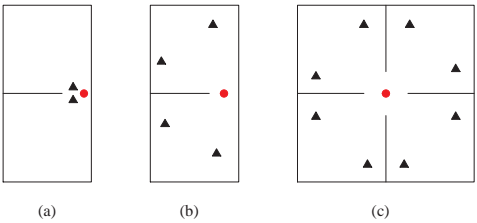}
    \caption{Trilateration requires many more lamps than MFLP does
    for a full coverage of the environment. Small triangles and red
    dots represent the lamps. In all cases, MFLP
    needs only a single lamp (red dot).}
    \label{fig:lateration}
\end{figure}

The trilateration method assumes a single sensor, with three (or
more) lamps as position references~\cite{pharos}. To reduce
unknowns, the sensing face is assumed to be placed horizontally.
Assume the sensor is located at $(x,y,z)$, and there are three
non-collinear lamps at distinct positions $(x_i, y_i, z_i), i =
0,1,2$, then the following system of equations can be established
with which one can solve for the solution

 \begin{equation}%\label{eq:rss-model}
    \frac{k}{d_i^3} \cdot |z-z_i| \cdot f_{\omega}\Big(\arccos \frac{z-z_i}{d_i}\Big) = s,
\end{equation}
where $d_i = \sqrt{(x-x_i)^2+(y-y_i)^2+(z-z_i)^2}, k>0$, and $s$ is
the RSS.
%$$\frac
%{(z-z_i)^2}{((x-x_i)^2+(y-y_i)^2+(z-z_i)^2)^2}
%= s, i = 0,1,2$$ %Trilateration requires the
%lamps should not be too close to each other or collinear. In the
%former case, the lamps may look to the sensor as if they were a
%single source of light, thus giving rise to large errors. In the
%latter case, the receiver will be confused as to which side of the
%line of lamps it is located, since there are two valid solutions at
%both sides of that line.

While simplifying the design at the receiver side, trilateration
shifts cost to the transmitter side. In theory, trilateration
requires three times as many lamps as MFLP does. In practice
however, the difference can be much higher.
Figure~\ref{fig:lateration}(a) shows a scenario where two adjacent
rooms share a wall and a relatively narrow gate. With trilateration,
every point needs to see at least three lamps. For a minimum
deployment cost, the three lamps could be deployed near the gate.
However, trilateration further requires the lamps not be close by or
collinear, or huge errors or position ambiguities may arise. This
makes it necessary to place at least two additional lamps in each
room (Figure~\ref{fig:lateration}(b)). Thus, trilateration ends up
using five times as many lamps as used by MFLP.
Figures~\ref{fig:lateration}(c) further shows a case where
trilateration needs nine times as many lamps. In a real-world
environment, obstructions may appear in different forms, but a
similar comparison can be drawn between the two approaches in terms
of light coverage and deployment density.

Therefore, when the positioning system is deployed from scratch and
lamp deployment cost is a primary concern, the MFLP approach appears
to be a more economical solution. On the other hand, when there
already exist dense lamps (as in some public places such as shopping
malls, airport terminals, etc) that can be re-used for positioning,
and when a small size of receiver is preferred, the trilateration
approach might be the choice. Since the trilateration approach is
well studied and understood, we shall concentrate on the MFLP
approach in the following sections.

\section{LIPS Receiver Design}\label{sec:lips-embedded}

In this section we describe a design for a dedicated LIPS receiver
based on the MFLP principle.

\subsection{Number and placement of light sensors}

\begin{figure}[tb]
    \centering
    \includegraphics[width=0.45\textwidth]{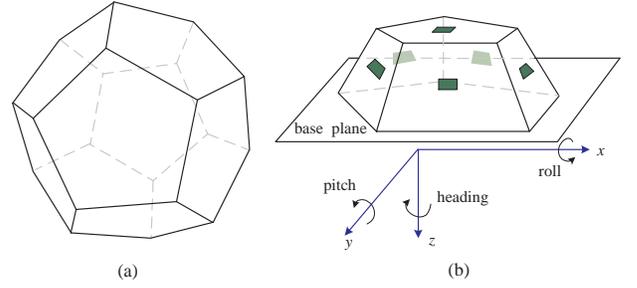}
    \caption{(a) Dodecahedron (conceptual model). (b) Half dodecahedron (implementation model)
    on which six light sensors, represented by green blocks, are installed.}
    \label{fig:dodecahedron}
\end{figure}

We say that a sensor face can `see' a point $p$ if there is a line
of sight between $p$ and all points on that sensing face. Although a
receiver can be positioned with only three sensors at particular
places, visibility of those sensors to a lamp may be lost when the
receiver is moving around. Thus, we need more than three sensors to
support positioning everywhere (assuming full coverage of light on
the ground). Toward that goal we need to answer two questions: how
many sensors do we need, and how to place them?

For implementation convenience, we focus on a regular polyhedron
framework on which the sensors are to be placed. Our choice is a
dodecahedron model, as shown in Figure~\ref{fig:dodecahedron}(a), in
which 12 regular pentagonal faces each host a sensor. This model
possesses three desirable properties:

\begin{enumerate}
  \item {\em Tri-face visibility}: Any point $p$ in the space beyond a short
  distance from the dodecahedron can see at least three faces of that
  dodecahedron;
  \item {\em Tri-face linear independence}: Any three of the faces
  seen by the above mentioned point $p$ are linearly independent;
  \item {\em Minimal faces}: Among all regular polyhedra, a regular dodecahedron
  has the fewest faces that satisfy the above two properties.
\end{enumerate}

Theorem~\ref{theo:tri-face} gives a more formal description of the
Tri-face visibility property. The theorem can be proved with basic
trigonometric operations and is thus omitted here.

\begin{theorem}\label{theo:tri-face}
Assume the edge length of a regular dodecahedron centered at the
origin is $a$, then for an arbitrary point at a distance $D$ from
the origin, at least three faces of the dodecahedron can see it if
$$D \geq \Big(\sqrt{1+\frac 2{5}\sqrt{5}} + \frac 1{2}\sqrt{\frac
5{2} + \frac{11}{10}\sqrt{5}}\Big)a \approx 2.49a.$$
\end{theorem}

In LIPS, $a$ is at the order of a few centimeters, so the
dodecahedron faces can be viewed as roughly passing through the
origin from the perspective of the lamp. Theorem~\ref{theo:tri-face}
means that if we place 12 sensors along the faces of a regular
dodecahedron, then the receiver can always be positioned, regardless
of the receiver's orientation.

The tri-face linear independence property can be easily verified by
examining the plane coefficients of the faces of a dodecahedron.
Finally, the minimal faces property can be proved by excluding the
regular polyhedra with fewer faces. For example, a cube-aligned
placement of sensors may be able to position a receiver sometimes,
as demonstrated in Section~\ref{subsec:linear-indep}, but it is easy
to pick a point in the space from which only a single face is
visible.

To further reduce the cost, we make a simplification to the
conceptual model by employing only a half of the dodecahedron, which
is fixed on a {\em base plane} (Figure~\ref{fig:dodecahedron}(b)) of
the receiver. This way we need only six sensors attached to the six
exposed faces. This does not affect the positioning ability as long
as lamps are hung above at ceilings and the half dodecahedron is
oriented upward.

\subsection{Sensing plane coefficients}

The coefficients of a sensing plane, namely $A$, $B$ and $C$, are
obtained from acceleration and magnetic sensors. Following the
aircraft convention, we use three attitude angles, {\em pitch}, {\em
roll} and {\em heading}, to describe a receiver centered coordinate
system. Figure~\ref{fig:dodecahedron}(b) shows the coordinate
system, in which $x$, $y$ and $z$ are defined as forward/right/down
based on the right-hand rule, and the three attitude angles are
referenced to the local horizontal plane which is perpendicular to
the earth's gravity.

\begin{figure}[tb]
    \centering
    \shortstack{
            \includegraphics[width=0.15\textwidth]{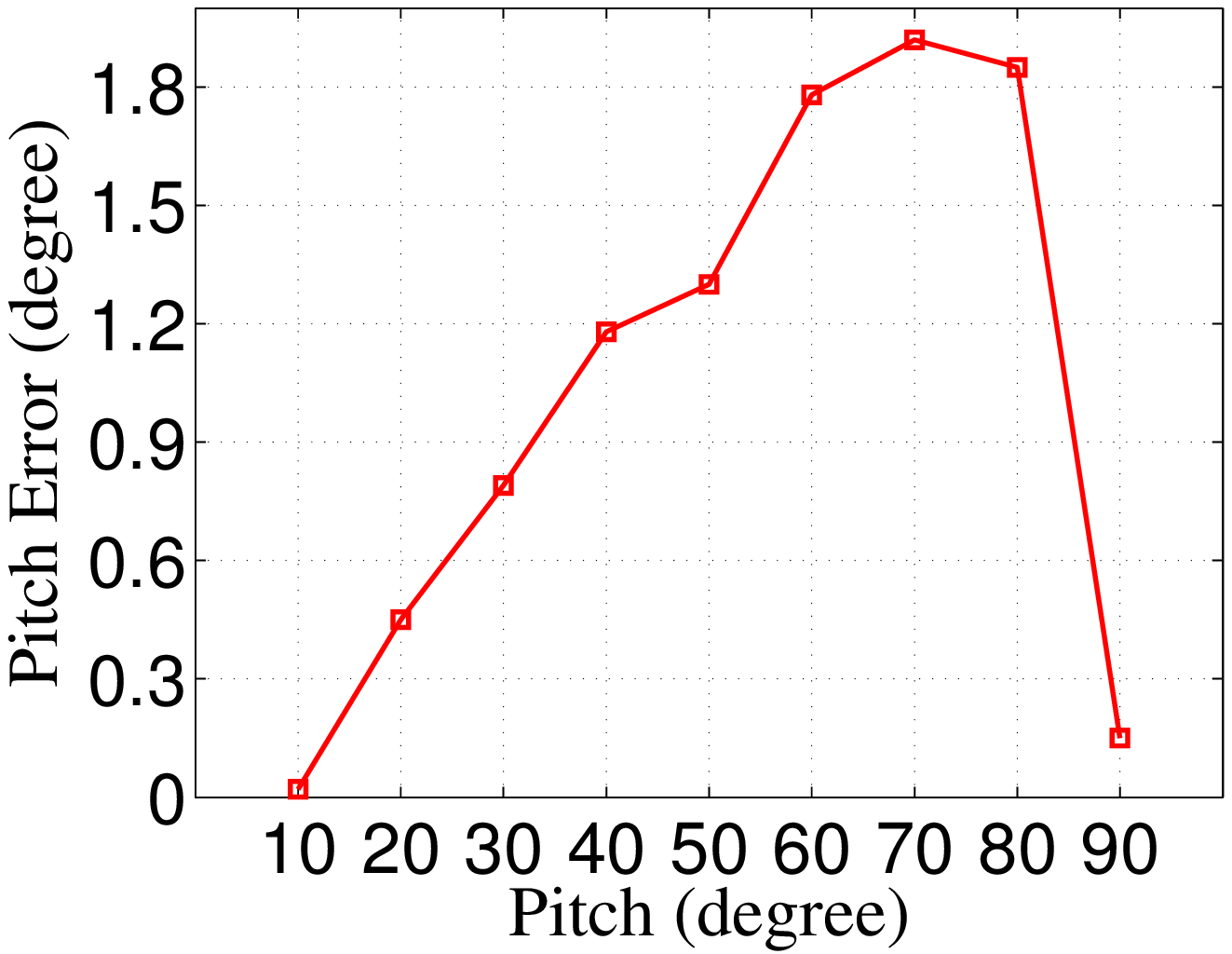}\\
            {(a)}
    }
    \shortstack{
            \includegraphics[width=0.15\textwidth]{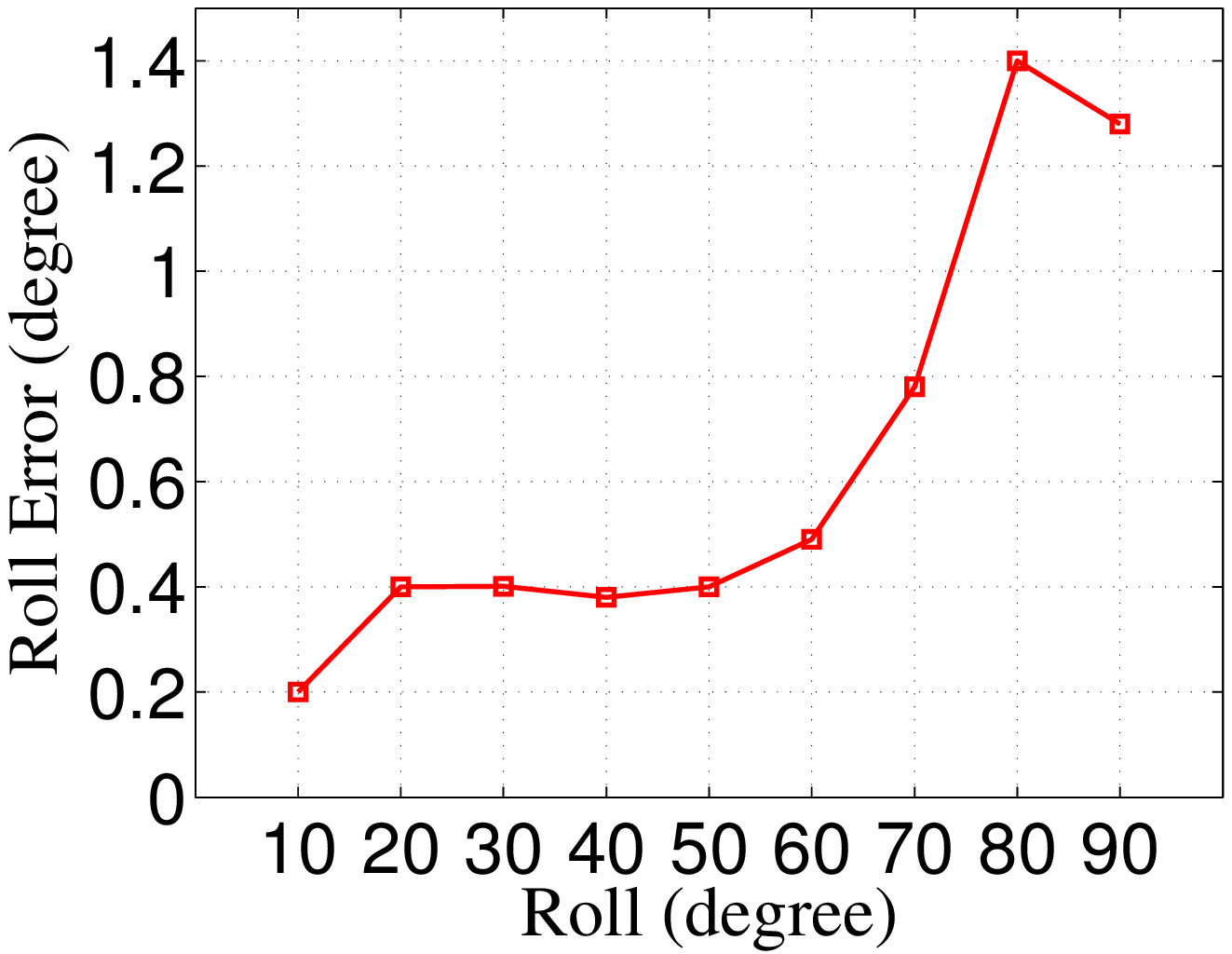}\\
            {(b)}
    }
    \shortstack{
            \includegraphics[width=0.15\textwidth]{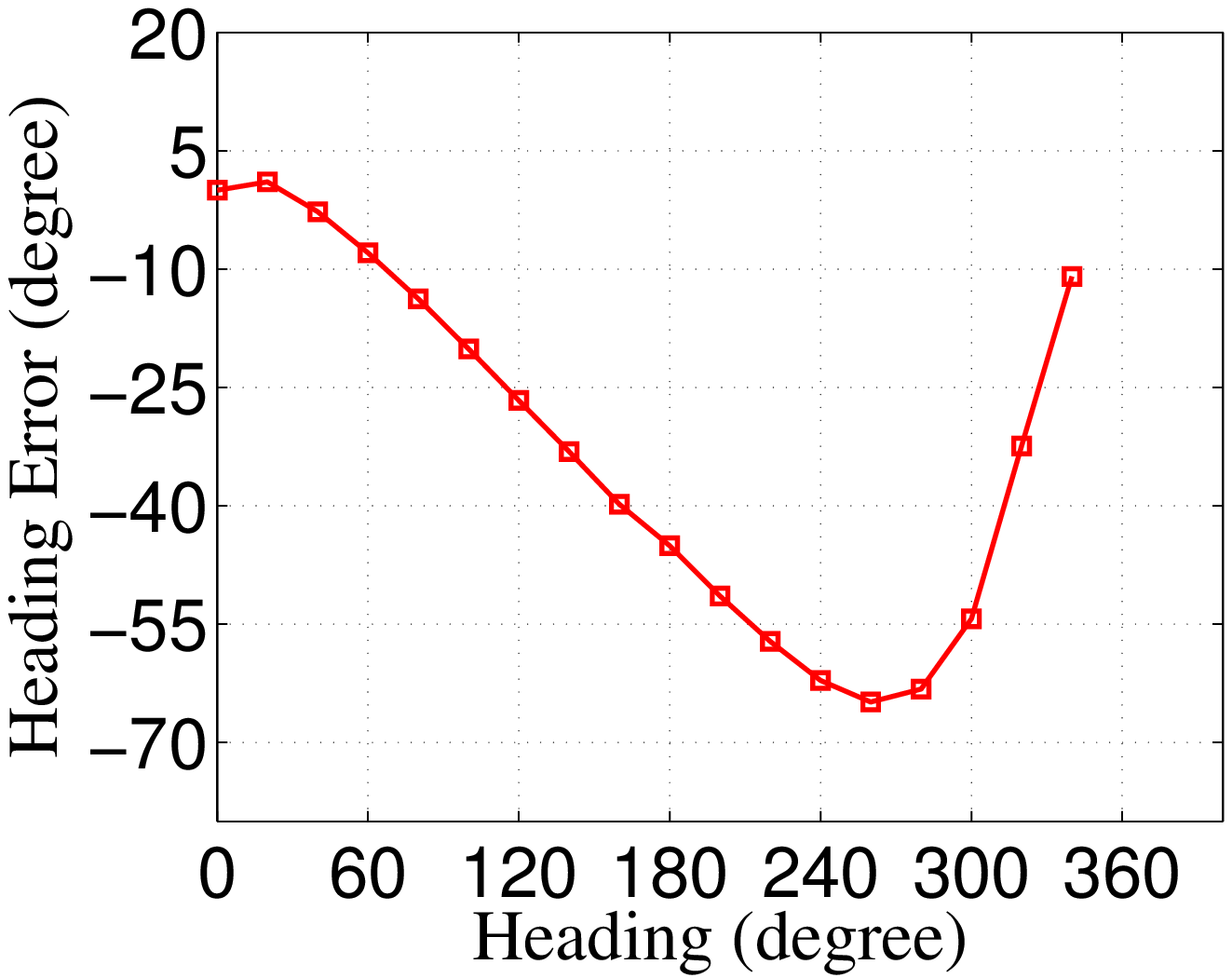}\\
            {(c)}
    }
    \caption{\label{fig:errors-angle} Errors of pitch, roll, and heading, produced by an acceleration sensor
    and a magnetic sensor.}

\end{figure}

Denote the pitch, roll, and heading by $\theta_p, \theta_r$ and
$\theta_h$. The first two angles can be obtained with an
acceleration sensor, which produces three components of the gravity
along the $x$, $y$, and $z$ axes. Comparing these components against
the acceleration of gravity can give the two angles.
Figures~\ref{fig:errors-angle}(a)(b) show that these two angles can
be measured with accuracy to 2 degrees. A standard way of obtaining
$\theta_h$ is using an electronic compass. However, a compass is
very susceptible to interferences. To confirm this, we rotate a
compass containing a magnetic sensor and obtain the heading error
with respect to a calibrated electronic compass.
Figure~\ref{fig:errors-angle}(c) shows that the raw measurement of
heading can vastly deviate from the true value, with errors up to 60
degrees. We discuss how to calibrate a compass in the next section.

The three angles $\theta_p, \theta_r$ and $\theta_h$ entirely
determine the orientation of the sensing plane, thus the
coefficients $A, B$ and $C$. We use the rotation
matrix~\cite{rotation-matrix} to convert the angles to $A, B$ and
$C$.

\begin{figure}[tb]
    \centering
    \shortstack{
            \includegraphics[width=0.23\textwidth]{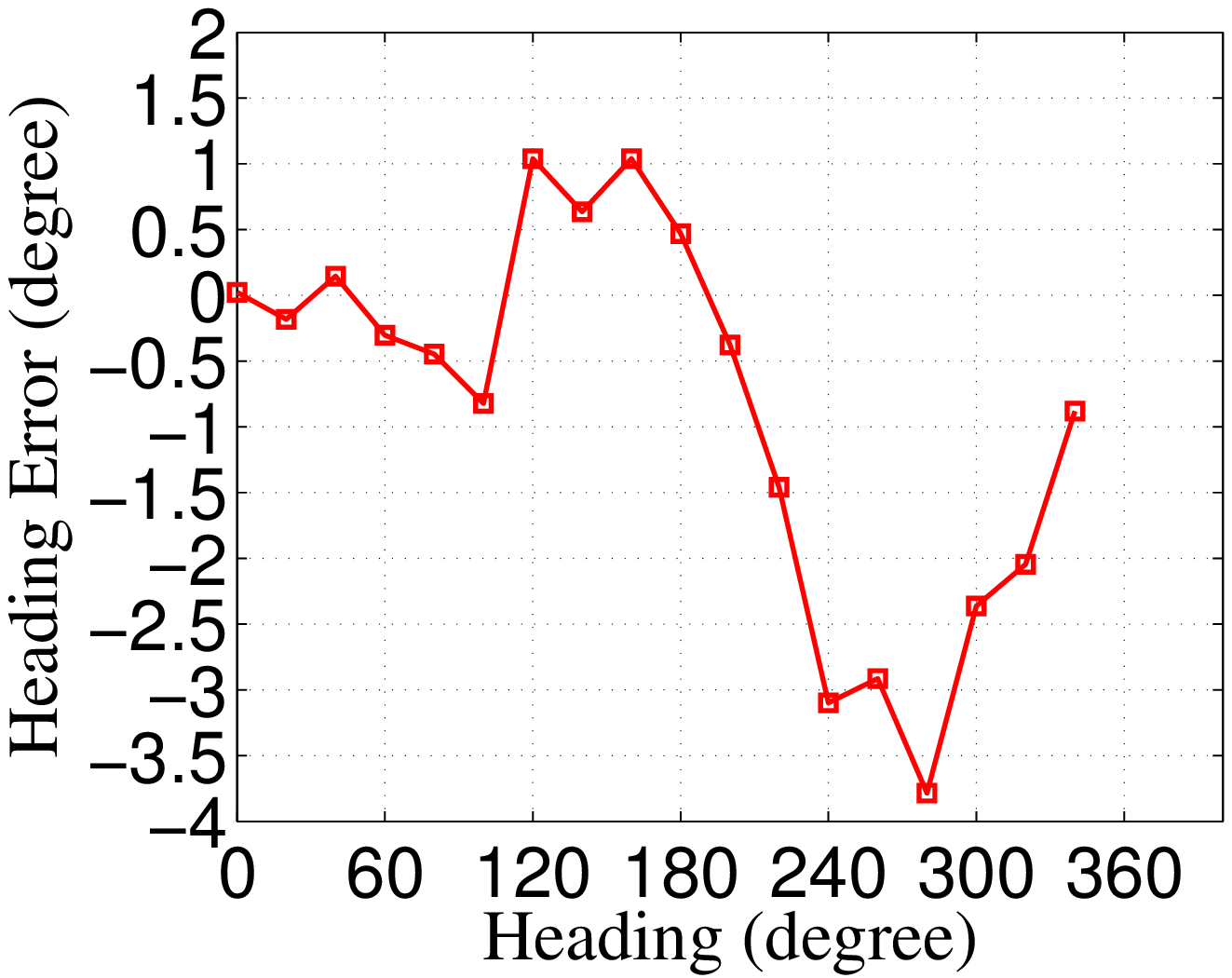}\\
            {(a)}
    }
    \shortstack{
            \includegraphics[width=0.23\textwidth]{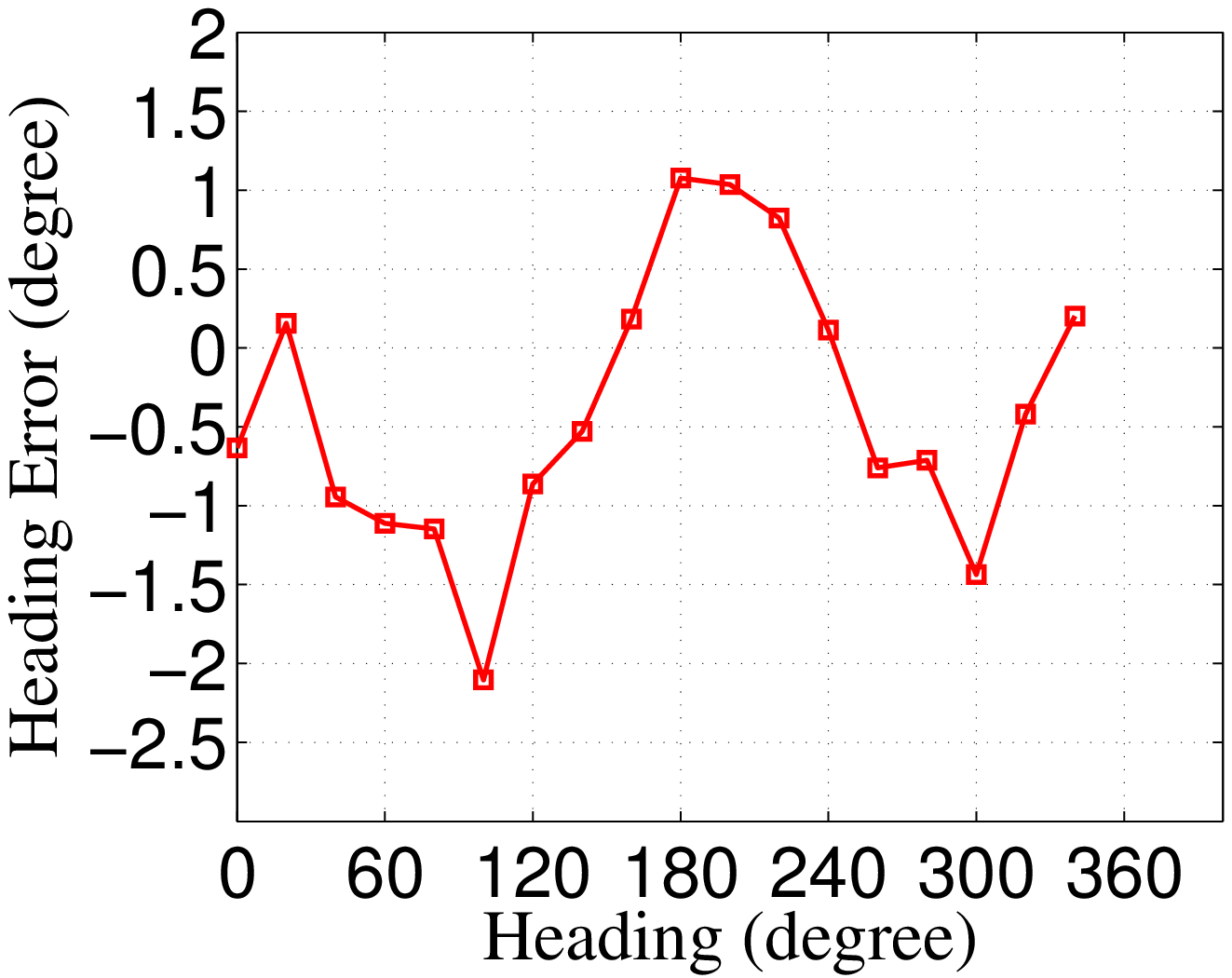}\\
            {(b)}
    }
    \caption{\label{fig:heading-angles} Heading errors under auto-calibration. (a) Half dodecahedron calibration. (b) Circular
    calibration.}
\end{figure}

%
%\begin{proof}
%If the edge length of a regular dodecahedron is $a$, the radius of a
%circumscribed sphere (one that touches the dodecahedron at all
%vertices) is $r_u = \frac{\sqrt{3}}{4}(1+\sqrt{5})a$. Furthermore,
%\end{proof}

\subsection{Accurate heading measurement}

Magnetic sensor is known to be vulnerable to environmental
interferences, because the Earth's magnetic field is a weak signal.
A mobile receiver carried around may experience distinct distortion
patterns at different locations. This property is exploited
by~\cite{magnetism} to enable receiver positioning, but causes
trouble to a system that needs accurate heading information in real
time. The general approach to auto-calibration of a compass requires
multiple types of sensors to provide redundant measurements to
correct heading errors. The redundant information can be from
optical trackers~\cite{auto-optical}, inertial
sensors~\cite{auto-gyro}, visual analysis~\cite{auto-visual}. In our
context, these techniques are unsuitable because they either involve
extra infrastructure or does not provide bounded errors in a long
period of time~\cite{auto-gyro}.

LIPS uses a new auto-calibration technique for heading measurement.
The idea is inspired by the standard manual calibration method, in
which sufficient 3D rotation or several full round 2D rotations are
performed to collect magnetic field strengths. Ideally the readings
along the three axes should form a sphere centered at the origin. In
practice, environmental interferences will distort the sphere,
resulting in a tilted ellipsoid~\cite{compass-note}. Given a set of
manually collected magnetic field data, the least square fitting
method can be used to discover the parameters of the ellipsoid,
which are then applied to correct errors of raw measurements.
%
%
%which can be described as the following equation:
%
%\begin{eqnarray*}
%\frac{(x-x_0)^2}{a^2}+\frac{(y-y_0)^2}{b^2}+\frac{(z-z_0)^2}{c^2}+\frac{(x-x_0)(y-y_0)}{d^2}
%& \\+\frac{(x-x_0)(z-z_0)}{e^2}+\frac{(y-y_0)(z-z_0)}{f^2}=R^2&,
%\end{eqnarray*}
%where $x_0, y_0, z_0$ are the offsets caused by hard-iron
%distortion, $x,y,z$ are magnetic sensor raw data, $a,b,c$ are the
%semi-axes lengths, $d, e, f$ are cross axis effect to make the
%ellipsoid tilted, $R$ is a constant of the earth's magnetic field
%strength.

LIPS avoids the manual operation by using multiple magnetic sensors
that are placed in a spherical or circular layout. These sensors can
produce a number of magnetism readings at once, thus provide a
sparse sampling of the needed magnetic data. We have experimented
with various numbers of sensors and found six sensors strike an
acceptable balance between measurement accuracy and cost. The
spherical layout is approximated with a half dodecahedron. In this
experiment, we manually collected the readings along the six faces
using the same magnetic sensor. Figure~\ref{fig:heading-angles}(a)
gives the heading errors with the half dodecahedron calibration at
the same position that produced the errors in
Figure~\ref{fig:errors-angle}(c). We can see that the original error
of 60+ degrees is now reduced to around 4 degrees.

The six magnetic sensors can also be placed in a circular pattern
when the receiver is placed horizontally. Before performing the
collective calibration, the various magnetic sensors are
individually calibrated to achieve a consistent effect by rotating
the board in an interference-free environment and collecting the
readings from individual sensors. On a horizontal plane, the
reference model of the magnetic field strength should be an ellipse
instead of an ellipsoid, which means that the fitting process deals
with fewer parameters. As a result, the heading accuracy after
calibration will be more accurate than with the half dodecahedron
calibration. Figure~\ref{fig:heading-angles}(b) shows that the
heading error has now drops to around 2 degrees.

\section{Prototype implementation} \label{sec:prototype}

We have implemented a dedicated LIPS receiver and a smartphone based
receiver. This section details the implementation.

\subsection{The CompEye receiver}\label{subsec:lips-dedicated}

\begin{figure}[tb]
    \centering
    \includegraphics[width=0.3\textwidth]{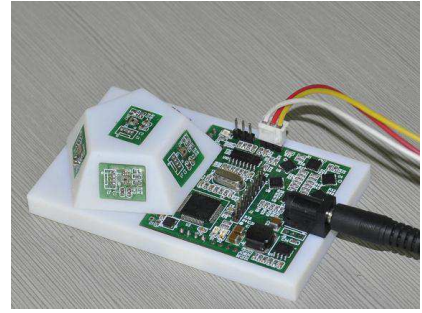}
    \caption{The CompEye receiver, sized 9.5 cm$\times$5.7 cm. The
    six light sensors are embedded in a half-dodecahedron model, and
    are connected to the main board from within the model.}
    \label{fig:lips-embedded}
\end{figure}

\begin{figure}[tb]
    \centering
    \includegraphics[width=0.28\textwidth]{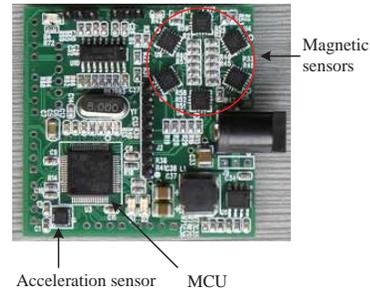}
    \caption{Main board of LIPS receiver, sized 5 cm$\times$5 cm.}
    \label{fig:board}\vspace{-10pt}
\end{figure}

Figure~\ref{fig:lips-embedded} shows the dedicated LIPS receiver,
called CompEye (compound eye), for position tracking applications.
The receiver consists of two parts: the main board, shown in
Figure~\ref{fig:board}, and the light sensing component. The former
is a circuit board integrating an STM32F103RC microcontroller (MCU),
six AKM8975 magnetic sensors, and an ST LIS33DE accelerometer. The
magnetic sensors are arranged along a circle for heading
calibration. The light sensing component comprises six ISL29023
light sensors fixed on the surfaces of a half-dodecahedron model;
each sensor is linked to the main board individually. In the
prototype design, the receiver is powered by an external battery,
and uses a wireless serial adapter for transmitting data to a
server. In the future, the power module will be replaced by a
lithium battery like one used by a mobile phone, and the WiFi module
will be integrated into the main board.

The MCU samples each light sensor at a rate of 640 Hz, and performs
FFT transformations to extract the light intensities of surrounding
light sources (identified by peaks in the frequency domain). The MCU
also samples each magnetic sensor at a rate of 20 Hz, and performs
calibration, first individually and then collectively, to obtain the
current heading of the receiver. The tilt and pitch are calculated
from the reading of the acceleration sensor. The MCU maintains a
sliding-window for each sensor, and performs the above calculations
every $\Delta$ seconds (e.g., $\Delta=0.3$), and sends the light
intensities, heading, title, and pitch to the sever. The sever
stores a digital map of the physical environment and of the deployed
lamps, with which it solves for the receiver's position using a
least square optimization algorithm.

\begin{figure*}[tb]
    \centering
    \shortstack{
            \includegraphics[width=0.25\textwidth]{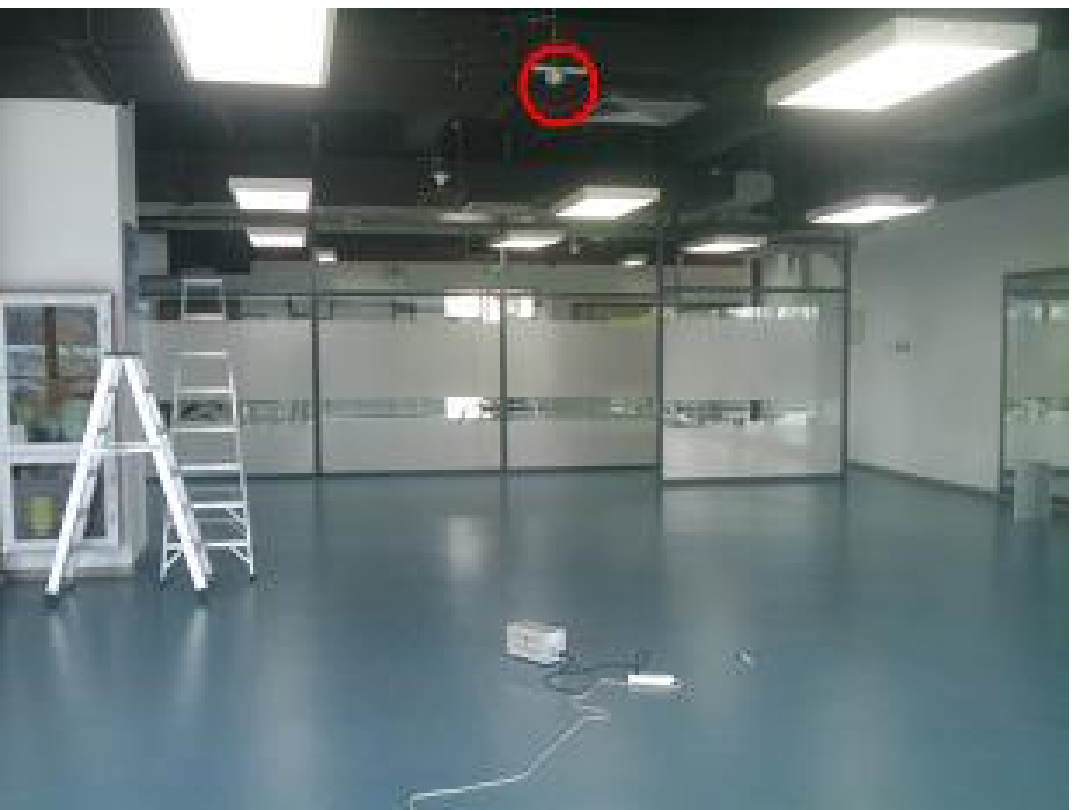}\\
            {{\small (a) Empty-room, 2nd floor, bldg. A}}
    }\quad\quad\hspace{12pt}%\quad
    \shortstack{
            \includegraphics[width=0.25\textwidth]{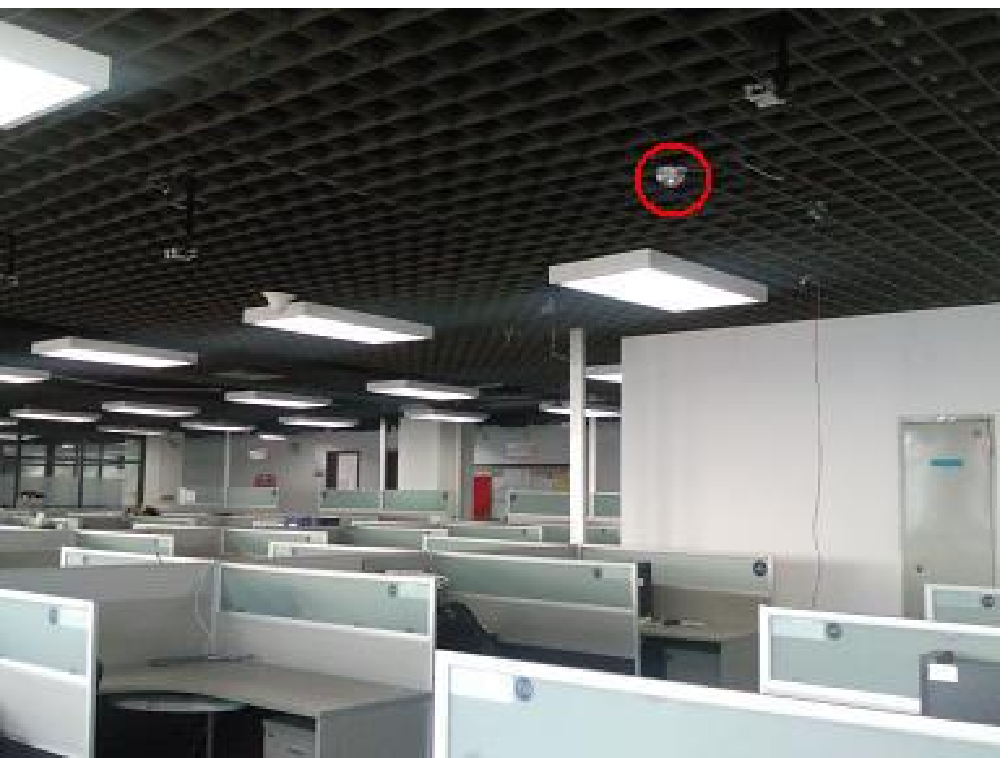}\\
            {{\small (b) Office, 9th floor, bldg. A}}
    }\\\quad\quad%\quad
    \shortstack{
            \includegraphics[width=0.25\textwidth]{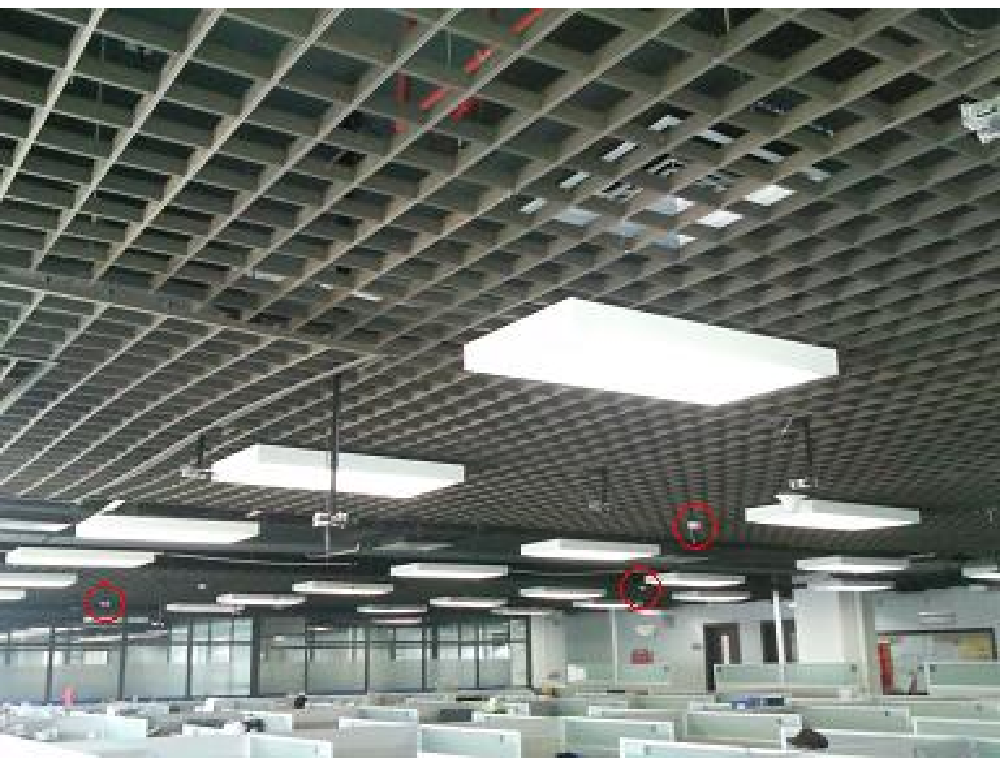}\\
            {{\small (c) Three-lamps, 9th floor, bldg. A}}
    }\quad\hspace{5pt}%\quad
    \shortstack{
            \includegraphics[width=0.25\textwidth]{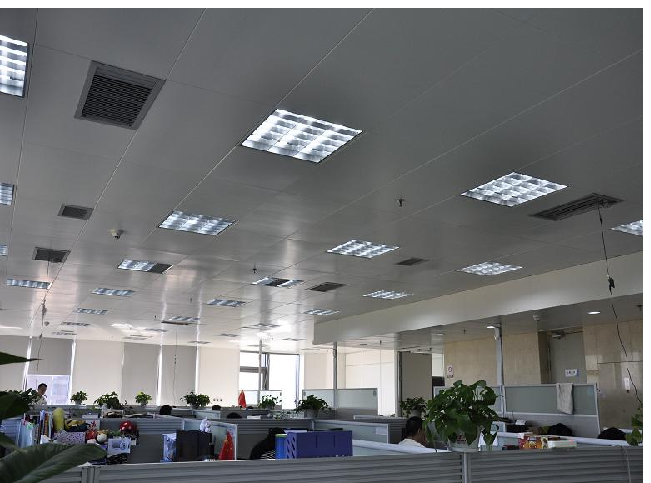}\\
            {{\small (d) Three-lamps-mobile, 8th floor, bldg. B}}
    }
    \caption{Experimental scenarios for LIPS prototype. (a) Empty-room: An empty
    room with a single IR LED lamp. (b) Office:
    An office environment with a single IR LED lamp.
    (c) Three-lamps: An office environment with three IR LED
    lamps flashing at rates 55 Hz, 65 Hz, and 75 Hz.
    (d) Three-lamps-mobile: Another office environment with three IR LED
    lamps flashing at rates 55 Hz, 65 Hz, and 75 Hz; the receiver is being moved.}
    \label{fig:scenarios}\vspace{-10pt}
\end{figure*}

\subsection{Smartphone}\label{subsec:lips-phone}

We used a Samsung Omnia II GT-I8000 smartphone to evaluate the
design. We installed an Android system with Linux kernel 2.6.32,
which allowed us to configure the light sensor in the driver to
sense infrared light instead of the default visible light.

After the configuration the phone can give correct readings of IR
intensity, as verified against a stand-alone sensor. However, under
periodical sampling, the sequence of readings produced by the
Android interface contains many uneven gaps in time, making the
frequency analysis difficult. This is because the OS kernel contains
a routine to smoothen sensed data and filter out readings with only
small changes. This treatment seriously affects frequency analysis.
We modified the {\small \verb"input_defuzz_abs_event()"} function in
{\small \verb"input.c"} to disable the smoothing procedure for the
light sensor, and re-built the kernel, after which the sensor
started working normally.

The smartphone can be positioned with the trilateration method,
since there is only one sensor. We evaluate its effectiveness in the
next section.

%The smartphone can be positioned with either the MFLP or
%trilateration method. With the MFLP method, the user needs to rotate
%the phone about the three axes, namely $x, y$ and $z$, to calibrate
%the compass for more accurate heading measurement. During the
%rotation, the light intensities as well as the pitch/roll angles can
%be collected from the Android API. These measurements provide
%necessary data for computing a position. However this approach does
%not work for continuous positioning applications, due to the
%requirement of manual operation. Therefore we focus on the
%trilateration method for a smartphone.

\section{Experiments}\label{sec:experiments}

%\subsection{Positioning with three faces or three Lamps}

We conduct experiments in three static scenarios and one mobile
scenario. We mainly consider IR LED lamps as they can be
incrementally deployed without interfering with the existing
lighting design of an environment.

\begin{enumerate}
  \item In this scenario, labeled {\em Empty-room}
(Figure~\ref{fig:scenarios}(a)), an IR LED lamp is hung on the
ceiling of an empty room, where there is little obstruction and the
floor is reflective. We selected 50 points of interest with rough
uniformity across the sensing area for test.
  \item The second scenario, {\em Office}
(Figure~\ref{fig:scenarios}(b)), is a crowded office environment
with dense cubicles, which create complex conditions for light
reflection. Again a single IR LED lamp was used, and 50 points, both
on the floor and on the desk, were selected for positioning.
  \item The third scenario, {\em Three-lamps}
(Figure~\ref{fig:scenarios}(c)), is similar to the second one,
except that there were three lamps used that flash at rates 55 Hz,
65 Hz, and 75 Hz. We deliberately made the lamps' sensing areas
overlap more than necessary to examine the effect of increased face
exposure to light signals. In this scenario, 98 points of interest
were chosen, with a bias to the overlapped areas. Throughout the
test, the receiver had its base plane placed horizontally on the
floor or the desk.
  \item The fourth scenario, {\em Three-lamps-mobile}
  (Figure~\ref{fig:scenarios}(d)), consists of three lamps placed in
  a similar layout as in the {\em Three-lamps} scenario. This
  experiment is conducted in a different building which creates more
  diversity of experimental conditions. The receiver is moved along
  three straight lines while position data is collected. We compare
  the collected traces against the ground truth to check the
  system's mobile performance.
\end{enumerate}

\begin{figure*}[th]
    \centering
    {\small
    \shortstack{
            \includegraphics[width=0.45\textwidth]{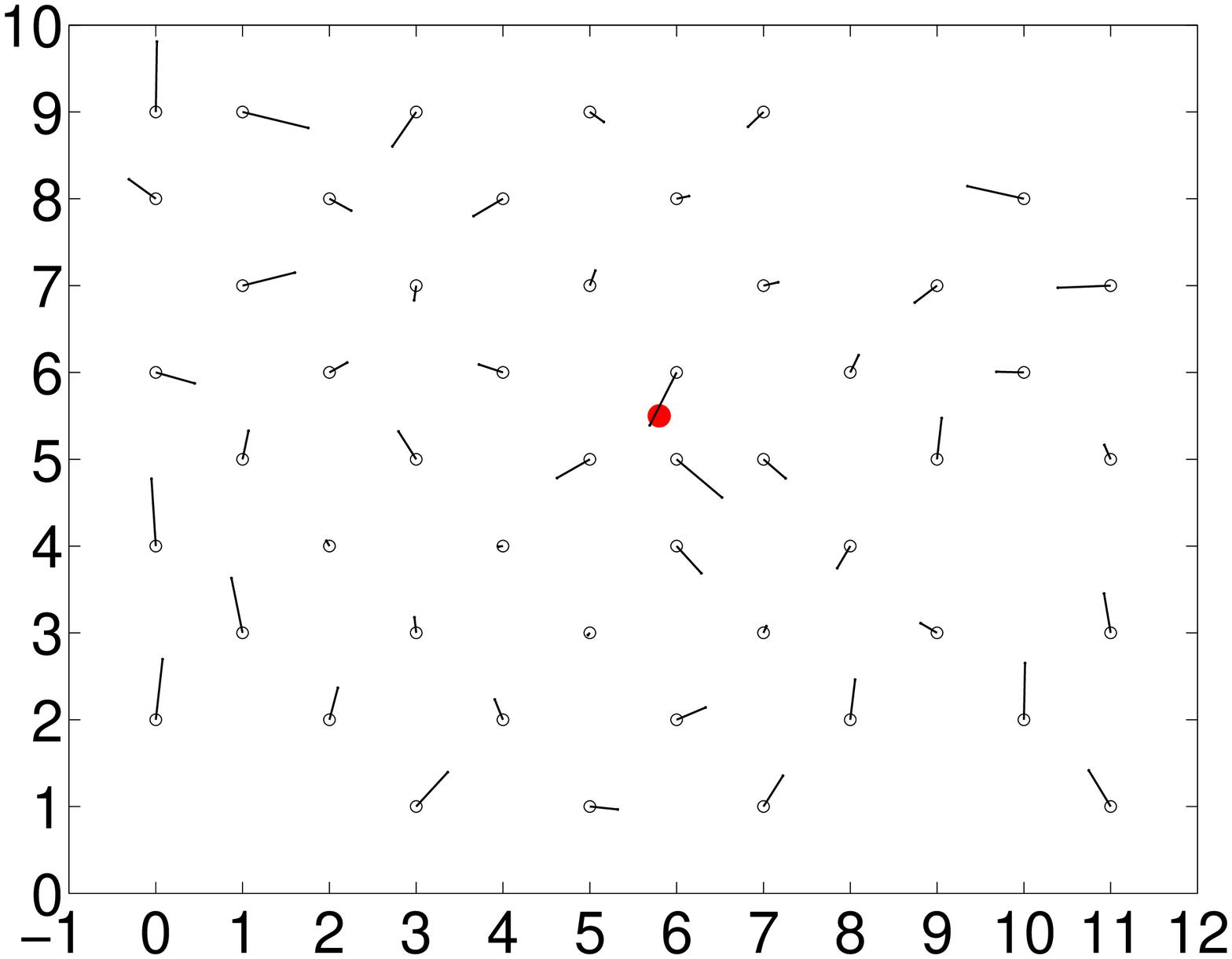}\\
            {(a) Empty-room}
    }\hspace{-10pt}
    \shortstack{
            \includegraphics[width=0.42\textwidth]{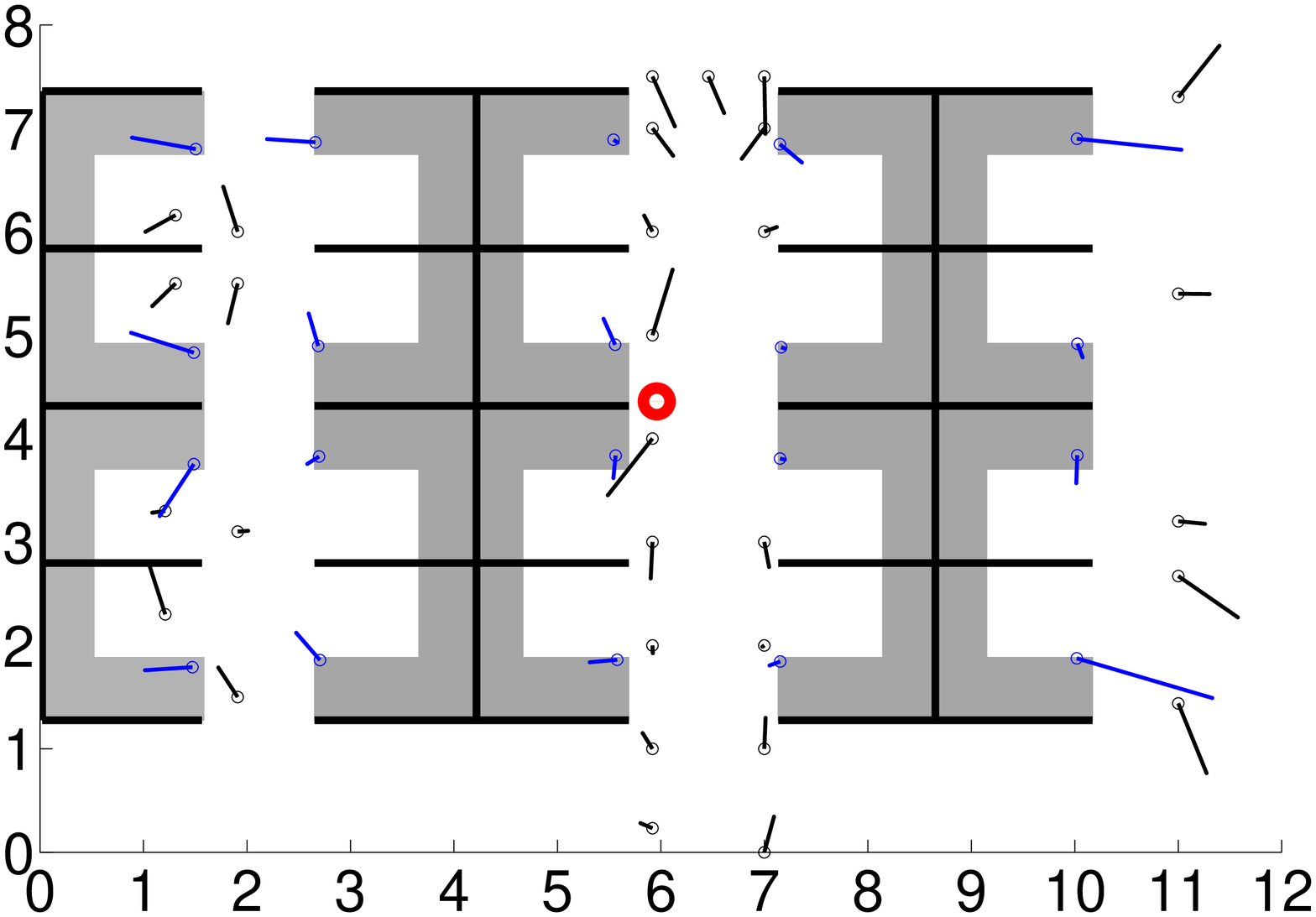}\\
            {(b) Office}
    }\\
    \shortstack{
            \includegraphics[width=0.4\textwidth]{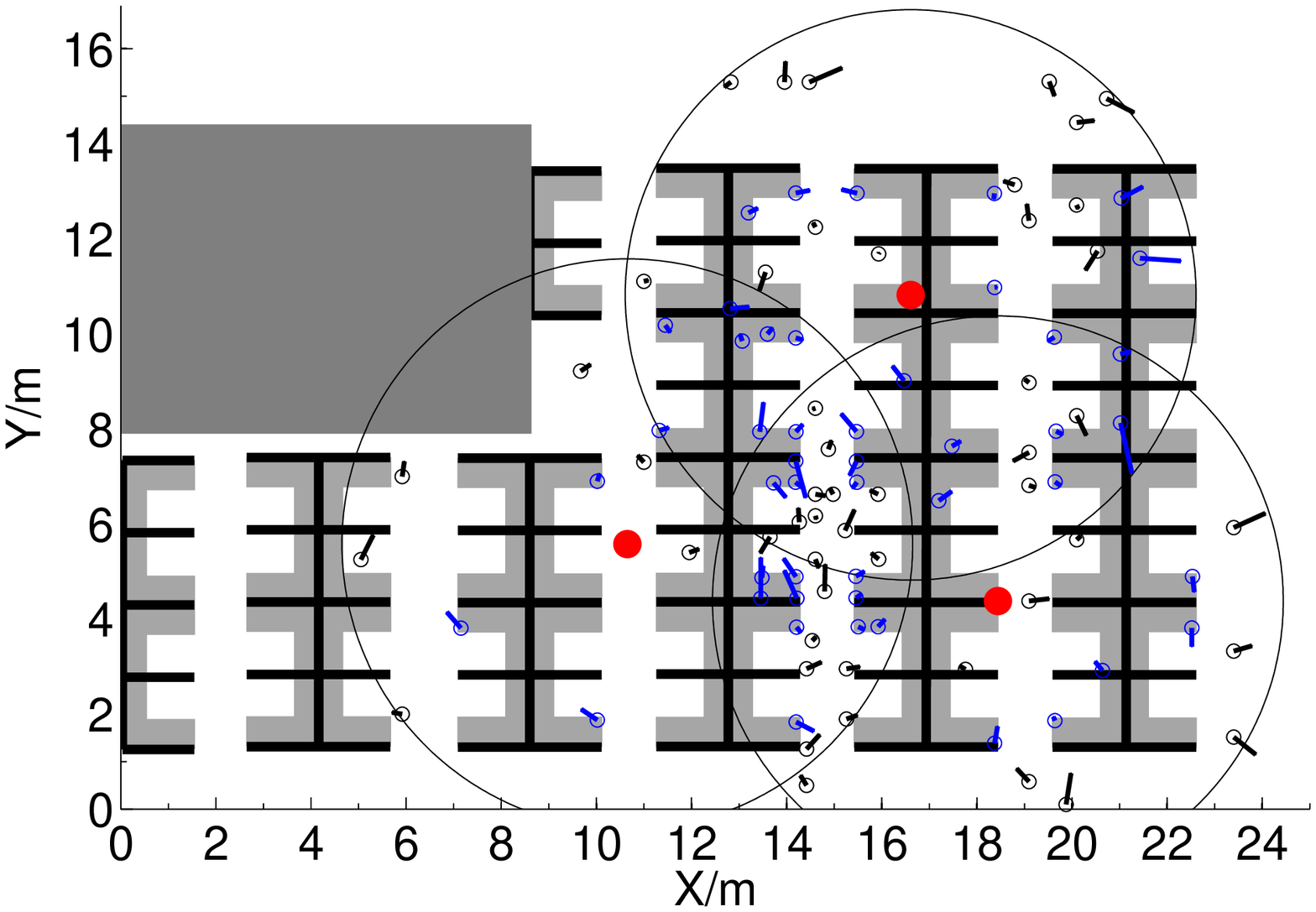}\\
            {(c) Three-lamps}
    }
    \shortstack{
            \includegraphics[width=0.4\textwidth]{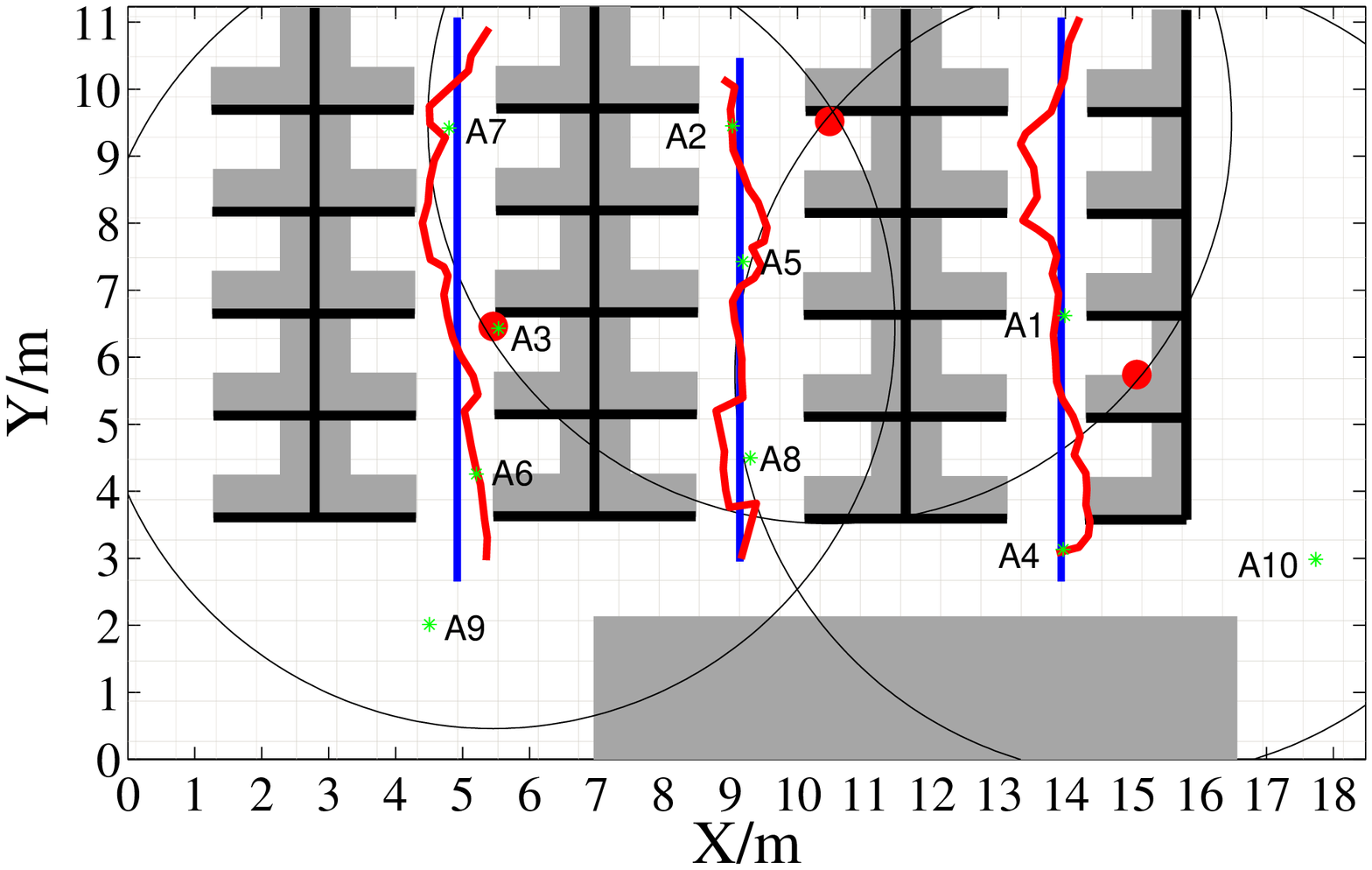}\\
            {(d) Three-lamps-mobile}
    }
    \caption{Positioning results of a CompEye receiver. Gray areas represent desks or
    obstacles, red dots lamps, and small circles test spots. (a) Empty-room, size 12m$\times$10m, (b) Office,
    size 12m$\times$10m, (c) Three-lamps, size 24m$\times$16m. (d) Three-lamps-mobile, 18m$\times$12m. }
    \label{fig:pos-errors}
    \vspace{-10pt}
    }
\end{figure*}

\begin{table} [h]
\centering{\small
\begin{tabular}{|l|c|c|c|c|} \hline

&\multicolumn{4}{|c|}{Error statistics (m)}\\ \hline\hline

Test cases  & Median & Mean & Max  & Stdev\\
\hline

Empty room, CompEye & 0.36 & 0.39  & 0.79 & 0.20\\ \hline

Office, CompEye  & 0.33 & 0.36  & 0.73 & 0.20\\ \hline

Three-lamps, CompEye  & 0.32 & 0.32  & 1.08 & 0.20\\
\hline

Three-lamps, smartphone  &  0.39 & 0.44  & 1.05 &  0.29\\
\hline

\end{tabular}
} \caption{Position error statistics in different scenarios. Note
that the smartphone case assumes a trilateration method, which
requires denser deployment of lights than the other cases.}
\label{tab:errors}%\vspace{-10pt}
\end{table}

\subsection{Position accuracy}

\vspace{5pt}\textbf{CompEye receiver.} When there is no obstruction,
a LIPS receiver can normally find at least three faces visible to a
lamp in range; normally it can find four. We have found that the
additional face generally does not improve the positioning result,
because of its small incident angle $\mu$ that produces a very low
RSS, potentially introducing increased errors from the tail of
$f_d(d)$. Therefore, for a particular lamp, we always choose the
three faces with the strongest RSS for position calculation. When
the receiver is covered by two or three lamps, it chooses the lamp
that imposes a stronger average RSS (over three faces) for
positioning.

Figures~\ref{fig:pos-errors} (a,b,c) show the maps of the various
environments as well as the positioning results. In the figures, the
red dots represent lamps, small dots true positions, and a line
segment connects a true position to its corresponding result of
positioning. The length of a line segment is thus proportional to
the positioning error. In all cases, the median and average errors
are below 0.4 meters (Table~\ref{tab:errors}). The consistence is
also reflected in the small standard deviation. Notice that quite
different environmental conditions, including floor/wall reflection,
ambient temperature and ambient light intensity (implied by time of
day), etc, are contained in these scenarios. These variations cause
unnoticeable difference in the positioning accuracy, suggesting that
LIPS is robust to environmental differences.

Figure~\ref{fig:pos-errors}(d) shows the mobile performance of the
LIPS receiver. In this experiment, the receiver is placed on a small
cart moving along the central lines of the three corridors. The
positions of the receiver are generated at an approximate interval
of 0.3 seconds. The red lines show the position traces of the
receiver; the cells are of size 0.6m $\times$ 0.6m. It can be seen
that the positioned traces stay within 0.6m of the true trace lines,
with an average much lower than 0.6m.

\vspace{5pt}\textbf{Smartphone.} We also experimented with the
trilateration method using a smartphone. A total of 33 points were
randomly picked with rough uniformity in the intersection area of
the three lamps. The phone was horizontally placed when collecting
the RSS. The error statistics are given in the last row of
Table~\ref{tab:errors}. It can be seen that the average error now
goes up to 0.44 meters, which is worse than those of the CompEye
receiver. This is mainly because of the absorbtion of light by the
phone screen and the blocking of phone body, especially when the
incident angle is small.

\subsection{Position stability}

It is normal for a positioning system to generate position
oscillation due to instability of signal propagation, environmental
interferences, and various errors in the system. The degree of
oscillation is an important impact factor for user's experience.
Though multiple position samples could be collected to increase
stability, in delay sensitive applications oscillation is still
likely to cause feelings of uncertainty and visual discomfort.

\begin{figure}[tb]
    \centering
    \includegraphics[width=0.35\textwidth]{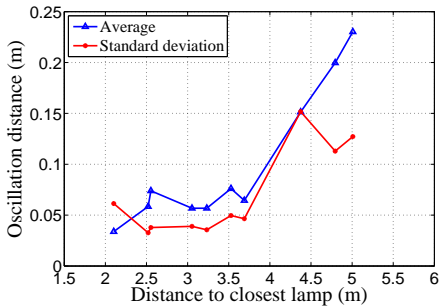}
    \caption{Oscillation distance of ten test spots in the Three-lamps-mobile scenario.}
    \label{fig:fluctuation}
\end{figure}

Given a spot in the field, we define the positioning system's {\em
oscillation distance} as the average distance of the produced
position samples from the centroid of those samples. In the
Three-lamps-mobile scenario, we choose 10 test spots A1, A2, ...,
A10 with different distance to their closest lamps, and for each
test spot, approximately 300 position samples are collected to
obtain an average and a standard deviation.
Figure~\ref{fig:fluctuation} depicts the results against the
distance to closest lamp. First, it can be seen that the averages
remain below 0.25m, indicating that the positions are quite stable.
Second, the oscillation distance generally increases with the spot's
distance to closest lamp, due to the increased signal/noise ratio at
a larger distance.

\begin{figure}[h]
    \centering
    {\footnotesize
    %\quad
    \shortstack{
            \includegraphics[width=0.23\textwidth]{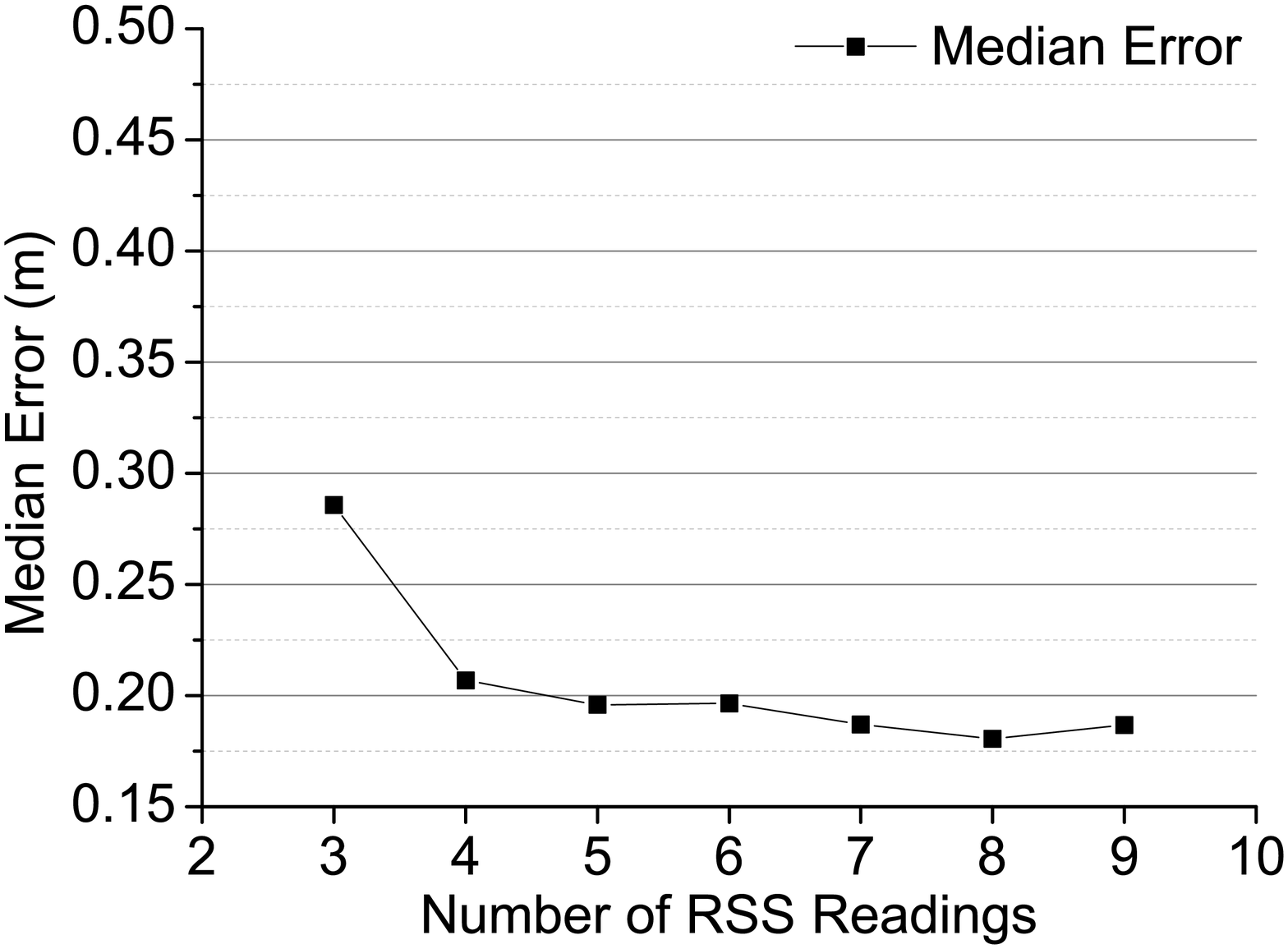}\\
            {(a)}
    }
    %\quad
    \shortstack{
            \includegraphics[width=0.23\textwidth]{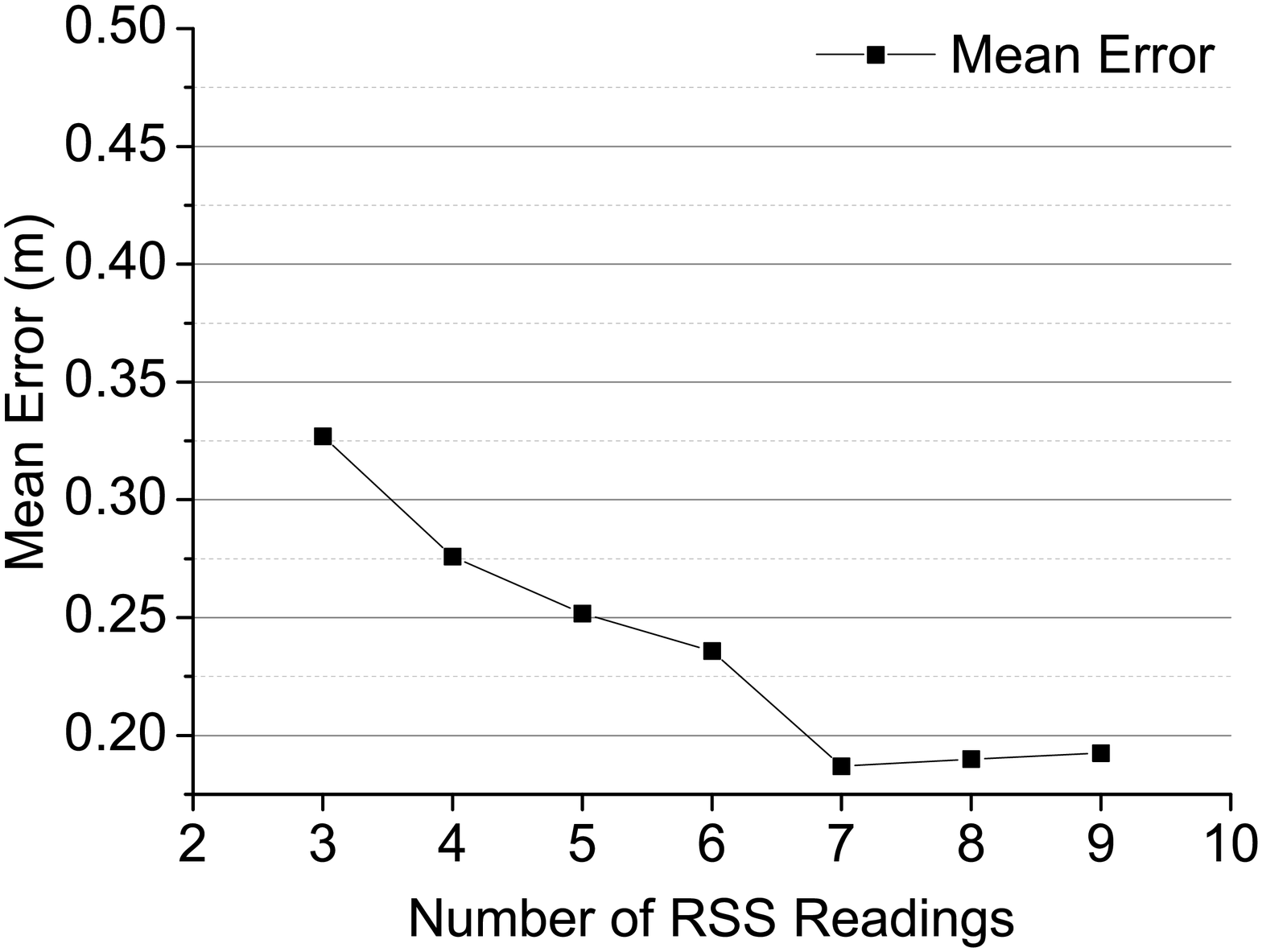}\\
            {(b)}
    }
    \shortstack{
            \includegraphics[width=0.23\textwidth]{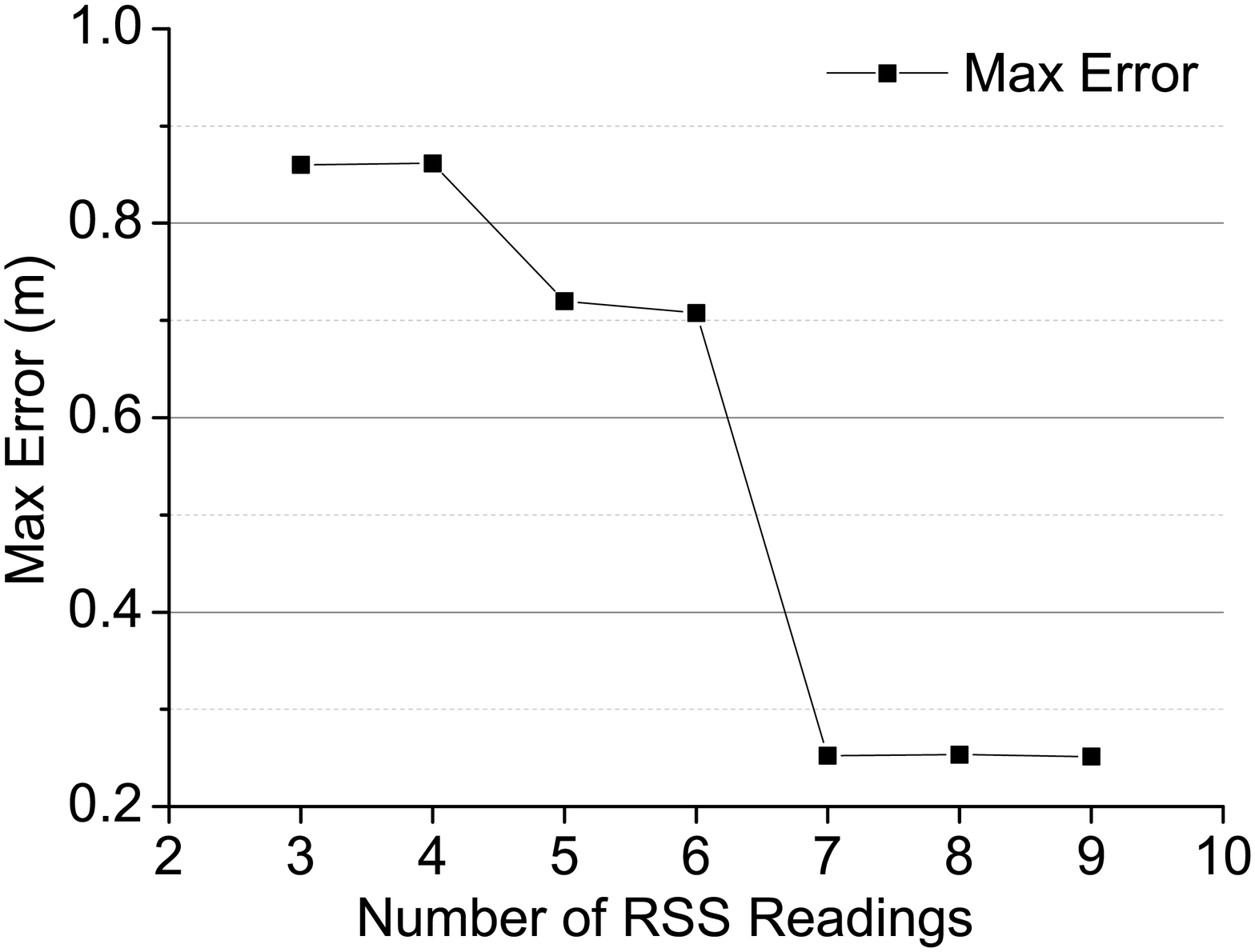}\\
            {(c)}
    }
    \shortstack{
            \includegraphics[width=0.23\textwidth]{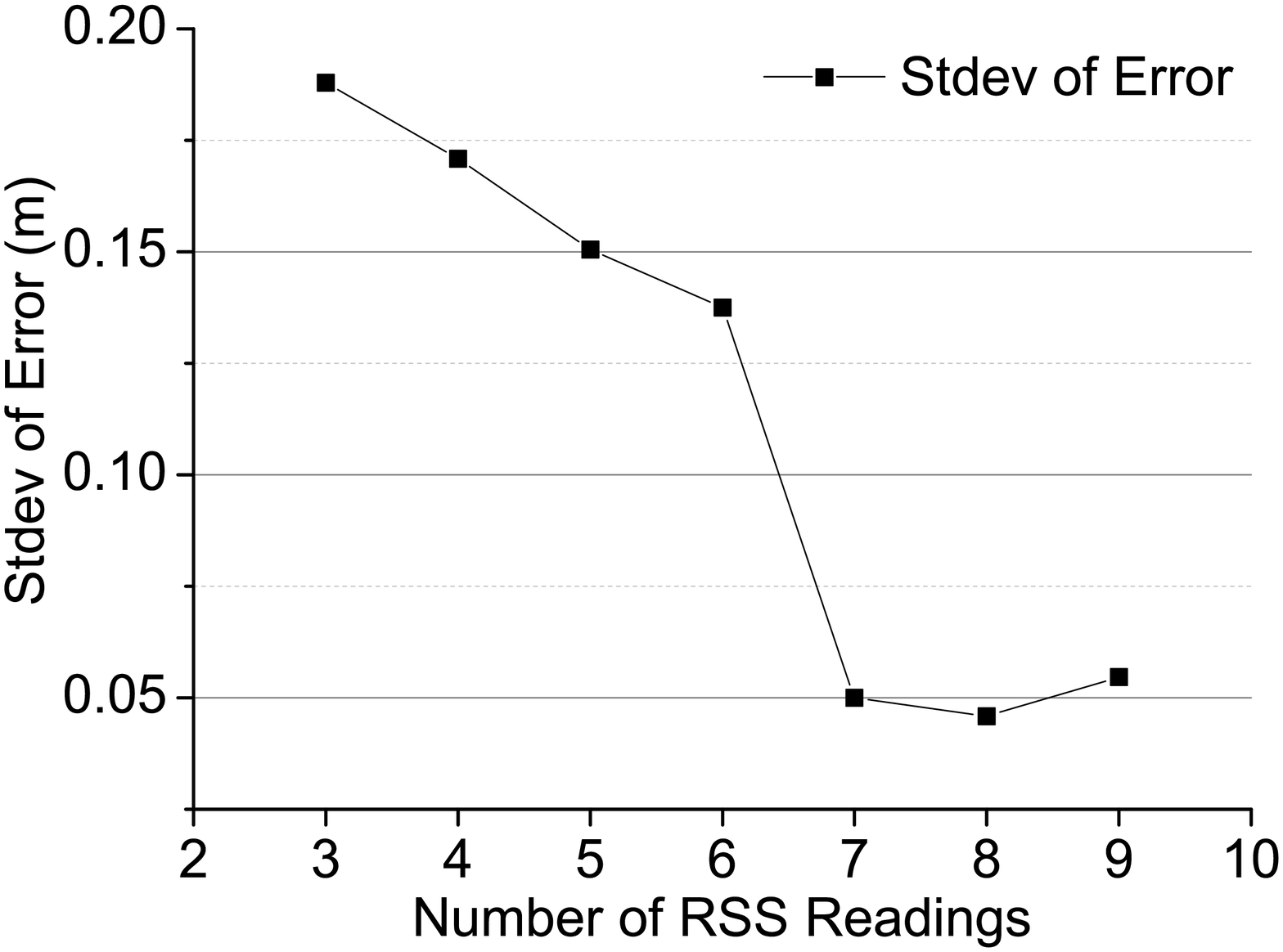}\\
            {(d)}
    }
    \caption{\label{fig:beyond-three} Positioning errors of the CompEye receiver for various
number of RSS readings.}
    }
\end{figure}
%
%\begin{table} [tb]
%\centering
%\begin{tabular}{|c|c|c|c|c|} \hline
%Num. of RSSs  & Median & Mean & Stdev & Max \\ \hline\hline
%
%3  & 0.29 & 0.33 & 0.19 & 0.86 \\ \hline
%
%4  & 0.21 & 0.28 & 0.17 & 0.86 \\ \hline
%
%5  & 0.20 & 0.25 & 0.15 & 0.72\\ \hline
%
%6  &  0.20 & 0.24 &  0.14 & 0.71 \\ \hline
%
%7  & 0.19 & 0.19 &  0.05 & 0.25 \\ \hline
%
%8 & 0.18 & 0.19 &  0.04 & 0.25 \\ \hline
%
%9 &  0.19 & 0.19 &  0.05 & 0.25 \\ \hline
%\end{tabular}
%\caption{Positioning errors of dedicated LIP receiver for various
%number of RSSs.} \label{tab:beyond-three}
%\end{table}

\subsection{Sensitivity analysis}

\vspace{5pt} \textbf{Impact of lamp density.} In the three-lamps
scenario, there are 98 points for testing, out of which 61 are
covered by a single lamp, 30 by two, and 7 by three. Thus, a
receiver may have more than three faces visible to a lamp, and a
sensing face may see multiple lamps. Let $m$ denote the number of
RSS readings a receiver collects. Since the receiver records only
three highest RSSs with respect to a lamp, and there are only three
lamps, we have $3 \leq m \leq 9$. In fact, each RSS will result in
an equation~\ref{eq:rss-model}, and we have already shown that three
such equations can lead to a position. Here we want to see if using
more than three RSSs will be beneficial. To that end, we pick $m'$
highest RSSs for each point $p$, where the receiver has $m \geq m'$,
and establish an system of $m'$ equations, by which we solve for
positions.

Figure~\ref{fig:beyond-three} shows how position errors are affected
by the number of RSS readings, $m$. It can be seen that in general,
increasing $m$ results in improved accuracy, reducing average errors
to below 0.2 meters. In particular, the maximum error and standard
deviation of error drop quite sharply with $m$, suggesting the
significant benefit of multiple lamps in reducing outliers and
improving positioning stability. The declining trend starts
flattening after $m>7$, implying a limiting accuracy of around 0.19
meters under the present RSS model.

Note that the case of more than three faces provides a
generalization of MFLP, which essentially mixes the principles of
MFLP and trilateration.

\vspace{5pt}\textbf{Error sensitivity.} In this test, we examine the
sensitivity of positioning quality to two main sources of error: RSS
error and heading error. We introduce artificial perturbations to
the measured data, and then calculate position errors. For each RSS
reading, a multiplicative perturbation $(1\pm\epsilon)$ is applied,
where $0 \leq \epsilon \leq 20\%$ is a parameter and the sign is
chosen randomly. For each heading measurement, an additive
perturbation $\epsilon_h \in [-10^{\circ}, 10^{\circ}]$ is applied.
In our experiments, the fluctuation of both measures was rarely
found to exceed half of the assumed ranges.

\begin{figure}[h]
    \centering
    {\footnotesize
    %\quad
    \shortstack{
            \includegraphics[width=0.23\textwidth]{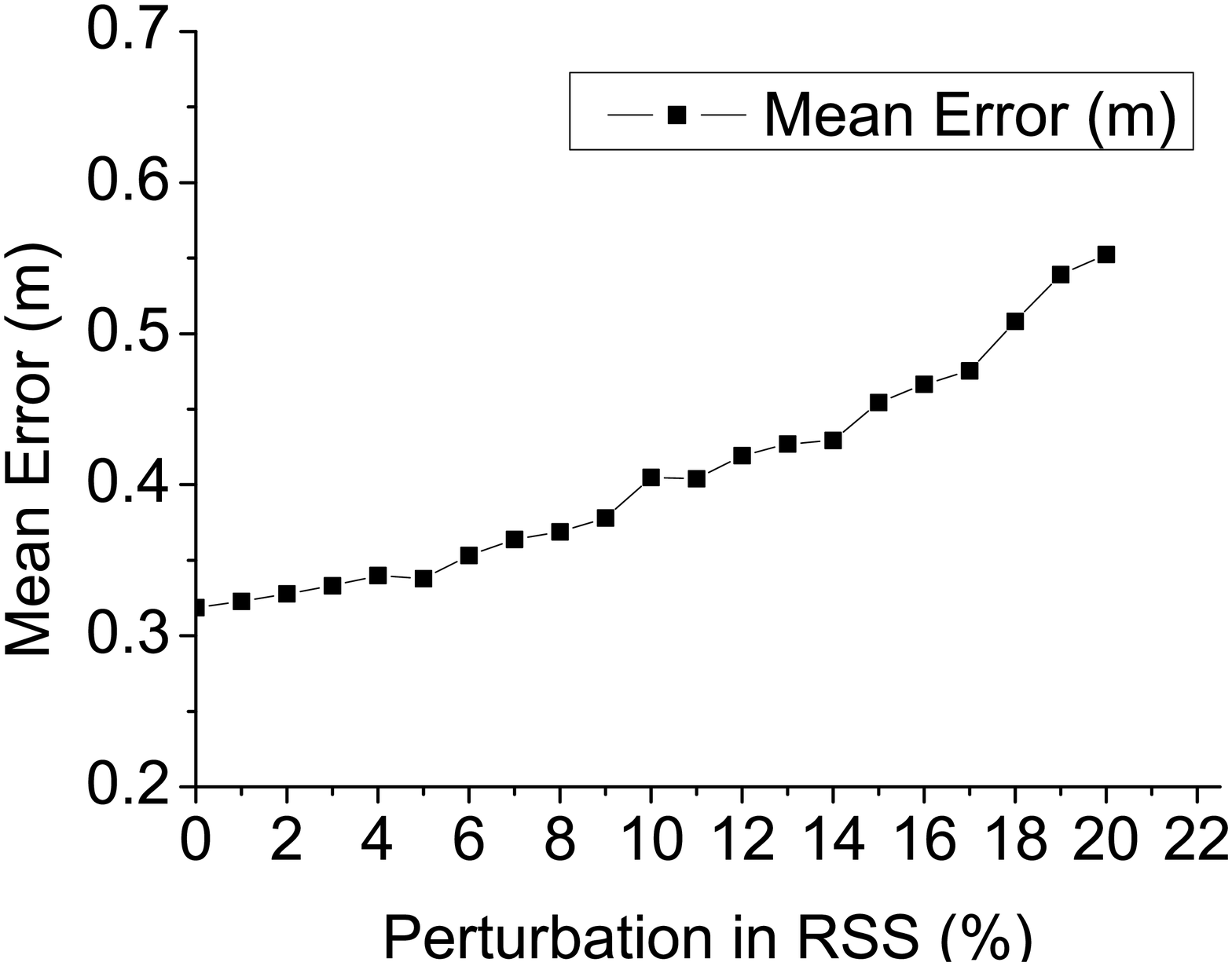}\\
            {(a)}
    }
    %\quad
    \shortstack{
            \includegraphics[width=0.23\textwidth]{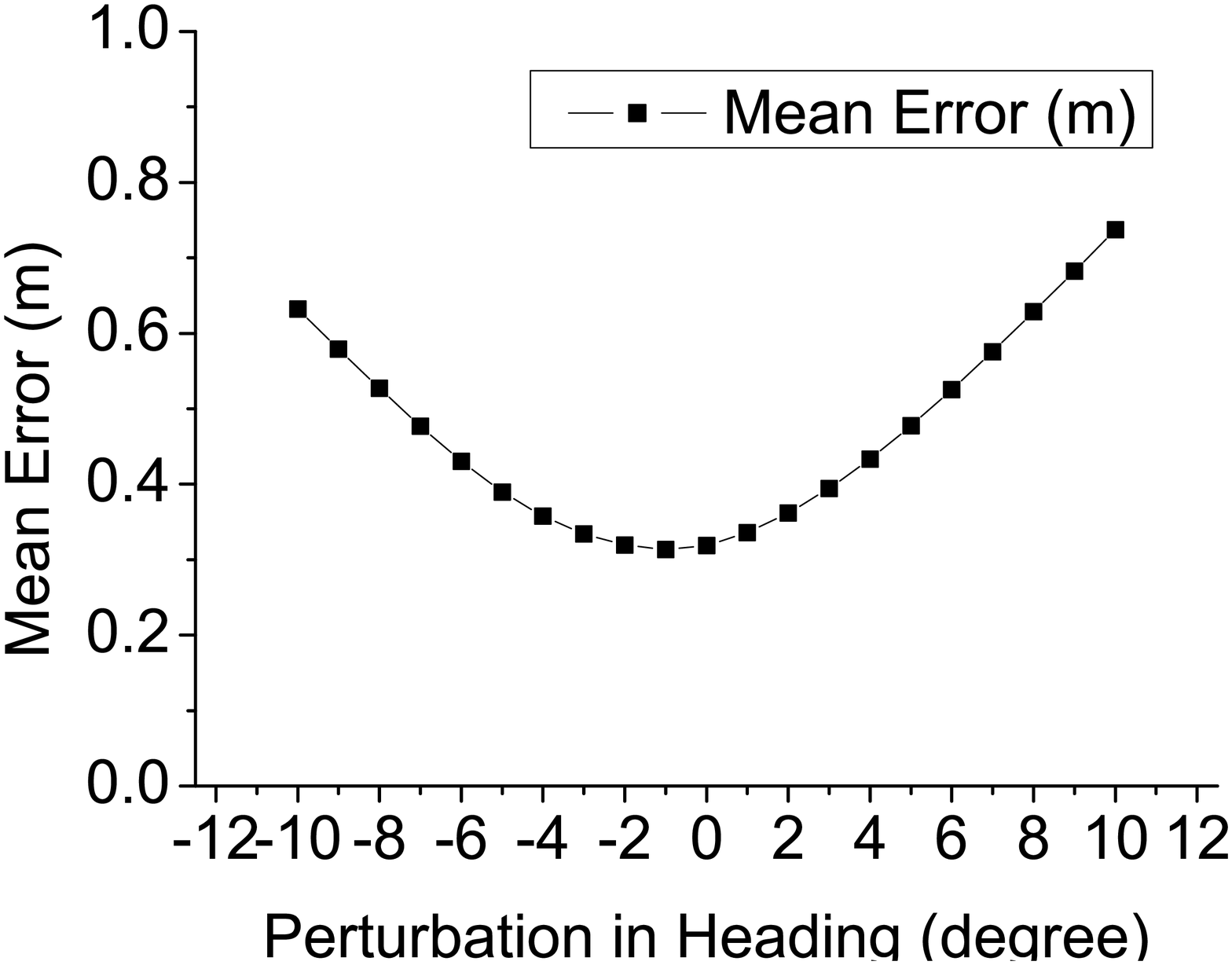}\\
            {(b)}
    }
    \caption{\label{fig:sensitivity} Position errors changing with perturbations to RSS and heading.}
    \vspace{-10pt}
    }
\end{figure}

Figure~\ref{fig:sensitivity} shows how positioning accuracy changes
with the perturbations in the Office scenario. As expected, the
average errors increase with larger perturbations. The increasing
trend, however, is smooth and mild, having average errors well under
1 meter even for unrealistically large perturbations.

\section{Discussion}\label{sec:discussion}

\textbf{Power consumption.} The LIPS receiver consumes power mainly
in three tasks: sensing, computation, and transmission. Each
ISL29023 light sensor has a power overhead of 70$\mu A\cdot$ 3.6 V =
0.25 mW in continuous sensing mode~\cite{isl29023}; each magnetic
sensor uses less than 0.2 mW at 20 Hz sampling rate, and a LIS33DE
accelerometer uses less than 1 mW ~\cite{LIS33DE}. Overall, the
sensing part accounts for less than 4 mW. The MCU and the networking
module normally have much higher power consumption rate. However,
their actual power consumption depends on the position sampling rate
as specified by the application. The computation and transmission
tasks can thus be performed with a low duty cycle to reduce power
consumption.

In terms of power overhead by the sensing task itself, LIPS consumes
merely 4 mW. It takes less than one second to obtain a relatively
stable position from a cold start, which amounts to about 4 mJ
energy. By contrast, a WiFi scan used by the WiFi-based positioning
system consumes 0.55 J~\cite{energy-trade-off}, which is two orders
of magnitude higher than LIPS's cost.

\textbf{Limitations.} The current design of LIPS system does not
support other forms of light sources. For example, our RSS model
does not apply to the widely used tube-shaped lamps. In principle,
the point-source based model should be extended to a line-segment
model. We have done some preliminary experiments with LED tubes, and
found that the RSS characteristics remain stable, though the
relations between RSS and distance/angle become more complicated.
This means that a geometry based positioning method is still
possible, though its mathematical foundation needs further
investigation.

Currently, the full scale range of an ISL29023 light sensor is set
to 1000 lux. In a typical indoor environment (e.g., office
lighting), the daylight's illuminance is less than 500
lux~\cite{Lux}, which is well within the sensor's capability.
However, the sensor quickly gets saturated under direct sunlight,
for example in the glass-ceilinged atria of some modern buildings,
making the receiver unable to position itself. This problem can be
mitigated by raising the sensing range (to a maximum of 64,000), at
the cost of reduced sensitivity or data resolution, which implies
reduced positioning accuracy. For higher adaptability while
retaining the accuracy, we will need to seek more powerful light
sensors to work for more challenging circumstances.

\section{Related work}\label{sec:related}

\vspace{5pt} \textbf{Positioning principle.} Most existing indoor
positioning systems follow one of three basic principles: proximity
detection, fingerprint matching, and trilateration/angulation. The
proximity detection approach~\cite{bluetooth,landmarc} localizes a
receiver simply with the positions of signal sources that can be
sensed by the receiver. Fingerprint matching~\cite{horus,radar}
further employs signal strength to obtain more accurate positions.
The last approach, trilateration/angulation, can produce highly
accurate positions when ranging/angulation ability is available on
the signal transmitter or receiver. Various methods such as
time-of-arrival (TOA), time difference of arrival (TDOA), angle of
arrival (AOA), and two-way sensing~\cite{push-limit}, can be used to
obtain distance or angle, based on which a position can be
calculated.

LIPS follows a principle similar to the third one, with an important
difference. In LIPS, the light sensor is simultaneous sensitive to
distance and incident angle of light signal, yet there is no way to
obtain either measure with the RSS alone, rendering the traditional
trilateration and angulation methods inapplicable. In LIPS, only
when several sensors' readings are collected can one solve for the
receiver's position, without explicit knowledge of distance and
angle.

Light signal based indoor positioning has been studied in a number
of previous works; see for example~\cite{bytelight,
vlc-positioning,vc-multi-optical,vlc-elec-letters,vlc-globecom}. The
visible light communication (VLC) technology~\cite{vlc-positioning}
uses cameras on mobile devices to capture flashing patterns from
programmed LED lamps, whose positions are then used to estimate the
user's position. The ByteLight solution~\cite{bytelight} appears to
follow a similar approach. The Pharos~\cite{pharos} system also uses
light signal and is based on the classic trilateration
method~\cite{pharos}. In~\cite{vc-multi-optical}, a positioning
system is designed that uses a single transmitter and multiple
receivers based on visible light produced by white LEDs. The
multiple receivers are independent and thus the positioning
principle behind fundamentally differs from ours.

%
%A general approach to positioning is to establish a one-to-one
%mapping between location and received pattern of signals from one or
%more sources. The pattern can be simply the existence of a
%signal~\cite{bluetooth,landmarc}, signal
%strength~\cite{horus,radar}, frequency
%characteristics~\cite{acoustic-background,surround-sense}, or other
%forms~\cite{magnetism}. More sophisticated methods try to obtain
%richer information from the received signals, such as distance and
%angle, using principles such as time-of-arrival (TOA), time
%difference of arrival (TDOA), angle of arrival (AOA), and two-way
%sensing~\cite{push-limit}. Based on the measured distance or angle,
%a multilateration or angulation~\cite{aoa} method can be used to
%compute a position of the receiver.

\vspace{5pt} \textbf{Infrastructure dependence and deployment cost.}
Early indoor positioning systems use dedicated devices such as
Bluetooth~\cite{bluetooth}, RFID~\cite{landmarc}, or sensor nodes to
realize positioning via proximity detection, but on the other hand
require a relatively dense deployment of signal transmitters. Higher
positioning accuracy is provided by more accurate ranging techniques
such as a combination of radio and sound signals~\cite{cricket},
Ultra Wide Band (UWB) technology~\cite{uwb-book}, etc. These
techniques require synchronization and coordination among signal
transmitters or expensive devices, which greatly increase the
overall cost.

A class of positioning techniques require little or no dedicated
infrastructure. They use ambient signals such as cellular
radio~\cite{celluar}, FM radio~\cite{fm-based},
magnetism~\cite{magnetism}, ultrasound~\cite{ultrasound} to create
position fingerprints. These solutions often provide only coarse
grained positioning, for example at room level
granularity~\cite{ultrasound}) or work for only special environments
(e.g., steel rich buildings~\cite{magnetism}). The most notable
technique in this class is the WiFi RSS fingerprint based
scheme~\cite{horus,radar} (the WiFi scheme). It exploits
pre-existing AP hotspots to create signal fingerprints. Depending on
the density and placement of APs, the positioning
accuracy varies. % from below 1 meter to more than 10 meters~\cite{empirical-self}. %Overall, this
%approach is appealing for its low cost and accessibility due to the
%widespread deployment of Wifi infrastructure. %When a relatively high
%accuracy (e.g., $<1$m) is desired while existing APs are cannot
%create sufficiently strong fingerprints, dedicated APs should be
%installed.

Infrastructure cost depends not only on hardware investment, but
also on human effort in site survey. The latter factor is a serious
concern for fingerprinting based solutions, represented by the WiFi
scheme. Recently researchers have proposed various schemes that use
little or zero explicit human effort~\cite{zee,lifs,
no-site-survey}. These techniques leverage inertial sensors on
smartphones to automatically infer a user's real position while
collecting RSS data. While they represent significant advances
toward low-cost deployment, the solutions are not generic, since
they require the environments to possess special structural
characteristics. When the indoor environment contains large free
spaces (e.g., in a factory or a large warehouse), or has a symmetric
layout, position ambiguity may arise which prevents exact positions
from being inferred from the user's trace.

LIPS uses off-the-shelf LED lamps as signal sources, with a cheap
microcontroller attached to each lamp to control flashing. On the
receiver side, a dedicated receiver uses a number of light and
magnetic sensors, each costing no more than a few US dollars.
%
%The proximity based positioning schemes~\cite{bluetooth} often
%achieve accuracy to a few meters, are are highly dependent on the
%density of signal emitters. Several UWB positioning
%systems~\cite{uwb-book} are reported to achieve an accuracy of 15
%cm, using dedicated and expensive signal emitters.

\vspace{5pt} \textbf{Accuracy and stability.} Different applications
require different levels of positioning accuracy. For example,
in-building pedestrian route guidance may work well with an accuracy
to a few meters, while automated handling requires positioning
accuracy within 1 cm~\cite{uwb-book}. The WiFi scheme mostly offers
an accuracy of 1 to 3 meters, and is known to be instable in
positioning quality, since the RSS at a fixed position varies
significantly over time, and is sensitive to obstacle presence,
receiver orientation, and type of device. Also the distinctiveness
of fingerprints depends heavily on AP density~\cite{limits}. It is
reported that the same scheme can produce drastically different
performance across different environments. For example, the classic
RADAR scheme is found to generate median accuracies of 1.3m and 5m
in a small and a large buildings, respectively~\cite{no-pain}. An
empirical study~\cite{empirical-self} shows that an algorithm can
yield 5$\times$ worse accuracy than in its original test
environment. LIPS is based on light intensity, which is relatively
stable once separated from ambient lights, thus the positioning
quality is better guaranteed than the WiFi scheme does.

\section{Conclusion}\label{sec:conclusion}

We have presented a light intensity based positioning system, {\em
LIPS}, for indoor environments. LIPS exploits ordinary lighting
devices such as LED lamps as signal transmitters, and uses light
sensors as signal receivers. Several light sensors on a receiver can
jointly determine the receiver's position with the measured RSS. The
main contribution of LIPS is that it explores a new way of indoor
positioning, with fairly high accuracy and high stability. The
design is fingerprint free, requiring little human intervention
other than the establishment of an RSS model. In the future we will
extend the design to more complex lighting environments.

\section*{APPENDIX A: Proof of Theorem~\ref{theo:unique}}

%\subsection{Proof of
%Theorem~\ref{theo:unique}}\label{app:proof-uniqueness}
%\vspace{15pt}
\begin{proof}
Consider three linearly independent sensing planes, $A_i x + B_i y +
C_i z = 0, 1 \leq i \leq 3$, that generate three nonzero RSS values
$s_1, s_2$ and $s_3$. Substituting these variables into
Eq.~\ref{eq:rss-model} we can get a system of nonlinear equations:
%\begin{equation}
%  \left\{
%   \begin{aligned}
%   \frac{A'_1x+B'_1y+C'_1z}{(x^2+y^2+z^2)^{3/2}} h\Big(\arccos \frac{z}{\sqrt{x^2+y^2+z^2}}\Big) = \frac{s'_1}{k}\label{EQ_1}\\
%   \frac{A'_2x+B'_2y+C'_2z}{(x^2+y^2+z^2)^{3/2}} h\Big(\arccos \frac{z}{\sqrt{x^2+y^2+z^2}}\Big) = \frac{s'_2}{k} \label{EQ_2}\\
%   \frac{A'_3x+B'_3y+C'_3z}{(x^2+y^2+z^2)^{3/2}} h\Big(\arccos \frac{z}{\sqrt{x^2+y^2+z^2}}\Big) =
%   \frac{s'_3}{k} \label{EQ_3}
%   \end{aligned}
%   \right.
%  \end{equation}%
{\small
\begin{numcases}{}
   \frac{A'_1x+B'_1y+C'_1z}{(x^2+y^2+z^2)^{3/2}} f_{\omega}\Big(\arccos \frac{z}{\sqrt{x^2+y^2+z^2}}\Big) = \frac{s'_1}{k}\label{EQ_1}\\
   \frac{A'_2x+B'_2y+C'_2z}{(x^2+y^2+z^2)^{3/2}} f_{\omega}\Big(\arccos \frac{z}{\sqrt{x^2+y^2+z^2}}\Big) = \frac{s'_2}{k} \label{EQ_2}\\
   \frac{A'_3x+B'_3y+C'_3z}{(x^2+y^2+z^2)^{3/2}} f_{\omega}\Big(\arccos \frac{z}{\sqrt{x^2+y^2+z^2}}\Big) =
   \frac{s'_3}{k} \label{EQ_3}
\end{numcases}
} where
$A_i'=A_i/\sqrt{A_i^2+B_i^2+C_i^2}$,$B_i'=B_i/\sqrt{A_i^2+B_i^2+C_i^2}$,
$C_i'=C_i/\sqrt{A_i^2+B_i^2+C_i^2},$ and $s_i' = s_i \in (0,s_m/k]$
or $s_i' = -s_i \in [-s_m/k, 0)$ depending on the symbol of
$A'_ix+B'_iy+C'_iz$. Performing dividing among these equations gives
{\small
\begin{numcases}{}
   (A'_1-\frac{s_2}{s_1}A'_2)x+(B'_1-\frac{s_2}{s_1}B'_2)y+(C'_1-\frac{s_2}{s_1}C'_2)z = 0
   \label{eq:eq3}\\
   (A'_1-\frac{s_3}{s_1}A'_3)x+(B'_1-\frac{s_3}{s_1}B'_3)y+(C'_1-\frac{s_2}{s_1}C'_3)z =
   0\label{eq:eq4}
  \end{numcases}
} It can be shown that Eq.~\ref{eq:eq3} and Eq.~\ref{eq:eq4} are
linearly independent, otherwise it can be verified that the vectors
$(A_i,B_i,C_i)$ are linearly dependent, which contradicts with the
assumption. With this linear independence, we can represent $x$ and
$y$ with $z$ as $x=c_1z$ and $y=c_2z$, where $c_1$ and $c_2$ are
functions of $s_i, A_i, B_i$ and $C_i$. Substituting them into
Eq.~\ref{EQ_1} and using the fact $z>0$ can solve for $z$, which
then gives $x$ and $y$. Thus we obtain a unique solution of
$(x,y,z)$.
\end{proof}

\end{document}